\documentclass[12pt,authoryear,sort&compress]{article}
\usepackage{amsmath}
\usepackage{rotating}
\usepackage{scalefnt}
\usepackage[doublespacing]{setspace}
\usepackage[ignorehead, nohead]{geometry}
\usepackage{fancyhdr}

\usepackage{floatrow}

\usepackage{calc}
\newcommand{\linesperpagedesired}{31}
\newlength{\newbaselineskip}
\AtBeginDocument{%
	\setlength{\baselineskip}{\newbaselineskip}%
}
\usepackage{nowidow}

\usepackage[protrusion=true, expansion=false, kerning=true]{microtype}
\usepackage[T1]{fontenc}
\usepackage[scaled = 1.105]{newtxtext}
\usepackage[scaled = 1.105, smallerops, varg, upint, slantedGreek, noamssymbols]{newtxmath}
\usepackage[scaled = 0.90]{DejaVuSansMono}  
\usepackage{floatrow}

\usepackage[utf8]{inputenc}
\usepackage{graphicx}
\usepackage{color}
\usepackage{array}
\usepackage{url}
\usepackage{breakurl}
\usepackage[breaklinks]{hyperref}

\hypersetup{
	colorlinks=true,
	linkcolor=MidnightBlue,
	citecolor=MidnightBlue,
	filecolor=MidnightBlue,
	urlcolor=MidnightBlue,
	bookmarks=true,
	bookmarksnumbered=true,
	bookmarksopenlevel=2,
	pdfstartview=Fit,
	pdfpagelayout=SinglePage,
	plainpages=false,
	pdfpagelabels=true
}
\newcommand{\email}[1]{\href{mailto:#1}{\nolinkurl{#1}}}

\usepackage[svgnames]{xcolor}

\usepackage{tikz}
\usepackage{wrapfig}
\usepackage[round, colon, authoryear]{natbib}

\providecommand{\U}[1]{\protect\rule{.1in}{.1in}}

\DeclareMathOperator{\Cov}{Cov}

\newcommand{\bfgamma}{\boldsymbol{\gamma}}

\newcommand{\bftheta}{\boldsymbol{\theta}}

\DeclareMathOperator{\E}{\mathbb{E}}

\DeclareMathOperator{\X}{\mathbf{X}}
\newcommand{\x}{\boldsymbol{x}}  

\def\dmath#1#2{
	$$\lineskiplimit=1000pt \advance\lineskip by #1\jot 
	\mathsurround=0pt \tabskip=0pt plus 1000pt
	\everycr{\noalign{\penalty\interdisplaylinepenalty}}
	\halign to \displaywidth{
		\hfil$\displaystyle{##}$\tabskip=0pt&%
		\hfil $\displaystyle{{}##{}}$\hfil &%
		$\displaystyle{##}$\hfil \tabskip=0pt plus 1000pt minus 1000pt&%
		\refstepcounter{equation}\label{##}\llap{(\theequation)}\tabskip=0pt\cr
		\noalign{\ifdim \prevdepth>-1000pt \vskip -#1\jot\fi}
		#2\crcr}$$}

\newdimen\dummy
\dummy=\oddsidemargin
\newcolumntype{C}[1]{>{\centering\arraybackslash}p{#1}}
\usepackage{subfigure}

\usepackage{caption}
\usepackage{tabularx}
\usepackage{makecell}
\usepackage{booktabs}
\usepackage{siunitx}
\newcolumntype{T}[1]{S[table-format = #1, table-space-text-pre = {***}, table-space-text-post = {***}]}

\usepackage{abstract}

\begin{document}

	\begin{titlepage}
		\title{ \vspace{-2.2cm}Mental Health and Abortions among Young Women: Time-Varying Unobserved Heterogeneity, Health Behaviors, and Risky Decisions}

		\author{ Lena Janys\thanks{University of Bonn (Department of Economics), HCM and IZA, ljanys@uni-bonn.de} \qquad Bettina Siflinger\thanks{Tilburg University (Department of Econometrics \& OR), Netspar, CESIfo, b.m.siflinger@uvt.nl \newline \indent\indent Lena Janys was funded by the German Research Foundation (DFG) under Germany's Excellence Strategy --EXC-2126/1--390838866, under Germany's Excellence Strategy --EXC-2047/1 --390685813 and under the individual fellowship with grant-number 441253219. Bettina Siflinger acknowledges support by the German Research Foundation DFG through the SFB 884. We thank Otilia Boldea, Marieke Bos, Pavel \v{C}{\'i}\v{z}ek, Jason Fletcher, Joachim Freyberger, Holger Gerhardt, Lukas Kiessling, Tobias Klein, Nikolaus Schweizer, Thomas Siedler, Miriam W\"ust, Nicolas Ziebarth and participants at workshops and seminars at the University of Bonn, Hertie School Berlin, Hamburg Center for Health Economics, Tinbergen Institute, University of Hannover, Virtual Mental Health Seminar (VMESS), ``The Importance of Early-Life Circumstances: Shocks, Parents and Policies'' (Copenhagen), European Health Econometrics Workshop (Leuven), World Congress of the Econometric Society (Milan), International Health Economics Workshop (Mainz), the IAAE (Rotterdam) and the Annual Health Econometrics Workshop (Emory) for helpful comments and discussions. We thank Statistics Sweden and Socialstyrelsen for the data, and Hans-Martin von Gaudecker, M{\aa}rten Palme, Lars Gullikson and Alexander Paul for their efforts to make them accessible.}}

		\maketitle	\thispagestyle{empty}	
		
		\begin{abstract}
			\singlespacing{		
		
		\noindent In this paper, we provide causal evidence on abortions and risky health behaviors as determinants of mental health development among young women. Using administrative in- and outpatient records from Sweden, we apply a novel grouped fixed-effects estimator proposed by \cite{bonhomme2015grouped} to allow for time-varying unobserved heterogeneity. We show that the positive association obtained from standard estimators shrinks to zero once we control for grouped time-varying unobserved heterogeneity. We estimate the group profiles of unobserved heterogeneity, which reflect differences in unobserved risk to be diagnosed with a mental health condition and analyze mental health development and risky health behaviors other than unwanted pregnancies across groups. Our results suggest that these are determined by the same type of unobserved heterogeneity, which we attribute to the same unobserved process of decision-making. We develop and estimate a theoretical model of risky choices and mental health, in which mental health disparity across groups is generated by different degrees of self-control problems. Our findings imply that mental health concerns cannot be used to justify restrictive abortion policies. Moreover, potential self-control problems should be targeted as early as possible to combat future mental health consequences. 
		
\textbf{JEL-Codes:}: I12, I10, C23, D91
			} 
				
		\end{abstract}
		
	\end{titlepage}

\doublespacing
	
\section{Introduction}\label{sec:intro}\vspace{-0.5cm}
In recent years, economists have increasingly paid attention to mental health problems and their consequences, especially when occurring during adolescence and young adulthood \citep{biasi2018career, cuddy2020rules}. Mental health problems are often first diagnosed in early adulthood and are very pervasive, in particular among young women \citep[see][]{eatonprev}. In 2017, about 13--19\% of adolescents between 15-25 in the US experienced at least one major depressive episode \citep{depressionUS2017}. As pointed out by \cite{currie2020child} mental health problems can reflect deficits in non-cognitive skills that are crucial for human capital development and labor market outcomes in adulthood. Thus, knowing about potential determinants of mental problems is of first-order importance. 

{One possible determinant that is often discussed in connection with mental health problems is abortion. In the US, abortions for women aged 15-24 years account for almost 40\% of all abortions in 2017 \citep{cdcabortion2016}. As pointed out by \cite{reardon2018abortion} abortion is \emph{consistently associated} with elevated rates of mental illness compared to women without a history of abortion. While there are different perspectives on the interpretation of this association, there is hardly any evidence for a causal relationship. Yet, in many countries, the association between abortion and mental health seems to be sufficient for politicians to justify restrictions on abortion access such as waiting times, mandatory disclosures, or parental consent laws \citep{guttmacher2020}. }

This paper investigates the impact of having an abortion from an unwanted pregnancy on the incidence of mental health conditions in young women in Sweden. We use individual-level administrative panel data that includes the universe of inpatient and outpatient contacts with the healthcare system, including general practitioners and specialists. While most studies on mental health rely on inpatient records or prescription drug data as a proxy for diagnoses, our records contain detailed information on mental health diagnoses and abortions, thus providing a comprehensive picture of the prevalence of mental health conditions and abortions from unwanted pregnancies in the population. {Our primary measure of mental health is diagnoses on mood disorders which mainly consist of diagnoses on depression. We also analyze anxiety and fear-related disorders as important dimensions of mental health problems.} 
 	
In the absence of any policy variation in abortion legislation, identifying a causal effect is challenging. Traditional estimators using within-person variation such as event-study or individual-specific fixed-effects assume that individual unobserved heterogeneity is time-constant. In our application, this seems too restrictive, as it neglects that selection into abortion is dynamic. To address this issue, we use a grouped fixed-effects estimator, henceforth GFE, proposed by \cite{bonhomme2015grouped}. The basic idea of the GFE estimator is that individuals who share similar unobserved characteristics are clustered in groups. Within these groups, unobserved heterogeneity can vary with age, with no further restrictions on the functional form of these unobserved heterogeneity trajectories.  
	
We compare the results from the individual-specific fixed-effects (OLS FE) and the GFE estimator. The estimated OLS FE-coefficient for abortion is positive and highly statistically significant. By contrast, we estimate a precise zero effect of abortion on mental health diagnoses when using the GFE estimator. The significant difference in the estimated coefficients stresses the importance of accounting for time-varying unobserved heterogeneity \emph{in addition} to individual-specific time-constant fixed-effects. We also compare the identifying assumptions of the Differences-in-Differences (DiD) estimator under random treatment assignment with those of the GFE estimator, showing that the assumptions are not nested. Thus, the choice of estimator depends on the particular application.\footnote{Our estimated unobserved heterogeneity profiles would violate the parallel trends assumptions of the DiD even with randomized treatment assignment and thus fail to identify a causal effect, see Section \ref{subsec:GFE_DiD}.}

Since the GFE estimator is a fixed-effects estimator, we perform a within-person comparison to estimate causal effects. This implies that our estimates can answer questions about how much a variable of interest affects the outcome trajectory of an individual. In our case, we estimate the joint event of an unwanted pregnancy followed by an abortion. Because our estimated effect is close to zero, we can reasonably conclude that this adverse life event does not change the mental health trajectory of an affected woman.\footnote{An unwanted pregnancy could also be a neutral event in terms of mental health costs. Then, abortion restrictions would not affect mental health. Due to other costs of denying an abortion documented in the literature, abortion restrictions would have detrimental effects without improving mental health.} It implies that in the counterfactual where a woman is denied an abortion, we would expect her mental health to deteriorate unless we were willing to assume that continuing the unwanted pregnancy would \emph{improve} her mental health trajectory. Thus, an abortion can make up for the (potentially) adverse life event of an unwanted pregnancy as if it had never happened.

The GFE estimator requires the researcher to select the number of groups of time-varying unobserved heterogeneity. We employ several performance measures to select the correct number of groups and choose the GFE estimator with two groups as our main specification. The estimated unobserved mental health profiles differ considerably across groups in both scale and slope. While most young women share a relatively flat age profile of unobserved heterogeneity, about 6\% exhibit a profile that steeply increases with age. We interpret the profiles as the age-dependent, unobserved risk of developing mental health problems. This implies that the majority of women exhibit a low unobserved mental health risk as they age. By contrast, a small but significant share of women has a low mental health risk at age 16 that sharply accumulates as these women age. 


	To investigate the robustness of our main specification, we discuss alternative dynamic processes and implement several alternative estimators. We find no evidence for reverse causality or dynamic abortion effects. We moreover instrument abortion decisions with miscarriages, showing that abortions have no detrimental mental health effects.


We next address the question of what factors are potentially picked up by the profiles of unobserved mental health risks. Since abortions from unwanted pregnancies are primarily the result of a woman's decision to engage in unprotected sex, we link mental health and abortions to other risky health behaviors observable in our data, i.e., chlamydia infections, STD screenings, and alcohol intoxication. The correlation between these observed behaviors and abortion is substantial, but controlling for them does not alter the point estimates of abortion. Moreover, estimated coefficients of these other behaviors exhibit a similar pattern as the abortion coefficients across all considered specifications. Finally, we show that the estimated unobserved mental health risk profiles are strongly correlated with these behaviors. Overall, these results suggest that risky health behaviors are also outcomes of the same choice process as abortion, rather than omitted control variables.

We propose a model of inter-temporal choices and mental health to understand how dynamic decision-making may lead to diverging unobserved heterogeneity profiles. As discussed by \citet{o2001risky}, adolescents may engage in unprotected sexual activities because they place a much higher weight on immediate gratification than on the considerable costs they may face in the future. We thus model women's time preferences as quasi-hyperbolic to induce self-control problems. We link the model to our empirical results by allowing for two groups of women who vary by the degree of present bias. This leads to different trade-offs, decisions, and a different evolution of risky behaviors and mental health. The estimated parameters indicate significant heterogeneity in the present bias across groups, resulting in different mental health trajectories. 
	
Many studies have investigated fertility and economic outcomes of abortion (e.g. \citealp{currie1996restrictions}; \citealp{gruber1999abortion}; \citealp{pop2006impact}; \citealp{ananat2007abortion}; \citealp{ananat2009abortion}; \citealp{myers2017power}). Nevertheless, mental health consequences have been understudied by economists. The medical literature has found mixed conclusions on whether an abortion negatively impacts mental health.\footnote{\cite{reardon2018abortion} provides a detailed discussion of the medical literature on abortion and mental health.} To a large extent, these inconclusive results can be attributed to methodological issues of a difficult-to-study subject. Randomized controlled trials are ethically not feasible. Survey data often suffer from non-classical measurement error, under-reporting, and recall bias in the presence of stigma.\footnote{\cite{biggs2020perceived} show that in the US perceived abortion stigma at baseline is associated with higher self-reported psychological distress five years after an abortion.} Individual-level data is rarely available, even in countries where administrative data is widely used.\footnote{Two medical studies address some methodological issues using an event-study design and Danish healthcare registers. \cite{munk2011induced} find no evidence of an increased risk of mental disorders after a first-trimester induced abortion. \cite{steinberg2018examining} show that women who had a first-trimester induced abortion have higher rates of antidepressant use. Event-study approaches have the disadvantage of failing to identify key components of the model \citep{borusyak2017revisiting} and cannot account for time-varying unobserved heterogeneity. Thus, a causal interpretation is unlikely to be valid.} 

An innovative approach to quantify the effect of abortion denial on women's lives is the Turnaway Study.\footnote{The Turnaway Study collects individual longitudinal information of women who received an abortion and women who were denied an abortion due to ineligibility based on cut-off dates in the US. The study followed women over five years after the initial abortion encounter to collect information about health, well-being, education, and labor market outcomes \citep{miller2020happens}.} With this data, \cite{biggs2017women} find no effect of abortion on depression. However, there are two potential concerns with the Turnaway study: First, the treatment and control groups differ substantially in their observable characteristics. This raises concerns about potential differences in unobservables and endogenous selection into treatment and control groups. Second, the sample size is very small, and thus power is an issue, implying that effects would need to be very large to be detected. At least in \cite{biggs2017women}, this leads to very wide confidence intervals and inconclusive results.

In economics, studies analyzing abortion effects typically exploit changes in legislation for identification and focus on the US \citep[see, for instance,][]{currie1996restrictions, gruber1999abortion, ananat2007abortion, steingrimsdottir2016reproductive, fischer2018impacts, lindo2020far, miller2020economic}. \cite{myers2017power} uses state-level variation in access to the contraceptive pill and abortions to estimate the impact on fertility and marriage. She shows that while legalizing the pill for minors does not significantly affect these outcomes, abortion legalization had a considerable impact. Only a few studies have looked at changes in abortion legislation outside the US \citep{molland16, pop2006impact}. \cite{clarke2018abortion} examine the effect of abortion on health in Mexico, with mental health as a secondary outcome. Exploiting both progressive and regressive changes in abortion legislation, they show that the initial legalization resulted in a sharp decline in maternal morbidity but find no effect on mental health in either direction. However, the study uses inpatient postpartum depression as the only measure of mental health, limiting the scope of their result. A common limitation of the studies discussed above is that changes in legislation might be intertwined with changes in stigma, thus potentially violating the identifying assumption of the DiD estimation strategy. This may be particularly important when mental health is the outcome of interest \citep[see][]{biggs2020perceived}.

We complement this economics literature in several ways. Our analysis uses administrative records, covering all women in the region of Sk{\aa}ne over ten years. Hence, we observe all abortions from \textit{unwanted} pregnancies and  mental health diagnoses on the individual level. Our identification strategy does not rely on state- or cohort variation in abortion legalization, as the Swedish abortion policy has not changed since the early 1970s. Instead, we deal with unobserved heterogeneity in the abortion decision using a novel estimator -- the GFE estimator -- which allows for time-varying unobserved heterogeneity within groups of individuals \citep{bonhomme2015grouped}. Our analysis is carried out in Sweden, a country with virtually no restrictions on abortion or contraception, which minimizes the potentially confounding effects of abortion stigma on mental health. The joint analysis of abortions and other risky health behaviors highlights the importance of accounting for dynamic unobserved heterogeneity. In particular, we show that it is not sufficient to control for other behaviors in conventional individual fixed-effects models, as they may be driven by a similar underlying decision-making process as abortion decisions. 

Our theoretical model shows that heterogeneity in the degree of present bias is sufficient to explain heterogeneity in mental health trajectories. Using non-standard time-preferences is motivated by a large literature in behavioral- and health economics (for comprehensive reviews see \cite{cawley2011economics} in health economics; \cite{gruber2000risky} and \cite{frederick2002time} in behavioral economics). \citet{gruber2001addiction} is an early, highly influential paper showing that inconsistent time preferences can generate economic models which rationalize risky health behaviors. Among adolescents, present-biased preferences have been analyzed in the context of smoking or alcohol consumption \citep{sutter2013impatience}, and risky sexual behavior \citep{chesson2006discount}. Our theoretical model combines these insights and links them to results generated by a novel econometric estimation approach to illustrate the evolution of mental health among young women.

Finally, our study adds to a growing literature on the relationship between preferences, non-cognitive skills, and mental health. As pointed out by \cite{currie2020child}, mental health issues are an important determinant of human capital development as they reflect deficits in non-cognitive skills. \cite{heckman2006effects} show that non-cognitive skills play a substantial role in explaining adolescents' decisions to engage in risky behavior, such as marijuana use or illegal activities. Studying the relationship between time-inconsistent preferences, non-cognitive skills, and depression, \cite{cobb2020depression} show that self-control problems are strongly correlated with non-cognitive skills such as the internal locus of control and partly explain the depression gap in risky health behaviors among adults.\footnote{\cite{borghans2008economics} discuss how to incorporate preferences and personality traits in economic models.} While we cannot incorporate a link between non-cognitive skills and present biased preferences, our theoretical model illustrates how mental health develops as a consequence of dynamic decisions under preference heterogeneity.

Our work has several implications. First, the precisely estimated null-effect indicates that an abortion from an unintended pregnancy has no detrimental effect on mental health. Thus, mental health can not justify policies that impose restrictions on abortions. By contrast, they may even have unintended negative consequences if more restrictive policies lead to a stronger political and social stigmatization of abortions \citep[see, e.g.,][]{biggs2020perceived}.
Second, restricting abortion access seems inadvisable: there is previous evidence on adverse economic consequences of restrictive abortion policies \citep[see for instance][]{miller2020economic, miller2020happens, felkey2018restrictions, lindo2019waiting}. Our null results imply that unrestricted access to abortion does not lead to additional mental health costs. Taken together, restrictive abortion policies are thus unlikely to be welfare-enhancing. Third, the substantial differences in the estimated unobserved heterogeneity profiles between high-risk and low-risk women imply that general mental health screenings are unlikely very effective tools for combating mental illness in adolescents. Instead, interventions should target high-risk women at younger ages, using tools similar as in \cite{alan2018fostering} to reduce self-control problems and the likelihood to develop severe mental illnesses.\footnote{\cite{aizer2017review} discusses different approaches of reducing self-control problems among adolescents. Based on a model of skill formation, she argues that programs to be effective should be implemented in pre-school age as it allows to control the environment interacting with such investments.} By doing so, one may keep not only direct medical costs low but also reduce indirect costs of mental health disorders such as lower educational attainment and fewer earnings \citep{biasi2018career,currie2010child,fletcher2010adolescent}.

The paper is organized as follows. Section \ref{sec:inst} outlines the Swedish health care system and the abortion history in Sweden. In Section \ref{sec:data}, we describe the data and measures for mental health and abortion. Section \ref{sec:emp} introduces our empirical strategy, and Section \ref{sec:results} discusses our results and associated robustness checks. The theoretical model is presented in Section \ref{sec:model}. Section \ref{sec:conclusion} concludes.

\section{Institutional background}\label{sec:inst}
\subsection{The Swedish health care system}\label{subsec:HCsystem}
In Sweden, health care is primarily public and organized at the regional level. Within a region (e.g., Sk{\aa}ne), different municipalities have different health care centers (or primary care units) that house all out-patient care. Here, ``out-patient'' refers to all contacts with care providers that do not include at least one night's stay, i.e., all ambulatory care, such as visits to physicians, emergency care, nurses, or physiotherapists. In addition, it covers consultations by telephone. Typically, a small municipality has only one health care center. Larger cities have multiple centers. ``In-patient'' care, as opposed to out-patient care, refers to visits at health centers or hospitals that include at least one night's stay. 

Every individual is assigned to one health care center, usually the nearest one. When necessary, an individual goes to the center and is helped by the next available health care worker. There is no path dependence in the identity of the health care worker across consecutive contacts. Individuals are dealt with sequentially by the first available health care worker on a given day. The health care system is funded through a proportional regional income tax. Healthcare is free of charge, except for a small deductible capped at 900 SEK (about 117 USD) per year during our observation period.

\subsection{Abortions in Sweden}\label{subsec:abortion}
In Sweden, abortions were first legalized by the Abortion Act of 1938, guaranteeing access for limited cases. The act states that pregnancies may be terminated if the child's birth threatens the mother's life or health or if the child is expected to have severe malformations or mental deficiencies \citep{glass1938effectiveness}. The current version of the abortion act took effect in January 1975. It grants access to abortions on request until week 18 without any restrictions. Importantly, minors do not require parental consent to receive an abortion \citep{abortsweden09, abortsweden}. Thus, the decision to terminate a pregnancy is solely made by the pregnant woman regardless of her age.

In 1992, Sweden approved the ``abortion pill'' (mifepristone), which allows terminating a pregnancy at an early stage (at most 49--56 days after conception) without a hospital stay \citep{jones2002mifepristone}. Between weeks 9--13, abortions are conducted through surgical intervention. After week 13, an overnight stay at the hospital is required. Since the mid-1990s, the emergency contraceptive pill (ECP), also known as morning-after pill, has been available. In 2001, the ECP was approved to over-the-counter (OTC) purchase \citep{guleria2020emergency}. Figure \ref{app:agg_ab} in Appendix \ref{app_fig} shows the aggregate time trends in abortions by gestation week and age for Sk{\aa}ne and the whole of Sweden. There is a trend to substitute later abortions (week 9--11) with earlier abortions (before week 9) regardless of age. Besides, there is no discernible discontinuity around the date of OTC availability.

In 1999, 26.3 per 1,000 women had an abortion in the age group 20--24, and 19.0 per 1,000 women aged 19 and below. These numbers increased over time to 34.7 and 24.4 abortions per 1,000 women in these age groups. In Sk{\aa}ne, numbers are slightly lower, but with 33.9 and 22.3 abortions per 1,000 women, they are still very high \citep{abortsweden}, in particular, compared to other developed countries \citep{haegele2005ab}. 

\begin{figure}[t]
	\includegraphics[width=0.49\textwidth]{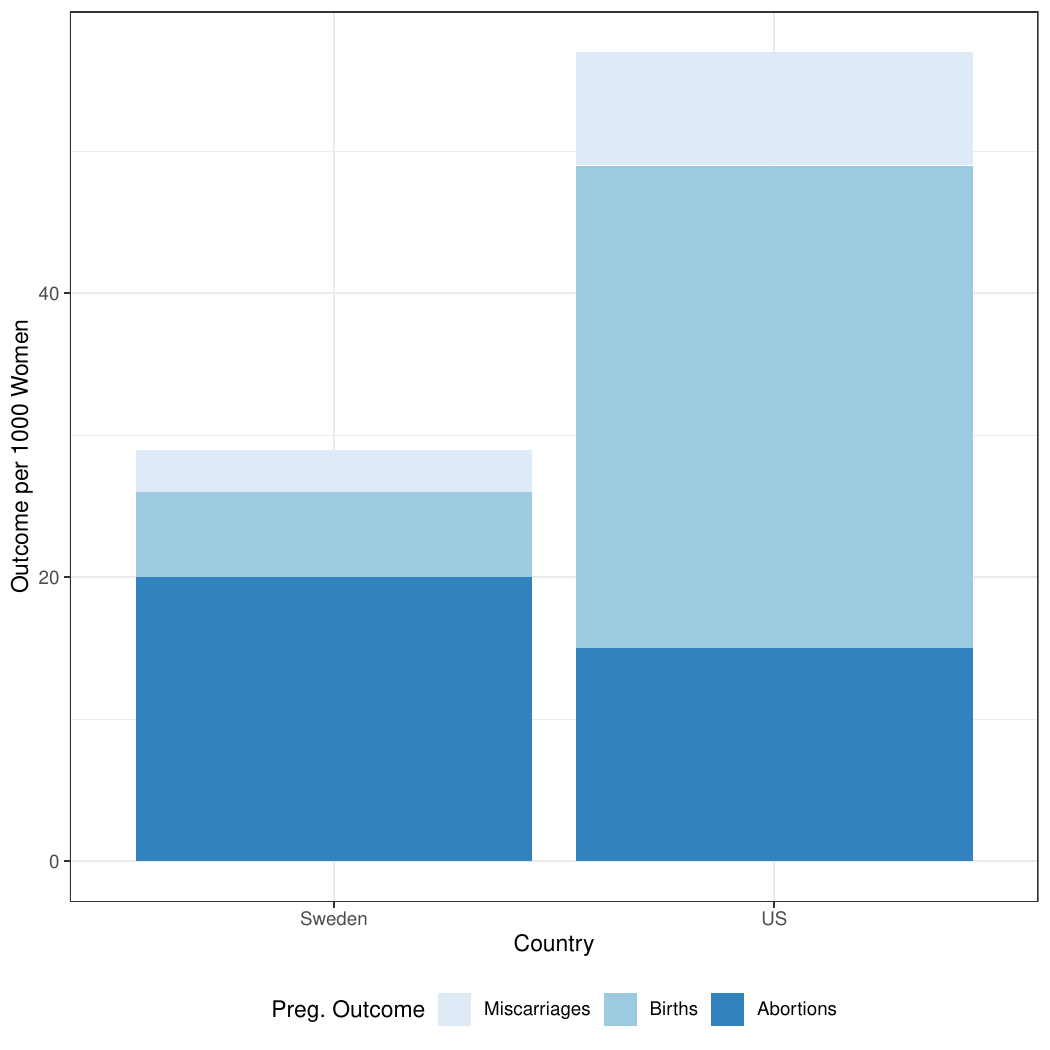}
	\caption{Pregnancy outcomes at ages 15--19, Sweden and US, 2010 \citep{sedgh2015adolescent}.}\label{fig:US_SWE}
\end{figure}

Figure \ref{fig:US_SWE} compares abortion rates and alternative birth outcomes among adolescents in Sweden to those in the US, a country in which access to abortion is more restricted in practice. Abortion rates are much higher in Sweden. However, teenage birth- and miscarriage rates in Sweden are only about 15\% and 20\% of those in the US.\footnote{In Sweden, teenage women rarely bear children. According to \cite{lager2012young}, approximately six children were born per 1,000 young women aged 15--19. 80\% of all pregnant women aged 15--19 and 41\% of all pregnant women aged 20--24 opted for abortion in 2009. These numbers are similar to our sample statistics. Among all 16--to 19-year-old pregnant women, 76\% opted for an abortion, while 19\% gave birth and 5\% had a miscarriage.}



What would we expect from restricting access to abortions in Sweden? According to the literature, abortions could be substituted by increased birth rates, abstinence, or higher contraceptive use. \citet{fischer2018impacts} show that proclivities for risky sexual behavior are not very sensitive to restrictive abortion policies, at least not among adolescents in the US. This is in line with the finding that abstinence-only sexual education programs are not effective in increasing abstinence \citep{santelli2017abstinence} or reducing birth rates \citep{kearney2015investigating}. Substituting abortions by higher contraceptive use is also unlikely to happen, at least not in Sweden, where contraception is widely available and easily accessible. \cite{sydsjo2014reimbursement} find no evidence that increased contraceptive use is associated with lower rates of induced abortions. Thus, introducing abortion restrictions in Sweden would most likely lead to an increase in teenage birth rates, all else being equal.

	 Abortion access may determine not only pregnancy outcomes but also the level of abortion stigma. Abortion stigma can be generated through negative judgments of the social environment and structurally through restrictive abortion policies of governments and institutions, which can increase social stigmatization. \cite{biggs2020perceived} shows that abortions are associated with stigma, which increases psychological distress in women. Restricting access to abortions may increase stigma and mental health problems in women seeking an abortion but leave the abortion effect itself unchanged. Thus, in a country with a very restrictive abortion policy and strong stigma, increased abortion access without reducing stigma may not immediately lead to the desired effect of reduced mental health problems. Instead, such policies could even increase mental disorders in the short run.

\section{Data}\label{sec:data}
\subsection{Description of different data registers}\label{subsec:desc}
Our empirical analysis is based on combined register data for Sk{\aa}ne, the third most populous and southernmost region in Sweden. It consists of individual-level longitudinal records from the intergenerational register, the Sk{\aa}ne inhabitant register, the income tax register, and the in- and out-patient registers. The in-and out-patient registers are from the ``patient administrative register systems'' administrated by the Regional Council of Sk{\aa}ne. A unique feature of our data is the detailed records of \textit{all} occurrences of in-patient \textit{and} out-patient care for all inhabitants of the region. The registers have previously been used by \citet{tertilt2015association}, \cite{nilsson2018health} and \cite{vandenBerg2018}. The health care registers are collected to determine the monetary streams from the region to the health care centers and hospitals.

In Sweden, each individual has a unique identifier that is used to record all contacts with the health care system and the general public administration, tax boards, employment offices, and other public agencies. We use the identifier to merge the health care registers to the LISA dataset, which combines several other registers.\footnote{LISA is the ``Integrated database for labor market research'', see \cite{SCB2009}.} LISA covers all persons born in Sweden between 1940 and 1985, their parents, and all their children \citep{meghir2005educational}. For individuals aged 16 and above, LISA provides a rich set of annual socio-economic information, such as employment status, incomes by type, level of education, or marital status. Further, the intergenerational register allows linking individuals to their children and parents. The merged dataset contains about 1 million individuals, which is the vast majority of inhabitants of Sk{\aa}ne in 1999--2008. From these data, we construct an annual panel data set which comprises all women born between 1983 and 1985 and living in Sk{\aa}ne between 1999--2008. We chose to select these birth cohorts to guarantee that we observe women aged 16 to 23 years in all periods.

\subsection{Diagnosis variables \& abortions}\label{subsec:var}
We define individual measures for mental health and abortions using ICD-10 diagnosis codes. Chapter 5 of the ICD-10 catalog comprises diagnosis codes for mental and behavioral disorders. The chapter is divided into 11 sub-chapters that classify diagnoses into, e.g., organic mental disorders, schizophrenia, affective, somatoform disorders, behavioral or developmental mental disorders. Our main outcome of interest is the diagnosis of mental health conditions defined by codes F30--F39 for mood disorders. These are the most common psychiatric diagnoses in young adults and include depression and manic episodes, bipolar affective disorders, and persistent mood disorders. Mood disorders, including their sub-categories and drugs prescribed against these conditions, have frequently been used in the medical literature to classify mental health conditions \citep[see, for instance,][]{steinberg2008abortion,biggs2017women,biggs2020perceived}.

\begin{figure}[t]
	\mbox{
		\subfigure[Mental health diagnoses by age]{\includegraphics*[width=0.49\textwidth]{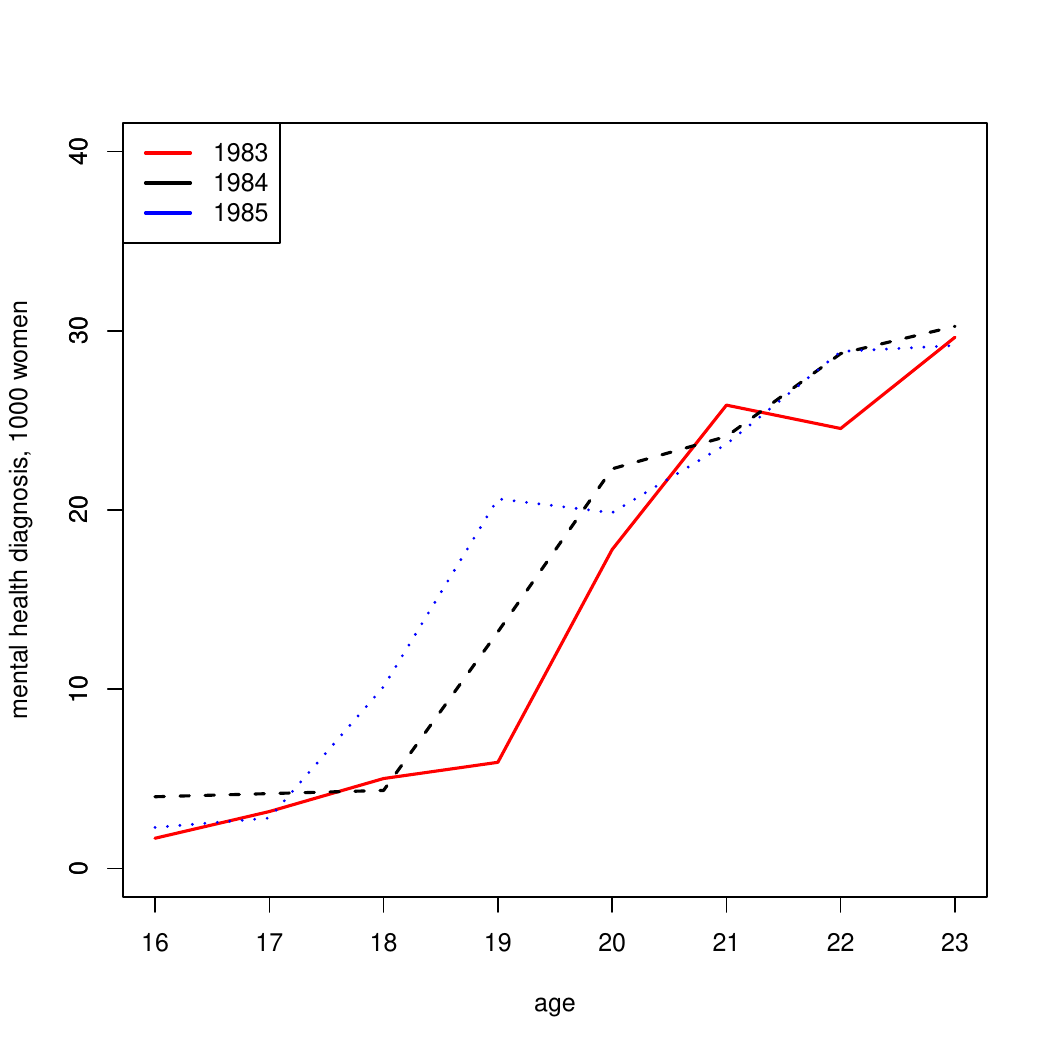}} 
		\subfigure[Abortion by age]{\includegraphics*[width=0.49\textwidth]{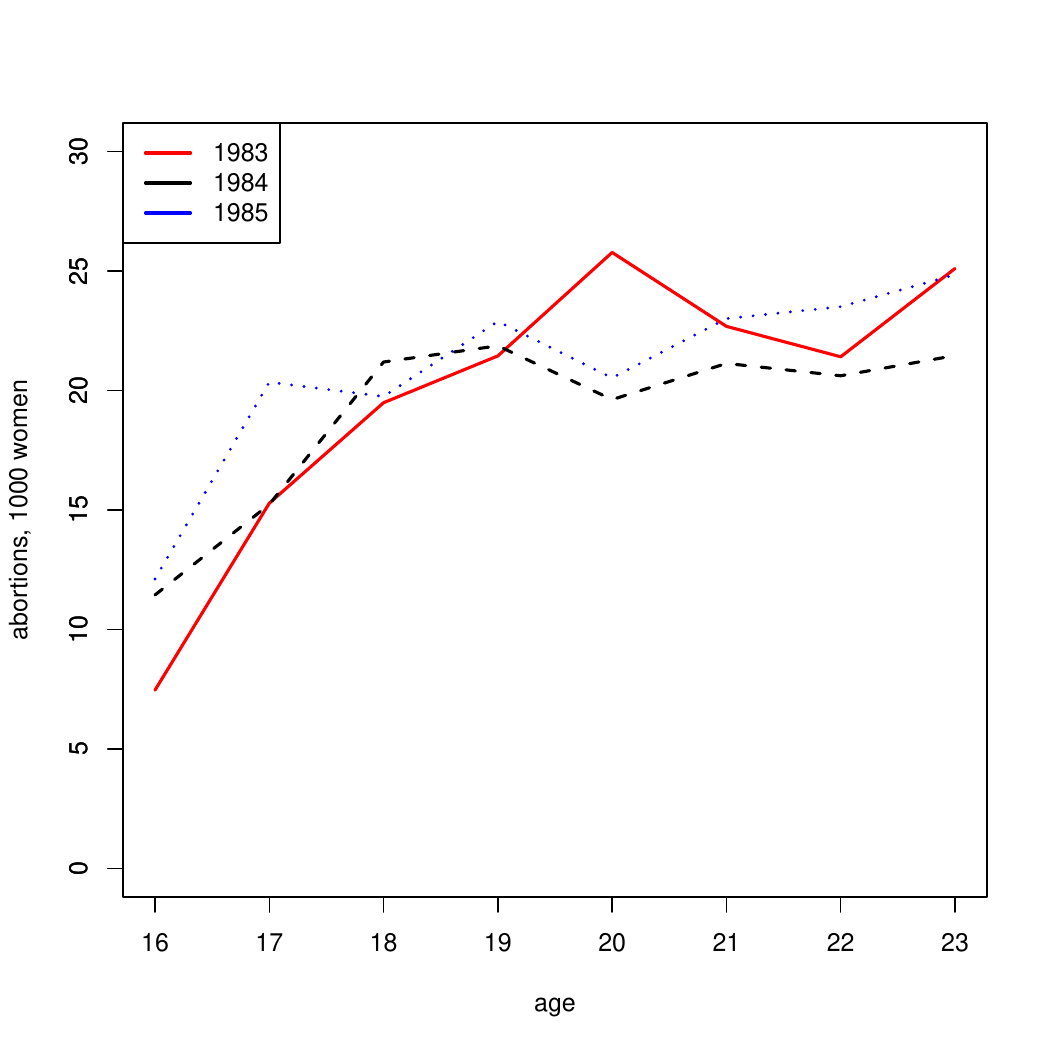}}}		
	\caption{Mental health diagnoses and abortion by age and cohort. }
	\label{fig_des}
\end{figure}

Figure \ref{fig_des}(a) shows the incidence of our mental health diagnoses per 1,000 women by age and birth cohort. Diagnoses are relatively low at age 16, with about 2-4 diagnoses per 1,000 women in these birth cohorts. From age 17, the numbers steadily increase to about 30 diagnoses in 1,000 women at age 23. Trends are similar across the three cohorts. 

In the subsequent analysis, we define mental health problems as an absorbing state (cumulative): once a woman is diagnosed with a mental disorder, she is classified as ill for the remaining observation period. This is motivated by the medical literature, which has shown that an episode of mood disorder, e.g., a depressive episode, among adolescents can last between a few months and several years \citep{eatonprev}. While short-term recovery rates are high, recurrence rates increase after 1--2 years, up to more than 50\% in the long run \citep[see, e.g.,][]{curry2011recovery}.

To measure abortions, we use pregnancy-related ICD-10 diagnosis codes. The codes O00--O08 refer to pregnancies with abortive outcomes.\footnote{Abortions can be complete or incomplete, with or without complications. We do not distinguish them.} The code O04 defines induced medical abortions. These can be surgical or pharmaceutical abortions as well as voluntary and medically-indicated terminations of pregnancy. The code Z64.0 defines an unwanted pregnancy. It includes women who later have an abortion, women who carried the pregnancy to term, or women who had a spontaneous abortion. We combine these two codes to define our measure of abortion as a medical abortion from an unwanted pregnancy.\footnote{The ICD-10 also codes spontaneous abortions/miscarriages (O03). Miscarriages are not the scope of our main analysis but will be used in a complementary analysis, see Subsection \ref{subsec:alternatives}.} Figure \ref{fig_des}(b) shows the incidence of abortions per 1,000 women by age and birth cohort. The cohorts exhibit similar trends in abortion rates. The rates sharply increase between ages 16--18 but remain roughly constant at later ages. The numbers in Figure \ref{fig_des}(b) correspond to those reported by \citet{abortsweden}.\footnote{Figure \ref{app:nr_abortions} in Appendix \ref{app_fig} plots the number of abortions after an unwanted pregnancy per woman in our age group. About 82\% receive one abortion between age 16--23, about 14\% receive two abortions, and about 3\% receive three. Less than 1\% of women undergo four or more abortions in this age group.} 

Table \ref{tab:des1} shows the descriptive statistics for all variables used in the empirical analysis. Our sample comprises 20,703 women aged 16--23 with an average of 19.5 years. Women are, on average, born in 1984, which implies that our birth cohorts are of similar size. As expected for such a young sample, most women are single, about 20\% are employed, and less than 30\% hold a college degree. The annual rate of abortions is about 2\%, and the incidence of mental health problems per year is about 1.6\%. In total, 10.6\% of women had an abortion, and 6.5\% had mental health problems during ages 16--23. Since our main estimation strategy requires a balanced panel, we construct two censoring indicators: one to flag missing observation periods and one to flag missing values. This balancing procedure leads to a final sample of $N\times T=165,624$ observations. %

\begin{table}[t!!!]
	\caption{Descriptive statistics for the three birth cohorts comprising our sample}
\label{tab:des1}%
	\begin{tabularx}{\textwidth}{@{} X C{1.5cm}C{1cm}C{1cm}C{1cm}C{1cm}}
		\toprule
		&       $N\times T$&        mean&          sd&         min&         max		\\
		\midrule
		\multicolumn{6}{@{} l}{{\textit{Mental health diagnoses and abortion}}}                                         \\		
		\quad Cum. mental health diagnoses (absorbing state)      		&      146,833&    .032&    	.175&           0&           1	\\
		\quad Mental health diagnoses (non-absorbing state)        	&      136,108&    .016&    	.126&           0&           1	\\
		\quad Abortion       						&      136,108&    .020&    	.140&           0&           1	\\
		\addlinespace
		\multicolumn{6}{@{} l}{{%
		\textit{Individual characteristics of women}}} \\	
		\quad Single              				&      134,464&    .989&    	.104&           0&           1	\\
		\quad Married					            &      134,464&    .010&    	.100&           0&           1	\\
		\quad Employed            				&      134,464&    .213&    	.410&           0&           1	\\
		\quad Log annual earnings              	&      134,177&    7.91&    	4.414&           0&    14.020	\\
		\quad College degree             			&      146,802&    .284&    	.451&           0&           1	\\
		\quad Age                 				&      165,624&    19.5&    	2.291&          16&          23	\\
		\quad Birth year          				&      165,624&		1984&    	.816&        1983&        1985	\\
		\quad Year                				&      165,624&    2003&    	2.432&        1999&        2008	\\
		\addlinespace
		\multicolumn{6}{@{} l}{{\textit{Individual characteristics of women's mother}}}                                         \\	
		\quad Employed							&      134,117&    .837&    	.370&           0&           1		\\
		\quad College degree           			&      156,760&    .364&   		.481&          	0&          1		\\
		\quad Log annual earnings            		&      134,064&    10.696&    	3.963&           0&     15.193		\\
		\quad Birth year        					&      164,992&    1955&    5.153&        1933&        1970		\\
		\quad Married					           	&      134,117&    .655&    	.475&           0&           1		\\		
		\quad Log disposable family income		&      133,937&    12.863&    	.656&           0&    18.479		\\
		\addlinespace
		\multicolumn{6}{@{} l}{{\textit{Individual characteristics of women's father}}}                                         \\	
		\quad Employed							&      131,163&    .846&    .361&           	0&           1	\\
		\quad College degree           			&      144,968&    .420&   	.494&          		0&          1		\\
		\quad Log annual earnings            		&      131,110&    11.04&  	4.071&          	0&    16.396		\\
		\quad Birth year        					&      164,512&    1952&    5.833&        1917&        1969		\\
		\bottomrule
	\end{tabularx}
\end{table}

\begin{figure}[t]
	\includegraphics[width=0.49\linewidth]{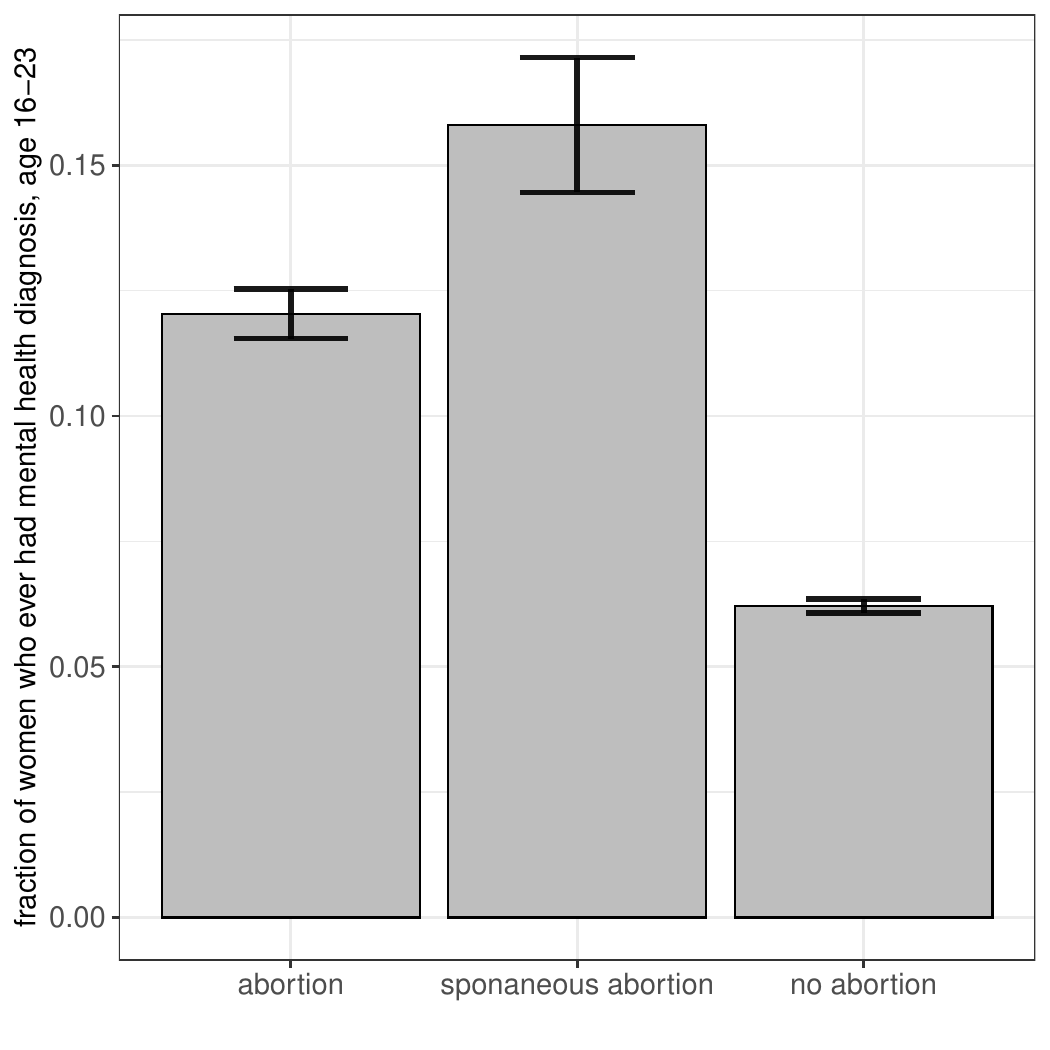}
	\caption{Share of women ever diagnosed with mental health disorders by (non)-abortive events.}\label{fig:bar}
\end{figure}

We also compare the incidence rates of mental health diagnoses by (non)abortive outcomes. Figure \ref{fig:bar} shows the fraction of women who were ever diagnosed with mental health problems among women who had an abortion after an unwanted pregnancy, experienced a miscarriage, or never had any abortion. Women with abortions are about twice as likely to be diagnosed with mental health problems than women without abortions. Women with a miscarriage have the highest incidence of mental health problems. Figure \ref{fig:bar} suggests that there is a relationship between abortions and mental health. This relationship is the topic of the coming sections.\footnote{Our measure of mental health does not comprise mental health issues without an official diagnosis. For our analysis to be valid, we must assume that women who have an abortion are not systematically more underdiagnosed than the female population. Our descriptive evidence does not indicate such an issue.}  %

\subsection{Underreporting of mental health issues}
Our health records capture the universe of inpatient and outpatient contacts with the healthcare system and all diagnoses made by health care professionals. This provides us with comprehensive data of all mental disorders in the region of Sk{\aa}ne. However, our data do not capture women who suffer from mild, non-clinical forms of mental disorders or do not seek out care. Thus, we may face an issue with underreporting of mild cases, which may lead to underestimating the impact of abortions on mental health problems.

We first validate our health records against survey information. Since we do not have survey data on depression, we compare the incidence of diagnoses on anxiety and fear-related disorders with self-reported anxiety and fear-related symptoms from the survey of living conditions (ULF) in 2008.\footnote{ULF data can be accessed from \url{https://www.scb.se/hitta-statistik/statistik-efter-amne/levnadsforhallanden/levnadsforhallanden/undersokningarna-av-levnadsforhallanden-ulf-silc/pong/tabell-och-diagram/halsa/halsa--fler-indikatorer/}.} ULF asks respondents whether they have experienced problems or symptoms of anxiety or fear. If a respondent answered the question with yes, she is asked whether the problems are mild or severe. In 2008, 27\% of women aged 16--24 had reported symptoms of anxiety or fear in the survey, and 7\% had experienced serious symptoms. In our sample, 10\% of women aged 16--23 were diagnosed with anxiety or fear-related disorders. Thus, our records may also capture less severe cases of anxiety.

ULF utilizes a single question to assess anxiety and fear-related problems, which may not reliably capture the complex nature of anxiety disorders \citep[see, for instance,][]{turon2019agreement}. More reliable information about mental disorders is obtained from self-administered screening tools or diagnostic interviews. \cite{olsson1997beck} and \cite{olsson1999adolescent} have investigated the prevalence of depression among 16--17-year-old high school students in the Swedish city of Uppsala. Depending on screening tools and cut-off values used, between 9 and 16 percent of women had depressive symptoms. When using diagnostic interviews, the lifetime depression prevalence among young women was 11.5\%. Screening 13--20-year-old youths in a clinical youth center at the university hospital in Uppsala in 2006, \cite{kristjansdottir2011could} find that about one-third of young women were screened for at least mild depression, and 12\% were screened for at least moderate levels of depression. In our data, about 8\% of women have received a depression diagnosis at age 16--23, which is only slightly lower than what was found with screening tools and diagnostic interviews. Thus, while our data may not capture all potential cases and non-cases of depression, they also do not seem to suffer from severe underreporting of mental disorders.

\section{Empirical strategy}\label{sec:emp}

 In this section, we present the empirical strategy to estimate the causal effect of abortion on mental health. We discuss the shortcomings of established linear methods, such as individual fixed-effects models, and introduce the grouped fixed-effects (GFE) estimator to overcome potential identification issues. We then discuss the identifying assumptions of the GFE estimator and compare them with the differences-in-differences (DiD) approach, one of the most popular methods for causal inference in applied microeconomics. 
 
A linear model that links abortion $A_{it}$ and mental health diagnosis $M_{it}$ is 
\begin{equation}\label{eq:ols_fe}
M_{it} = \xi A_{it} + \mathbf{\tilde{x}}_{it}^\prime\boldsymbol{\gamma} +\alpha_{it}+ \varepsilon_{it}, ~~~~~~ i = 1, \ldots, N;~ t=1, \ldots, T,
\end{equation}
where $\mathbf{\tilde{x}_{it}^\prime}$ comprises covariates for woman $i$ and her parents. $\varepsilon_{it}$ is an idiosyncratic error term with $\E[\varepsilon_{it}]=0$ and $\Cov(\X,\boldsymbol{\varepsilon})$. $\alpha_{it}$ is an unobserved individual-specific fixed-effect that varies across age. The parameter of interest is $\xi$, capturing the association between an abortion $A_{it}$ from an unwanted pregnancy and a mental health diagnosis $M_{it}$. 
 
Under the assumption that $\alpha_{i0}=\alpha_{i1}=...=\alpha_{iT}$ for all $i=1,...,N$, i.e., individual unobserved heterogeneity $\alpha_{it}$ is time constant, $\xi$ in Equation \eqref{eq:ols_fe} can be consistently estimated with a standard model with individual-specific, time-constant fixed-effects. Here, unobserved heterogeneity implies that decisions affecting both mental health development and abortion probabilities are independent over time. If this assumption is violated, estimates are biased. In our application, it seems plausible that $\alpha_{it}$ is dynamic: abortions from an unwanted pregnancy are outcomes of decisions that depend on past decisions and are determined by preferences. Thus, selection into abortions is likely dynamic, and a standard fixed-effects model fails to estimate a causal effect of abortion on mental health. Formally, for two time periods $t=0,1$, this implies that for two individuals, $j$ and $k$, with $\alpha_{j0}>\alpha_{k0}$, we get $\alpha_{j0}-\alpha_{k0}<\alpha_{j1}-\alpha_{k1}$. In general, an unobserved time-varying $\alpha_{it}$ is indistinguishable from $\varepsilon_{it}$ without further assumptions.

\subsection{Time-varying grouped fixed-effects estimator (GFE)}\label{subsec:gfe}
One solution to the problem described above is proposed by \cite{bonhomme2015grouped} who suggest clustering individuals with similar unobserved characteristics into a finite number of groups. This implies that women belonging to the same group share the same age profile of unobserved heterogeneity, 

\begin{equation}\label{eq:bm1}
M_{i t}=\xi A_{i t}+\tilde{\x}_{i t}^{\prime}\boldsymbol{\gamma}+\alpha_{g_{i} t}+\varepsilon_{i t},
\end{equation}
where $\alpha_{g_{i} t}$ represents time-varying, group-specific unobserved heterogeneity term for $g \in\{1, \ldots, G\}$ groups. The error term $\varepsilon_{it}$ may contain an individual-specific, time-constant fixed-effect $\alpha_i$, such that $\E[\varepsilon_{it}|\alpha_i]=0$. We write Equation \eqref{eq:bm1} more compactly by defining a parameter $ \boldsymbol{\theta}=\left( \xi,\boldsymbol{\gamma}\right)$ and a vector of regressors, $\x_{it}=\left(A_{it}, \tilde{\x}_{it} \right)$,
\begin{equation}\label{eq:bm2}
M_{i t}=\x_{i t}^{\prime} \boldsymbol{\theta}+\alpha_{g_{i} t}+\varepsilon_{i t}, \quad i=1, \ldots, N, t=1, \ldots, T.
\end{equation}

The GFE estimator is defined as the solution to 
\begin{equation}\label{eq:bm3}
(\widehat{\boldsymbol{\theta}}, \widehat{\alpha})=\underset{(\boldsymbol{\theta}, \alpha) \in \Theta \times \mathcal{A}^{G T}}{\operatorname{argmin}} \sum_{i=1}^{N} \sum_{t=1}^{T}\left(M_{i t}-\x_{i t}^{\prime} \boldsymbol{\theta}-\alpha_{\widehat{g}_{i}(\boldsymbol{\theta}, \alpha) t}\right)^{2},
\end{equation}
where $\widehat{g}_{i}(\boldsymbol{\theta}, \alpha)$ is the optimal group assignment determined by $$\widehat{g}_{i}(\boldsymbol{\theta}, \alpha)=\underset{g \in\{1, \ldots, G\}}{\operatorname{argmin}} \sum_{t=1}^{T}\left(M_{i t}-x_{i t}^{\prime} \boldsymbol{\theta}-\alpha_{g_i t}\right)^{2}.$$ 

For a given number of groups $G$, the estimator assigns individuals to groups via clustering and estimates the coefficients $\widehat{\boldsymbol{\theta}}$ as well as the group profiles $\widehat{\alpha}_{g_{i} t}$ in an iterative procedure.\footnote{A well-known issue with the GFE estimator is its sensitivity to the choice of initial values. To validate our results we randomly vary the seed and thus initial values. Our results are robust to different seed choices.} Standard errors are clustered at the individual level and obtained from analytical expressions in \cite{bonhomme2015grouped}.
 
In our main specification, we will also account for individual-specific, time-constant unobserved heterogeneity $\alpha_i$ by applying time demeaning. Thus, the solution is given as
\begin{equation}\label{eq:bm4}
(\widehat{\boldsymbol{\theta}}, \widehat{\alpha})=\underset{(\boldsymbol{\theta}, \alpha) \in \Theta \times \mathcal{A}^{G T}}{\operatorname{argmin}} \sum_{i=1}^{N} \sum_{t=1}^{T}\left(\dot{M}_{i t}-\dot{\x}_{it}^{\prime}\bftheta -\alpha_{\widehat{g}_{i}(\boldsymbol{\theta}, \alpha) t}\right)^{2},
\end{equation}
where $\dot{M}_{i t}=M_{i t}-\bar{M}_{i}$ and $\dot{\x}_{it}={\x}_{i t}-\bar{{\x}}_{i}$, and $\bar{M}_{i},\bar{\x}_{i}$ are time-demeaned quantities.

\subsection{Choosing the number of groups}\label{subsec:BIC}
The GFE estimator requires the researcher to choose the correct number of groups. Ideally, we obtain this number by data-driven methods. Yet, selecting the correct number is non-trivial as the choice of information criterion depends on the data generating process. This is a well-known problem when information criteria are used for model selection, see \cite{choi2019model} and \cite{bai2002determining}. The number of groups selected by an information criterion is a function of the penalty whose size depends on the number of groups $G$ and the numbers of covariates $K$, individuals $N$ and time periods $T$. Thus, no single criterion will select the correct number of groups in all potential applications. 

\cite{bonhomme2015grouped} suggest a Bayesian information criterion (BIC), 
\begin{equation}\label{eq:BIC}
BIC(G)= \frac{1}{N T} \sum_{i=1}^{N} \sum_{t=1}^{T}\underbrace{\vphantom{\sum_{t}} \left(M_{i t}-\x_{i t}^{\prime} \widehat{\theta}^{(G)}-\widehat{\alpha}_{\widehat{g}_{i} t}^{(G)}\right)^{2}}_{{\substack{\\[-0.1em]\text{objective}}}}
+\underbrace{\vphantom{\sum_{t}} \widehat{\sigma}^{2} \frac{G T+N+K}{N T} \ln (N T)}_{{\substack{\\[-0.1em]\text{penalty}}}},
\end{equation}
where the penalty is the second part of Equation \eqref{eq:BIC}. The estimated error variance $\widehat{\sigma}^2$ is calculated using $G_{max}$, the maximum feasible number of groups chosen by the researcher.

In our simulation exercise, we show that this BIC chooses the correct number of groups if $N$ is not much larger than $T$. Otherwise, this BIC does not sufficiently discriminate between different numbers of groups.\footnote{This BIC only estimates $G$ consistently if $N$ and $T$ go to infinity at the same rate \citep{bonhomme2015grouped}. In our application, this BIC thus might overestimate the true number of groups.} As an alternative, we use a BIC with a modified penalty $\widehat{\sigma}^{2} \frac{G(T+N-G+K)}{N T} \ln (N T)$, which puts more weight on $G$. However, this alternative criterion tends to penalize too much. We will thus use both criteria together with other sensitivity checks to pick the number of groups. 

In recent work on factor models, \cite{moon2015linear} show that if both $N$ and $T$ grow to infinity, the limiting distribution of the least-squares estimator of the parameter of interest is robust to including \emph{additional} factors. While it is useful to understand whether this also holds for the GFE estimator, exploring this is beyond the scope of this paper. 

\subsection{Assumptions on time-varying unobserved heterogeneity }\label{subsec:GFE_DiD}

In this section, we discuss the key assumption on individual time-varying unobserved heterogeneity needed to identify causal effects with the GFE estimator. We compare this assumption to that of the DiD estimator, and discuss situations in which these assumptions can be maintained. For this illustration, we use a potential outcome framework notation. 

Let $\tilde{\alpha}_{it}$ be the time-varying unobserved treatment assignment and $\tilde{\alpha}_{it}=\alpha_{it}-\alpha_{g_it}-\alpha_i$. $\alpha_{g_it}$ are the group-specific profiles (see Section \ref{subsec:gfe}), and $\alpha_i$ is an individual-specific, time-constant fixed-effect. The key identifying assumption of the GFE estimator is that the expected value of mental health given that no abortion has taken place, denoted as $M_{i t}(0)$, should be the same regardless of the ``treatment assignment'', and given covariates, time and \emph{unobserved group effects $\alpha_{g_it}$}. Broadly speaking, the assumption states that $\alpha_{g_it}$ captures the relevant time-varying variation determining dynamic selection into treatment.
\begin{equation}
	\E\left[M_{i t}(0) \mid \alpha_{g_it},\alpha_{i}, \x_{i t},\tilde{\alpha}_{i t}\right]=\E\left[M_{i t}(0) \mid \alpha_{g_it}, \alpha_{i}, \x_{i t}\right], 
\end{equation}
where $\x_{i t}$ may contain covariates and a time indicator. Under the assumption of constant treatment effects, the conditional expectation of observation $i$ under treatment is
\begin{equation}
\E\left[M_{i t}(1) \mid \alpha_{g_it},\alpha_{i}, \x_{i t}, \tilde{\alpha}_{i t}\right]=\E\left[M_{i t}(0) \mid \alpha_{g_it},\alpha_{i}, \x_{i t}\right]+\xi.
\end{equation}

Further assuming a linear functional form of the conditional mean function leads to 
\begin{equation}
M_{i t}=\alpha_{i}+\alpha_{g_it}+\xi A_{i t}+\x_{i t}^{\prime}\bfgamma+\varepsilon_{i t}.
\end{equation}

The $\operatorname{DiD}$ estimator relies on a similar set of assumptions about the potential outcomes under treatment $M_{i t}(1)$ and under non-treatment $M_{i t}(0)$. The main difference to the GFE estimator is the restrictions imposed on time-varying unobserved heterogeneity. The reason is that the identification of a causal effect with the $\operatorname{DiD}$ estimator relies on group differences in a before and after comparison (conditional on treatment assignment). 

Suppose we have two groups $s \in \{ 0,1\}$, where $s=0$ indicates the control group and $s=1$ is the treatment group. We assume that 

\begin{equation}
	\E\left[M_{i t}(0) \mid \alpha_{st}, \x_{i t},\tilde{\alpha}_{i t}\right]=\E\left[M_{i t}(0) \mid \alpha_{st}, \x_{i t}\right],
\end{equation}
where $\alpha_{s{t}}$ is time-varying unobserved heterogeneity that can only vary between the treatment and control group, i.e. $\tilde{\alpha}_{it}=\alpha_{it}-\alpha_{st}$. The difference in the time trends $\alpha_{s{t}}$ is constrained to be constant. This restriction is necessary to fulfill the parallel-trends assumption used in $\operatorname{DiD}$ estimation. In practice, this restricts all individuals in the treatment and control group to have parallel unobserved heterogeneity profiles.

The crucial difference in the identifying assumptions of the GFE and the $\operatorname{DiD}$ estimators is the restriction on the time-varying unobserved heterogeneity: the GFE estimator puts no restrictions on $\alpha_{g_it}$ but restricts the number of distinct profiles. The $\operatorname{DiD}$ estimator allows all individuals to be on individual slopes, but only within treatment and control group.\footnote{Even if the treatment assignment was random, the time-varying unobserved heterogeneity in the population is restricted by the parallel trends assumption. Suppose both, treatment and control group, contain two different types of individuals with non-parallel unobserved heterogeneity profiles in different proportions. Then the $\operatorname{DiD}$ estimator fails to recover the true treatment effect, even with random assignment, because the parallel trends assumption is violated. For further discussion see \cite{lechner2011estimation}.} 

The identifying assumption of the $\operatorname{DiD}$ estimator discussed above apply to situations in which the treatment assignment is random. With non-random treatment assignment and when using individual-level panel data, the identifying assumptions of the $\operatorname{DiD}$ estimator and the standard individual-specific fixed-effects estimator are basically identical. 

\section{Results}\label{sec:results}

In this section, we first present our main results on the effect of abortion on mental health obtained from different estimators. We then determine the optimal number of groups and present the group-specific unobserved heterogeneity age profiles $\widehat{\alpha}_{\hat{g}t}$. Because the GFE estimator is relatively new and has not been used extensively in empirical work, we provide a detailed simulation framework.\footnote{We also use our simulation to validate the inference results in our setting, since the asymptotic results in \cite{bonhomme2015grouped} only apply for large $N,T$.} In Appendix \ref{app_sim}, we introduce a data generating process based on Equation \eqref{eq:bm1} that matches the key characteristics of our data. We will refer to our simulation exercise when interpreting certain aspects of our estimation strategy, and we also validate specification choices made in the empirical model.

\subsection{Effect of abortion}\label{subsec:effabort}

We estimate the parameter of interest $\xi$ from Equation \eqref{eq:ols_fe} using three different estimators that impose different assumptions on $\alpha_{it}$: the standard OLS estimator, $\widehat{\xi}^\textup{OLS}$, i.e., $\E[\alpha_{it}]=0$; the OLS estimator with individual-specific fixed-effects $\widehat{\xi}^\textup{FE}$, i.e., $\alpha_{it}=\alpha_i$ (OLS-FE); and the GFE estimator, $\widehat{\xi}^\textup{GFE}_G$ for $G=2,3,4$. In all specifications, we control for year fixed-effects. Thus, $\widehat{\xi}^\textup{GFE}_G$ for $G=1$ is equivalent to the estimate $\widehat{\xi}^\textup{FE}$.

\begin{figure}[t]
	\includegraphics[width=0.49\textwidth]{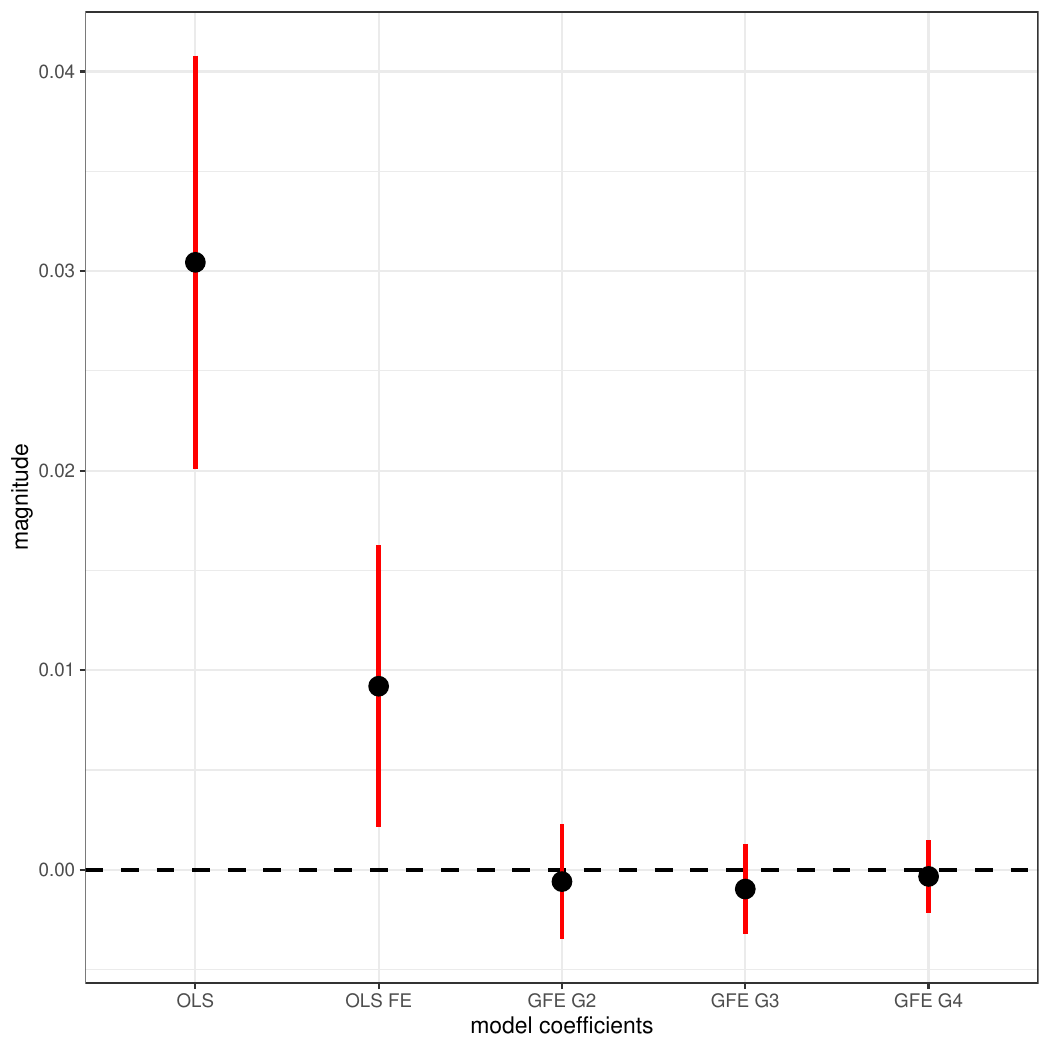}
	\caption{Plot of the estimated coefficients for the OLS estimator, the OLS estimator with individual-specific fixed-effects (OLS-FE), and the GFE estimator with $G=2,3,4$ groups.  }
	\label{fig_coeff}
\end{figure}

Figure \ref{fig_coeff} displays the coefficient estimates for $\widehat{\xi}^{\cdot}$ and the associated $95\%$ confidence intervals. The OLS estimate $\widehat{\xi}^\textup{OLS}$ is large and statistically significant, which is in line with positive associations found in previous studies. The OLS-FE estimate $\widehat{\xi}^\textup{FE}$ is about 70\% smaller than $\widehat{\xi}^\textup{OLS}$, but still positive and highly significant. The GFE estimates $\widehat{\xi}^\textup{GFE}_2$, $\widehat{\xi}^\textup{GFE}_3$ and $\widehat{\xi}^\textup{GFE}_4$ are very close to zero and precisely estimated. We attribute this precision to the large differences in the unobserved heterogeneity profiles. Accounting for these patterns drastically reduces the overall variance. Adding more groups further reduces the estimated standard errors. We observe a similar behavior in our simulations: due to the objective minimized by the estimator, we group individuals with similar time-varying unobserved characteristics, thus mechanically reducing variation when groups are added.

The GFE estimates are considerably smaller than $\widehat{\xi}^\textup{FE}$ and slightly negative.\footnote{The 95\% confidence intervals for the GFE estimates and the OLS FE estimate, $\widehat{\xi}^\textup{FE}$, ($[0.01627,0.00213]$) only marginally overlap for $\widehat{\xi}^\textup{GFE}_2$ ($[0.00331,-0.00348]$). We do not find any overlap in the 95\% confidence intervals for the estimated coefficients with $G=3$ and $G=4$ with that of $\widehat{\xi}^\textup{FE}$.} All GFE point estimates are very similar and lie within each other's 95\% confidence intervals. Our estimates are thus not very sensitive to the chosen number of groups. Our simulations confirm this: once we reach the correct number of groups, the estimated coefficient shrinks to around zero and remains stable when adding superfluous groups (see Figure \ref{fig:sim_coef}).

Our results have meaningful implications for the expected incidence of mental health diagnoses resulting from an abortion. At the sample mean, the OLS estimate predicts that an abortion increases the probability of mental health conditions from 3.2\% to 6.3\%, thus mental health problems almost double. The OLS estimate with individual fixed-effects is much smaller, but still predicts a significant increase in mental health problems by about 29\%, to 4.1\%. By contrast, the GFE estimator always predicts a marginal \emph{decrease} in the incidence of mental health problems. For $G=2$, for instance, the incidence of mental health issues slightly reduce to 3.1\% at the sample mean.\footnote{For $G=3$ the GFE estimate for abortions is -0.0010 and for $G=4$ the estimated coefficient is -0.0003. The estimated coefficients for all models can be found in Table \ref{app:allcoef} in Appendix \ref{app_tab}.}
 
These results illustrate that group-specific time-varying unobserved heterogeneity absorbs considerable variation that may otherwise be attributed to the effect of abortion on mental health. Ignoring time-varying unobserved heterogeneity would lead to severe overestimation of the true abortion effect. 

\subsection{Time profiles of group-specific unobserved heterogeneity}\label{sec:res_alpha}
We next address the question of the optimal number of groups. First, we describe how individuals are assigned to groups for an increasing number of groups. Second, we compute the BIC with two different penalties and discuss coefficient behaviors for different number of groups. Finally, we present the estimated profiles of unobserved heterogeneity.

\begin{figure}[t]
	\includegraphics[width=0.7\textwidth, height=0.49\textheight]{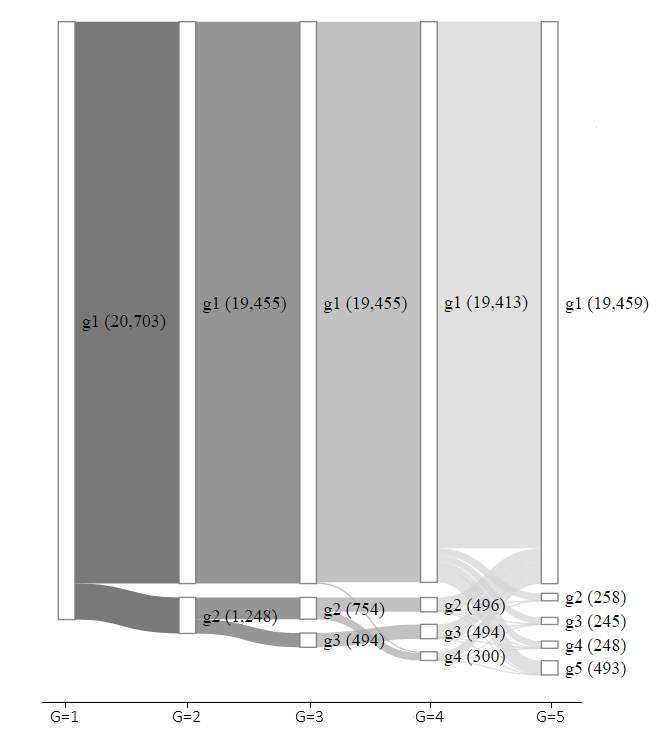}
	\floatfoot{\scriptsize{Note: white bars denote the groups $g$ for each $G$. The number of groups $g$ increases with $G$. Numbers in parentheses correspond to the number of women per group $g$ for 20,703 women.}}
	\caption{Group membership assignment of women for $G=1,2,3,4,5$.}\label{fig:bubble}
\end{figure}

Figure \ref{fig:bubble} shows how the GFE estimator assigns women to groups for $G=1,2,3,4,5$. White bars are nodes and correspond to group $g$ for each $G$. The gray-shaded connections illustrate the flow of women from one group to another when $G$ increases. We start with $G=1$ without grouped unobserved heterogeneity and all women being on individual time-constant trajectories. For $G=2$, the majority of women (93.9\%) are assigned to group $g1$, while 6.1\% are assigned to group $g2$. Setting $G=3$ results in a split of group $g2$ into the two subgroups $g2$ and $g3$. Nothing changes in group $g1$. For $G=4$, a new group $g4$ is formed mostly comprising former members of $g2$. A few women are reassigned from $g1$ to the new group $g4$. For $G=5$, the group assignment becomes rather chaotic. Former members of groups $g2$, $g3$, and $g4$ are assigned back to group $g1$; former members of different groups in $G=4$ are now grouped together; and women from $g1$ are now assigned to groups $g2$--$g5$. This movement pattern indicates that groups are not well-separated anymore for $G=5$ which is an assumption of the GFE estimator (see \cite{bonhomme2015grouped}, and Figure \ref{fig:sim_profile} in Appendix \ref{app_sim}). We thus conclude that the GFE estimator cannot deal with more than four groups in our application. 

\begin{figure}[t]
	\mbox{%
		\subfigure[BIC standard]{\includegraphics[width=0.49\textwidth]{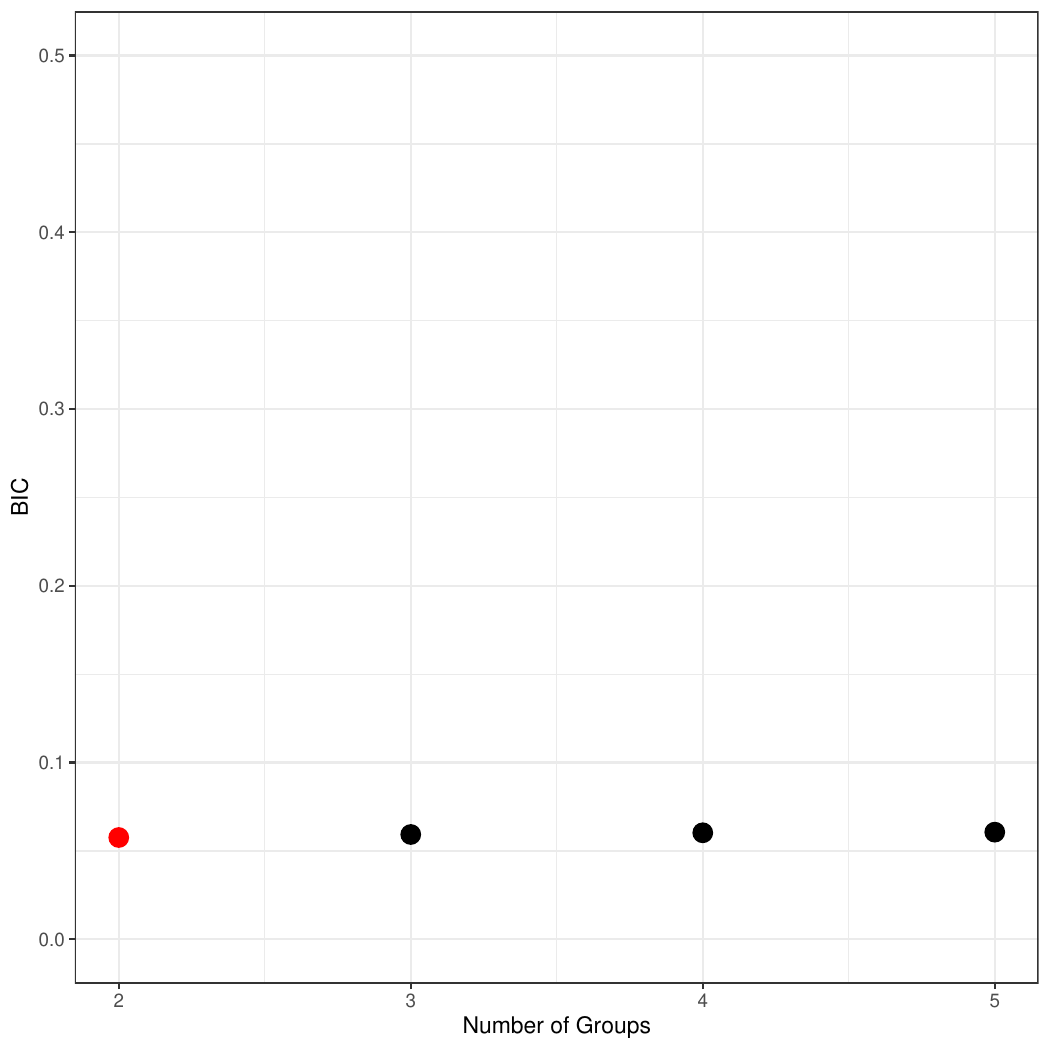}}
		\subfigure[BIC steeper penalty]{\includegraphics[width=0.49\textwidth]{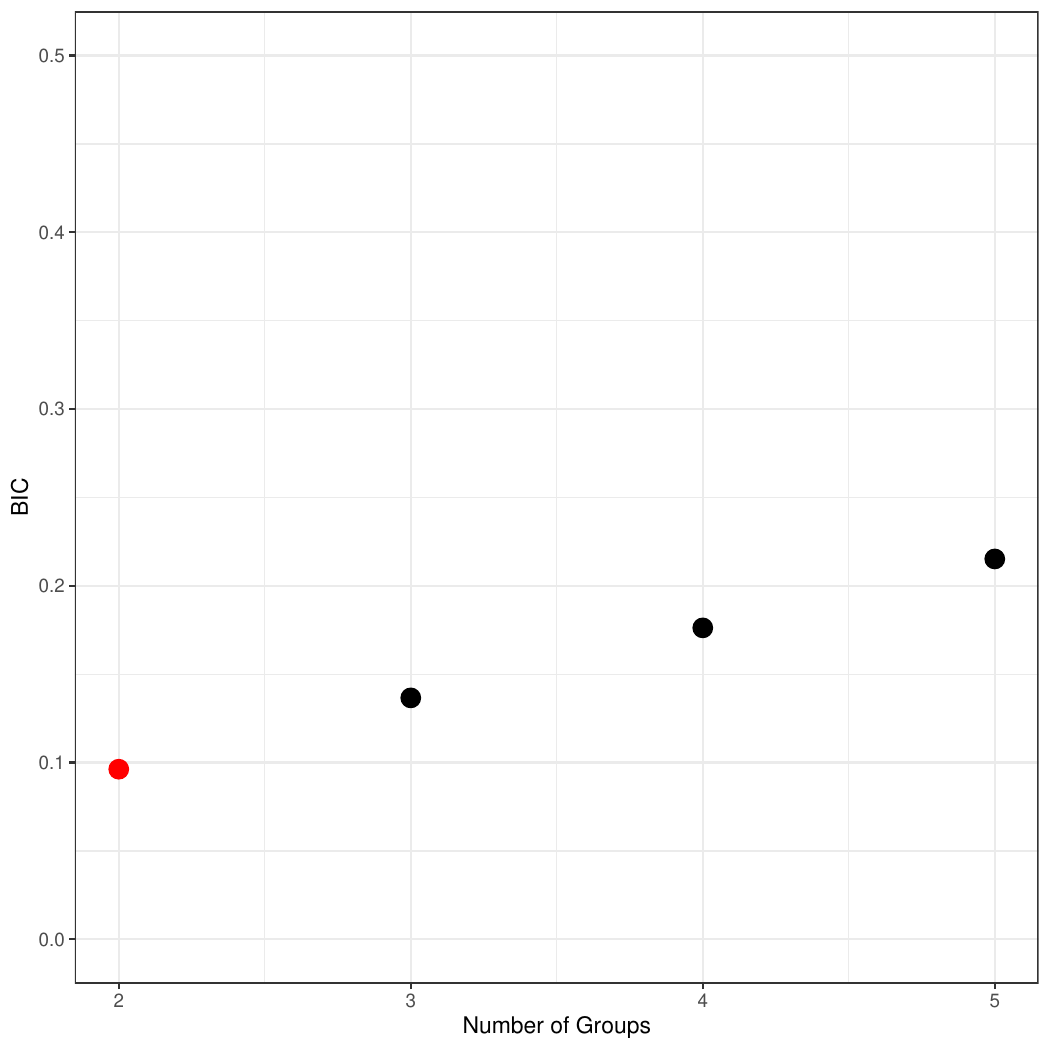}}} 
	\caption{Results for the two information criteria for $G=2-5$ and $G_{max}=10$. }
	\label{fig:BIC_app}
\end{figure}

We next determine the optimal $G$ using the two BIC from Section \ref{subsec:BIC}.\footnote{We set $G_{max}=10$, which is the highest number of groups where the algorithm converges reliably.} Figure \ref{fig:BIC_app} shows that both criteria are minimized at $G=2$ (highlighted in red). The standard BIC hardly varies with $G$, making a clear selection difficult (Figure \ref{fig:BIC_app}(a)). The BIC with the steeper penalty increases sharply in $G$ and is unambiguously minimized at $G=2$ (Figure \ref{fig:BIC_app}(b)). However, our simulations show that the performance of both BIC depends on the true DGP (see Figures \ref{fig:sim_BIC} and \ref{fig:sim_BIC1}). Thus we interpret these results with caution.

As shown in Figure \ref{fig_coeff}, the estimated GFE coefficients are stable after we reach $G=2$. In our simulations, we observe a similar coefficient behavior after reaching the true number of groups, suggesting that coefficient estimates are stable for any $G$ greater than the optimal $G$ (see Figure \ref{fig:sim_coef}). By combining the insights from group movements, the BIC, and the coefficient behavior, we conclude that the true number of groups is likely $G=2$.

\begin{figure}[t]
\includegraphics[width=0.49\textwidth]{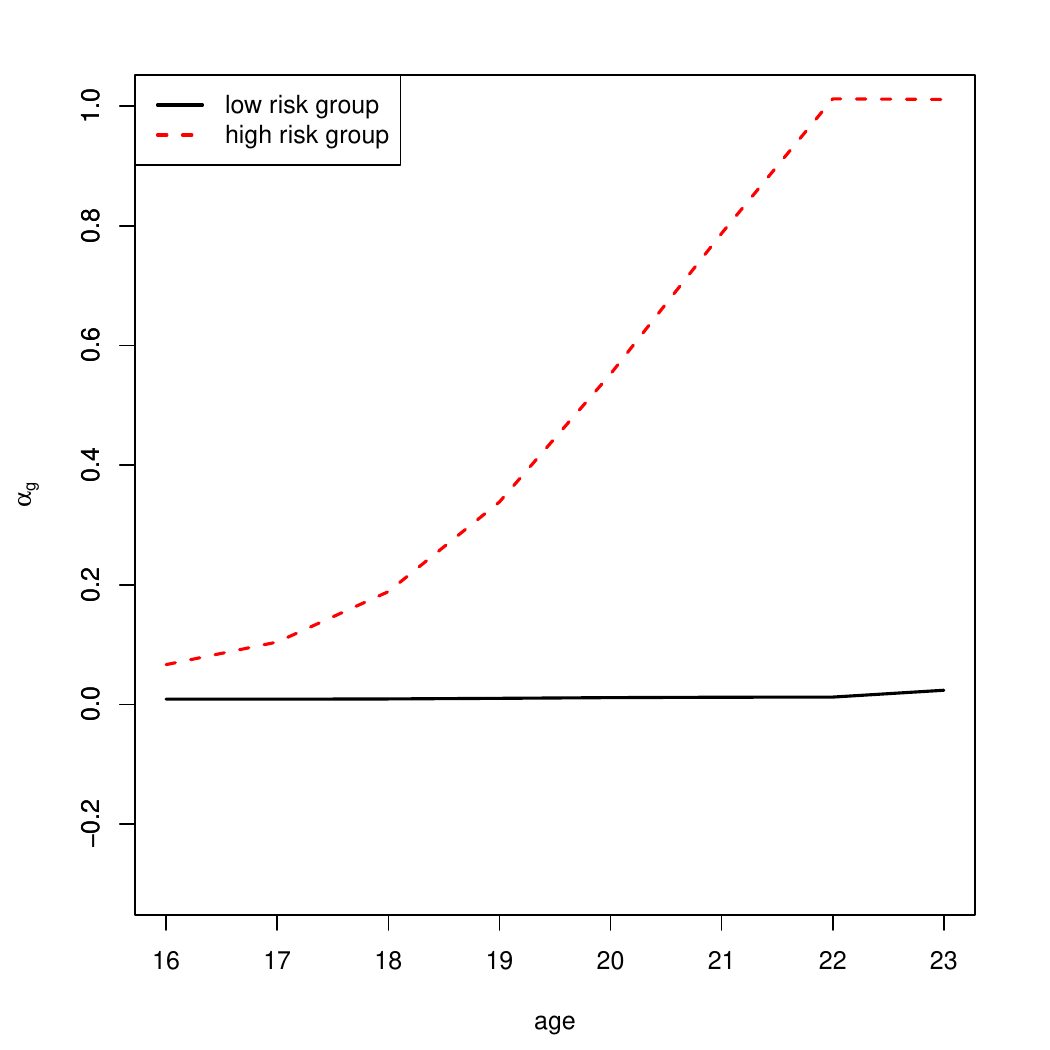}
	\caption{Profiles of unobserved mental health risk for $G=2$ groups.}\label{fig:profile2}
\end{figure}

Figure \ref{fig:profile2} presents the estimated unobserved mental health profiles, $\hat \alpha_{g}$, for $G=2$.\footnote{Figure \ref{fig:profile2} shows the profiles from the GFE without individual-specific time-constant fixed-effects. The profiles net individual-specific fixed-effects can be found in Figure \ref{app:profFE} in Appendix \ref{app_fig}.} The profiles for $G=3$ and $G=4$ are in Figure \ref{app:profG34}. The profiles for $G=2$ exhibit substantial heterogeneity across groups. The solid line represents a rather flat unobserved mental health trajectory. Women with this profile have a low unobserved mental health risk at all ages. We call these women the ``low-risk'' group. The dashed line represents the profile of women with an unobserved mental health risk that is low at age 16 but steeply rises with age. We call these women the ``high-risk'' group. The two profiles differ greatly in both intercept and slope, revealing considerable time-varying unobserved heterogeneity.\footnote{Due to a lack of variation in the group, the high-risk group profile remains rather flat after age 22.}

The group assignment is not only conditional on abortions, but on all covariates controlled for. This implies that our estimated profiles are net of this information. After all, if covariates were sufficient to describe the dynamics of the individual mental health trajectories, additionally controlling for unobserved time-varying heterogeneity would be redundant and the group profiles would be uninformative.\footnote{One could predict the group membership in a nonlinear way by higher order interactions of observed covariates with e.g. $L_2$-Boosting or other unsupervised learning methods.} Yet, it may be informative to compare observed individual characteristics between groups. Table \ref{app:group_char_socio} in Appendix \ref{app_tab} shows that women in the high-risk group have on average a lower socioeconomic background, such as lower parental earnings and higher parental unemployment rates. Also, these women were slightly younger when they terminated an unwanted pregnancy.

\subsection{Robustness checks: alternative dynamic processes and alternative measures of mental health}


Our findings suggest that the association between abortion and mental health disappears once we allow for group-specific time-varying unobserved heterogeneity. We also find that the estimated group-specific heterogeneity profiles starkly diverge with age. To rule out that dynamic processes other than time-varying unobserved heterogeneity cause these patterns, we employ several alternative identification strategies. Finally, we investigate whether abortion affects anxiety disorders and other dimensions of mental health.


\subsubsection{Alternative identification strategies and dynamic processes}\label{subsec:alternatives}
Several studies have shown that women who undergo an abortion have significantly more contacts for psychiatric care, show more symptoms of anxiety and have a history of receiving anti-psychotic or anti-anxiety medication \emph{before} the abortion takes place \citep[see, for instance,][]{steinberg2008abortion, munk2011induced, steinberg2018examining}. Thus, the probability of having an abortion from an unplanned pregnancy could be determined by a woman's past mental health condition. We address such potential issue with reverse causality by adding the first lag of our mental health measure to the right-hand side of Equation \eqref{eq:bm1}. To tackle the endogeneity in the lagged measure of mental health, we combine the GFE estimator for $G=2$ in first differences with an instrumental variable strategy, using the second lag of mental health as instrument \citep{anderson1982formulation}. Column (1) in Table \ref{AltIDent} shows that the estimated impact of abortion on mental health is close to zero and insignificant when controlling for past mental health. It suggests that time-varying unobserved heterogeneity captures a large share of the mental health dynamics. 

\begin{table}[t]
	\caption{Estimated coefficients for specifications with lagged mental health and instrument}
	\begin{tabularx}{\textwidth}{@{}X T{0.5}T{0.5}T{0.5}@{}} 
		\toprule
		&\multicolumn{1}{c}{Dynamics: lagged mental health} & \multicolumn{2}{c}{IV analysis}\\
		\cmidrule(lr){2-2} \cmidrule(lr){3-4}
		& {GFE, $G=2$} & {} & {}\\
		& {(1)}	& {(2)} 	& {(3)}	\\
		\midrule
		Abortions unwanted preg.&-0.0010	& -0.0116	& \\
								&(0.0038)	&(0.0231)	& \\
		All medical abortions	&			&			& 0.0010 \\
								&			&			& (0.0228) \\
		Mental health $t-1$		&1.0022***	&			& \\
								&(0.0324)	&			& \\[1em]
		Instruments&			\multicolumn{1}{c}{$MH_{t-2}$} & \multicolumn{2}{c}{$\widehat{Miscarriage}$}	\\[1em]
		First stage: Pr(Abortion) & & & \\
		Miscarriage				&			& -0.6751*** & -0.6808***\\				
								&			& (0.0187)	 & (0.0186)   \\[1em]
								
		Number women &\multicolumn{1}{c}{20,703}	& & \\					
		Observations &\multicolumn{1}{c}{144,921} 	& \multicolumn{1}{c}{4,912} & \multicolumn{1}{c}{4,679}\\ 
		\bottomrule
		\multicolumn{4}{@{}p{\linewidth}}{\scriptsize{Standard errors clustered on the individual level; *** $p< 0.01$, ** $p<0.05$, * $p<0.1$; Column (1): IV estimates in first differences using estimated group assignments from GFE with individual-specific fixed effects and the second lag of mental health diagnoses as instrument. Column (2): Estimated coefficient of abortion from an unintended pregnancy obtained from TSLS. The sample contains only women who either gave birth, had a miscarriage or had an abortion from an unintended pregnancy. Miscarriages are used as an instrument. Column (3): Estimated coefficient of all medical abortion obtained from TSLS. The sample contains only women who either gave birth, had a miscarriage or had a medical abortion. Miscarriages are used as an instrument. Control variables: woman: relationship status (single, in a relationship), log earnings, college degree, employed; mother: log earnings, employed, college degree, relationship status; father: log earnings, employed, college degree; log household disposable income; year fixed-effects, municipality FE, year of birth FE for woman/mother/father; indicator missing observations.}}					
	\end{tabularx}
	\label{AltIDent}
\end{table}

\begin{figure}[t]
	\mbox{%
		\includegraphics[width=0.49\textwidth]{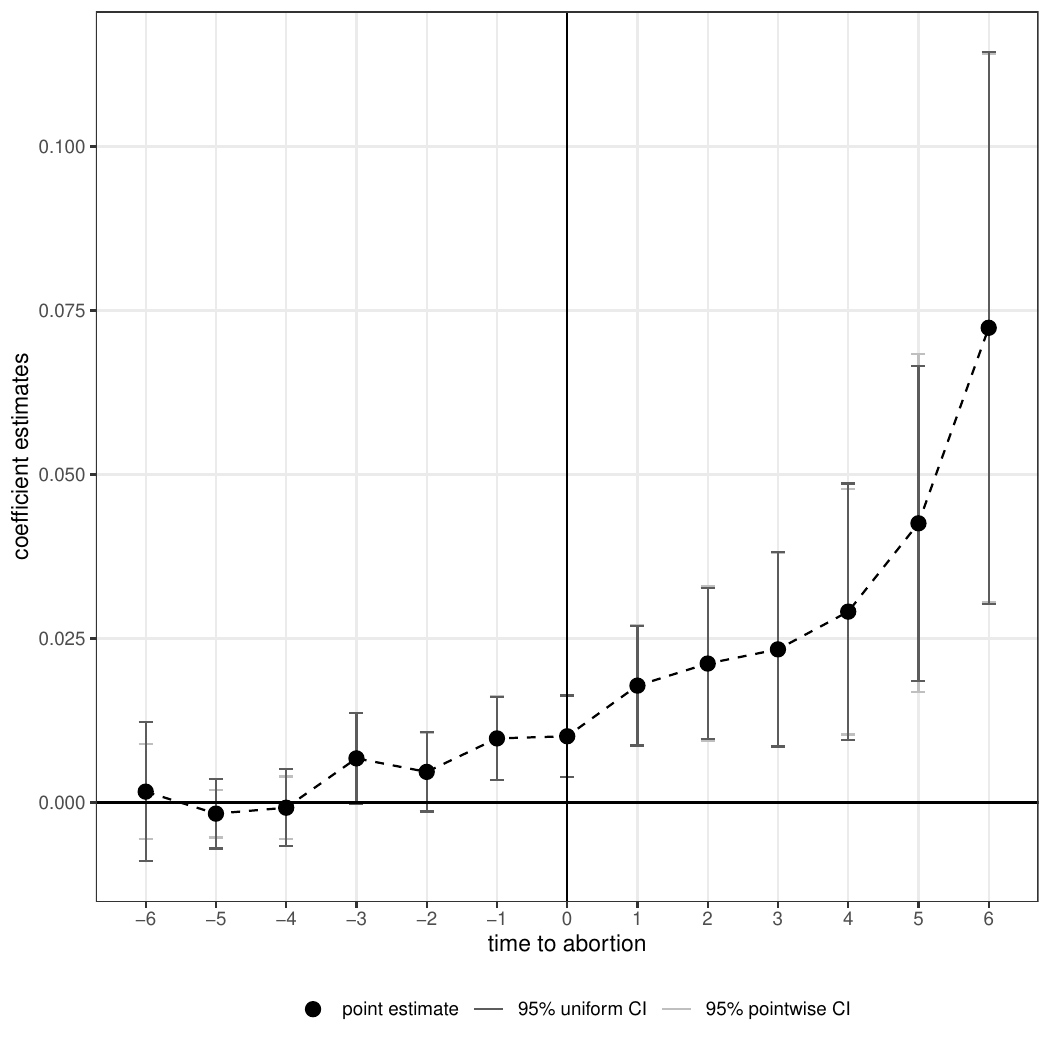}
	}
\floatfoot{\footnotesize{\emph{Note: The aggregated time-group ATT is $0.0309,\, se=0.0073.$}}}
	\caption{Estimated dynamic treatment effects of abortion on mental health.}
	\label{fig:eventstudy_coeff}
\end{figure}

Another dynamic process could be that abortion effects on mental health occur with some lag. In our empirical specification, the existence of such dynamic effects would downward bias the estimated GFE coefficients.\footnote{Our estimated GFE coefficient is downward biased if the dynamic effects have the same sign as the contemporaneous effect.}
To investigate potentially dynamic abortion effects, we plot mental health against the abortion event. Figure \ref{app:eventtime_bc}, Appendix \ref{app_fig}, shows that our unconditional mental health measure exhibits a strong time trend but develops very smoothly around the abortion event. Next, we apply the dynamic difference-in-difference (dynamic-DiD) estimator suggested by \cite{callaway2021difference}. This method estimates the group-time average treatment effect on the treated (ATT), extending the doubly robust DiD estimator of \cite{sant2020doubly} to multiple time periods. The main identifying assumption for the ATT is that parallel trends hold conditionally on covariates. Figure \ref{fig:eventstudy_coeff} shows the estimated group-time ATT if not-yet-treated observations are the control group.\footnote{We also estimated the dynamic-DiD using the never-treated as a control group. The results are very similar to those with the not-yet-treated control group and thus not presented here.} As in Figure \ref{app:eventtime_bc}, there is no discontinuity around the abortion event. However, the estimated coefficients follow a clear time trend, with some being significantly different from zero already several periods before the abortion event. This points towards a violation of the conditional parallel trends assumption. Pre-testing the parallel trends assumption further rejects the null hypothesis of parallel trends (Wald statistic  $p$-value$=0.001$). The rejection of conditional parallel trends is in line with our finding that the unobserved heterogeneity profiles starkly differ in a non-parallel way across groups.

Our empirical results could also be explained by age dynamics in abortion effects. If the effect of abortion on mental health depends on age, e.g., early abortions have stronger effects than late abortions, we might mistakenly attribute an early-abortion effect to age-varying unobserved heterogeneity, resulting in underestimating the abortion effect. Table \ref{app:agedependentAbortion}, Appendix \ref{app_tab}, displays the results from an OLS model with individual fixed-effects and age-dependent abortion. We do not find any significant age-dependent abortion effects. 

We finally apply an instrumental variables (IV) estimator to address general endogeneity concerns in the abortion \emph{decision}. We follow \cite{hotz1997bounding} and \cite{hotz2005teenage} who used miscarriages as instrument for teenage birth to investigate the long-term consequences of teenage childbearing. We adapt the estimator by using miscarriages as an instrument for abortion decisions. The instrument is valid if miscarriages are random and if the latent proportion in women with a miscarriage that would have had an abortion is equal to the observed proportion in the population \citep[see][]{hotz1997bounding}. To implement the IV strategy, we restrict the sample to women who either had an abortion or a miscarriage or gave birth at a given age. We construct two different samples: one that contains all medical abortions; and one that contains abortions from an unintended pregnancy. In the former sample, 59\% of women had an abortion and 35\% gave birth. Only a minority of 6\% had a miscarriage. The share is similar for the sample that only contains abortions from an unwanted pregnancy. Columns (2) and (3) in Table \ref{AltIDent} display the IV results. The first stage results suggest that miscarriages are reasonably relevant for having an abortion. The point estimates for abortion in the second stage are negative or close to zero and highly imprecisely estimated.\footnote{Both confidence intervals include the OLS point estimates obtained from these samples (0.0278 (0.011) for Column (2), 0.0245 (0.011) for Column (3)). As shown in \cite{young2021leverage} imprecisely estimated coefficients are a common feature of IV strategies.} Besides, there are concerns that miscarriages are a valid instrument. A significant share of miscarriages does not occur at random but is related to non-random, unobserved behavioral risk-factors which may be correlated with mental health problems \citep[see][]{rellstab2021effect}.



\subsubsection{Other dimensions of mental health problems} 

 In following the psychiatric literature, we consider as alternative measures of mental health issues: anxiety and fear-related disorders, bipolar disorders, depression and affective mood disorders without bipolar disorders \citep[see, for instance,][]{steinberg2008abortion,foster2015comparison,steinberg2018examining}.

Table \ref{AltMeasures} presents the corresponding estimated coefficients on abortion obtained from OLS without and with individual-specific fixed effects and from the GFE estimator with two groups and individual-specific fixed effects. Line E shows the estimated coefficient from our main specification. Regardless of the estimation strategy, the estimated association between an abortion due to an unplanned pregnancy is strongest for anxiety and fear-related disorders. However, the estimated GFE coefficient of abortion on anxiety disorders is only about 10 percent of the magnitude of the OLS FE coefficient in Column (2) and not significantly different from zero. The coefficient estimates for bipolar disorders are relatively small. The estimated coefficients for depression are similar in magnitude to those of our main specification, suggesting that depression is the primary driver. This is confirmed from the estimation results for mood disorders without bipolar disorders.




\begin{table}[t]
	\caption{Estimated coefficients for alternative measures of mental health, obtained from OLS without and without individual fixed-effects and GFE for $G=2$.}
	\begin{tabularx}{\textwidth}{@{}X T{0.5}T{0.5}T{0.5}@{}} 
		\toprule
		&\multicolumn{3}{c}{Estimated coefficients on abortion}\\
		\cmidrule(lr){2-4}
		& {OLS} & {OLS FE} & {GFE, $G=2$}\\
		Alternative outcomes & {(1)} & {(2)} & {(3)}  \\
		\midrule
		A. Anxiety \& fear-related disorders 	& 0.0523*** & 0.0157*** & 0.0017 \\
											& (0.0065) 	& (0.0047) 	& (0.0019) \\			
		B. Bipolar disorders 					& 0.0022* 	& 0.0015* 	& -0.0002 \\
											& (0.0012) 	& (0.0009) 	& (0.0002) \\
		C. Depression 							& 0.0283*** & 0.0089** 	& 0.0001  \\
											& (0.0051) & (0.0035) 	& (0.0014) \\
		D. Affective mood disorders 			& 0.0284*** & 0.0077** & -0.0004 \\
		without bipolar disorders			& (0.0052) & (0.0035)  & (0.0015) \\
		E. Main specification mental health	& 0.0306*** & 0.0092*** & -0.0006 \\ 
											& (0.0053)  & (0.0036) & (0.0015) \\
		Number women						&\multicolumn{3}{c}{20,703}  \\					
		Observations					 	&\multicolumn{3}{c}{165,624}\\ 
		\bottomrule
		\multicolumn{4}{@{}p{\linewidth}}{\scriptsize{Standard errors clustered on the individual level; *** $p< 0.01$, ** $p<0.05$, * $p<0.1$; Column (1): OLS regression of alternative measures of cumulative mental health diagnoses on abortion. Column (2): OLS regression with individual fixed-effects. Column (3): GFE estimation with $G=2$ groups and individual-specific fixed-effects. Control variables: woman: relationship status (single, in a relationship), log earnings, college degree, employed; mother: log earnings, employed, college degree, relationship status; father: log earnings, employed, college degree; log household disposable income; year fixed-effects, municipality FE, year of birth FE for woman/mother/father; indicator missing observations. }}					
	\end{tabularx}
	\label{AltMeasures}
\end{table}

\subsection{Sources of unobserved heterogeneity: Abortions, unwanted pregnancies and other risky behavior}\label{subsec:risky_beh}

A natural question is what factors are captured by profiles of unobserved mental health risk. While there may be several answers, one explanation is that the estimated profiles proxy common risky behaviors among young women. This includes unprotected sexual activity but also other risky behaviors such as drug- and alcohol consumption \citep{cawley2011economics}. In this case, controlling for such \emph{observed} behaviors would alter the estimated association between abortion and mental health. Alternatively, the profiles might absorb choice processes underlying different behaviors. In that case, observed risky behavior would result from a similar decision process. It would imply that (1) the association of abortion and mental health is robust to controlling for other risky behaviors; (2) the GFE estimate for abortion is unaffected, but estimates for other behaviors behave similarly to that for abortion; (3) other risky behaviors are contemporaneously correlated with the estimated unobserved heterogeneity profiles. We now assess how other risky behaviors are related to abortions, mental health, and the estimated unobserved heterogeneity profiles.

\subsubsection{Mental health, abortions and other risky health behaviors}
An important determinant for having an abortion is a woman's decision to engage in unprotected sexual activities, resulting in an unwanted pregnancy. Ex-ante, it is not clear whether an unwanted pregnancy reflects such a choice, including careless use of birth control, or whether it resulted from a random failure in contraception or sexual assault. In the latter cases, unwanted pregnancies are not in the choice set that may be captured by time-varying unobserved heterogeneity.\footnote{Even if all women face the same failure probability of contraception, one may still find a positive correlation between abortion probabilities and mental health diagnoses. For instance, women with mental health problems might start to have sex at earlier ages than women without mental health problems or have sex more frequently. Then this correlation would not be indicative of risky sexual behavior.} If abortions are outcomes of a woman's choice to engage in unprotected sex, then this may not only result in unwanted pregnancies but also other byproducts of unprotected sex. In our data, we observe a few other risky sexual and health behaviors \citep[e.g.][]{ markowitz2005investigation, cawley2011economics, mulligan16}: chlamydia infections and sexually transmittable disease (STD) screenings as risky sexual behavior, and excessive alcohol consumption as other risky behavior.\footnote{Chlamydia is the most frequently observed STD among young women (e.g., \cite{danielsson2012sexual}, \cite{STD2018EU} for Sweden, and \cite{STD2018US} for the US). Chlamydia infections are measured with ICD-10 codes A55,A56; STD screenings other than HIV are measured with the ICD-10 code Z113. Excessive alcohol consumption is measured with the ICD-10 code F110, ``acute drunkenness (in alcoholism)''.} 

  \begin{table}[t]
	\caption{Correlations between abortions and other risky health behavior among women aged 16--23 years}
	\begin{tabularx}{\textwidth}{@{} X T{4.5}@{\qquad}T{4.5}} \toprule
		&\multicolumn{2}{c}{Ever had}\\ 
		\cmidrule(lr){2-3}
		&{an abortion from} & {an unwanted}\\
		&{an unwanted pregnancy} & {pregnancy}\\
		\midrule
		Ever had a chlamydia diagnosis				& 0.145***	& 0.153*** \\
		& (0.014)	& (0.015)\\
		Ever had a STD screening					& 0.014*	& 0.025*** \\
		& (0.007)	& (0.008)\\
		Ever had a diagnosis of acute drunkenness 	& 0.098***	& 0.125*** \\
		& (0.024)	& (0.025)\\[1em]
		Sample mean in \%							& 10.6		& 12.6\\ 												  	
		Number women				&\multicolumn{2}{c}{20,703}  \\					
		Number observations			&\multicolumn{2}{c}{165,624} 	\\	\bottomrule
		\multicolumn{3}{@{}p{\linewidth}}{\scriptsize{Standard errors clustered on the individual level; *** $p< 0.01$, ** $p<0.05$, * $p<0.1$; OLS regressions with individual-specific FE of ever had an unwanted pregnancy on ever had a diagnosis on chlamydia/had an STD screening/diagnosis on excessive drinking. Control variables: woman: relationship status (single, in a relationship), log earnings, college degree, employed; mother: log earnings, employed, college degree, relationship status; father: log earnings, employed, college degree; log household disposable income; year fixed-effects, municipality FE, year of birth FE for woman/mother/father; indicator missing observations. }}					
	\end{tabularx}
	\label{sexualBeh}
\end{table}%

Table \ref{sexualBeh} shows that other risky behaviors are strongly correlated with abortions and unwanted pregnancies. Column (1) shows that women with chlamydia infection at age 16--23 have, on average, a 14.5 percentage points higher likelihood for an abortion, translating into a more than 130\% increase at the sample mean. STD screenings increase the likelihood for an abortion by 1.4 percentage points or 13\% at the sample mean. Excessive drinking at age 16--23 increases the probability of abortion by 9.8 percentage points or 92\% at the sample mean. Column (2) shows that the correlations between risky behaviors and unwanted pregnancies are somewhat stronger than for abortions but otherwise very similar.

We next examine whether other risky behaviors are omitted controls or whether they result from a similar choice process as abortions.\footnote{Abortions and unwanted pregnancies may follow differential selection. In our sample, 82\% of women with an unwanted pregnancy have an abortion. When regressing mental health on abortions and unwanted pregnancies, the abortion effect becomes small and insignificant, indicating no differential selection.} To this end, we re-estimate our main specifications and gradually add excessive drinking, chlamydia infections, and STD screenings. Columns (1)--(4) in Table \ref{MHRiskBeh} display the estimated coefficients for models with individual fixed-effects. All associations are positive, suggesting that these behaviors increase the probability of being diagnosed with mental health problems, but only the coefficient on STD screenings is significantly different from zero. Adding these behaviors as controls barely changes the estimated association between abortion and mental health. Column (5) presents the results from the GFE estimator for $G=2$. The added controls do not change the impact of abortion on mental health. The estimated GFE coefficients for other behaviors are 5--10 times smaller than in Column (4).  

\begin{table}[t]
	\caption{Estimated correlations between mental health development and risky behavior}
	\begin{tabularx}{\textwidth}{@{}X T{0.2}T{0.2}T{0.2}T{0.2}T{0.2}@{}} 
		\toprule
		&\multicolumn{5}{c}{Woman has mental health problems at 16--23}\\
		\cmidrule(lr){2-6}
		&\multicolumn{4}{c}{OLS FE} & {GFE}\\
		\cmidrule(lr){2-5} \cmidrule(lr){6-6}
		& {(1)} & {(2)} & {(3)} & {(4)} & {(5)} \\
		\midrule
		Abortion 					& 0.0092** 	& 0.0091**  & 0.0092** 	&  0.0091**	& -0.0006\\
		& (0.0036) 	& (0.0036) 	& (0.0036) 	&  (0.0036)	& (0.0015)\\	
		Acute drunkenness 			&  0.0084	& 			&  			& 0.0083  	& -0.0009\\
		& (0.0159)	& 			&     		& (0.0159) 	&  (0.0038)\\
		Chlamydia infection 		&  			& 0.0051		&  		& 0.0039  	&  0.0007 \\
		&  			& (0.0051)	&    		& (0.0050) 	& (0.0023)\\
		STD screening  				&  			&  			& 0.0084**	& 0.0081**	& 0.0006\\
		&  			&  			& (0.0035)	& (0.0035) 	& (0.0017)\\		
		Number women				&\multicolumn{5}{c}{20,703}  \\					
		Observations &\multicolumn{5}{c}{165,624}\\ 
		\bottomrule
		\multicolumn{6}{@{}p{\linewidth}}{\scriptsize{Standard errors clustered on the individual level; *** $p< 0.01$, ** $p<0.05$, * $p<0.1$; Columns (1)--(4): OLS regression of cumulative mental health diagnoses on abortion and current risky health behavior, controlling for individual-specific FE. Column (5): GFE estimation with $G=2$ groups and individual-specific FE. Control variables: woman: relationship status (single, in a relationship), log earnings, college degree, employed; mother: log earnings, employed, college degree, relationship status; father: log earnings, employed, college degree; log household disposable income; year fixed-effects, municipality FE, year of birth FE for woman/mother/father; indicator missing observations. }}					
	\end{tabularx}
	\label{MHRiskBeh}
\end{table}

The relationship between mental health and abortions could be influenced by past rather than current health behaviors \citep[e.g.][]{elkington2010psychological, hallfors2005comes}. Table \ref{app:mhrisk} in Appendix \ref{app_tab} shows the results when using lagged diagnoses on acute drunkenness, chlamydia infections, and STD screenings. The estimated associations between mental health and past health behaviors are strong and significant, but the abortion effect is again robust.
Overall, our results suggest that our observed health behaviors are unlikely to cause the omitted variable bias observed in OLS-FE regressions. Instead, these behaviors seem to result from similar decisions as abortions from unwanted pregnancies.

\subsubsection{Unobserved heterogeneity profiles and risky health behaviors} 

\begin{figure}[t]
	\mbox{%
		\subfigure[STD screening]{\includegraphics[width=0.32\textwidth]{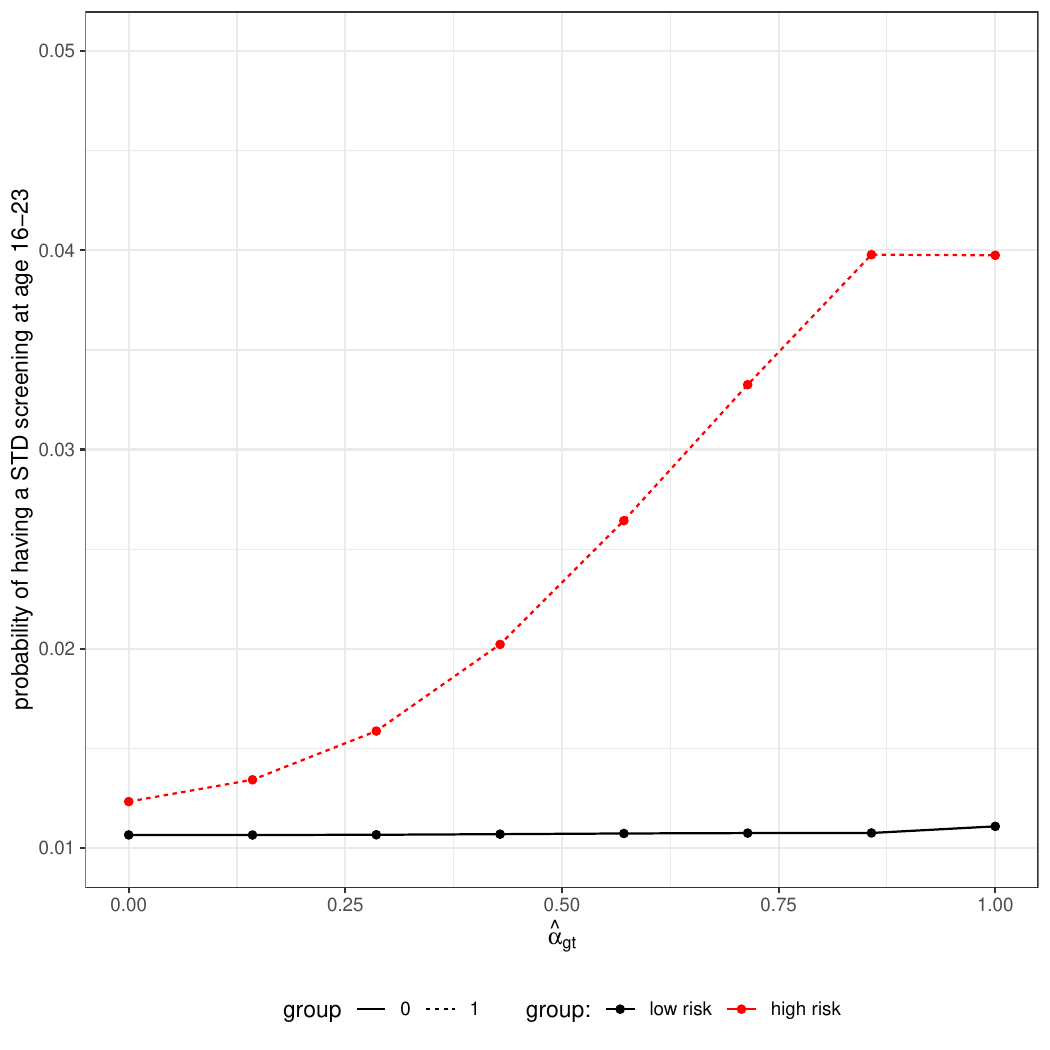}}
		\subfigure[Chlamydia infection]{\includegraphics[width=0.32\textwidth]{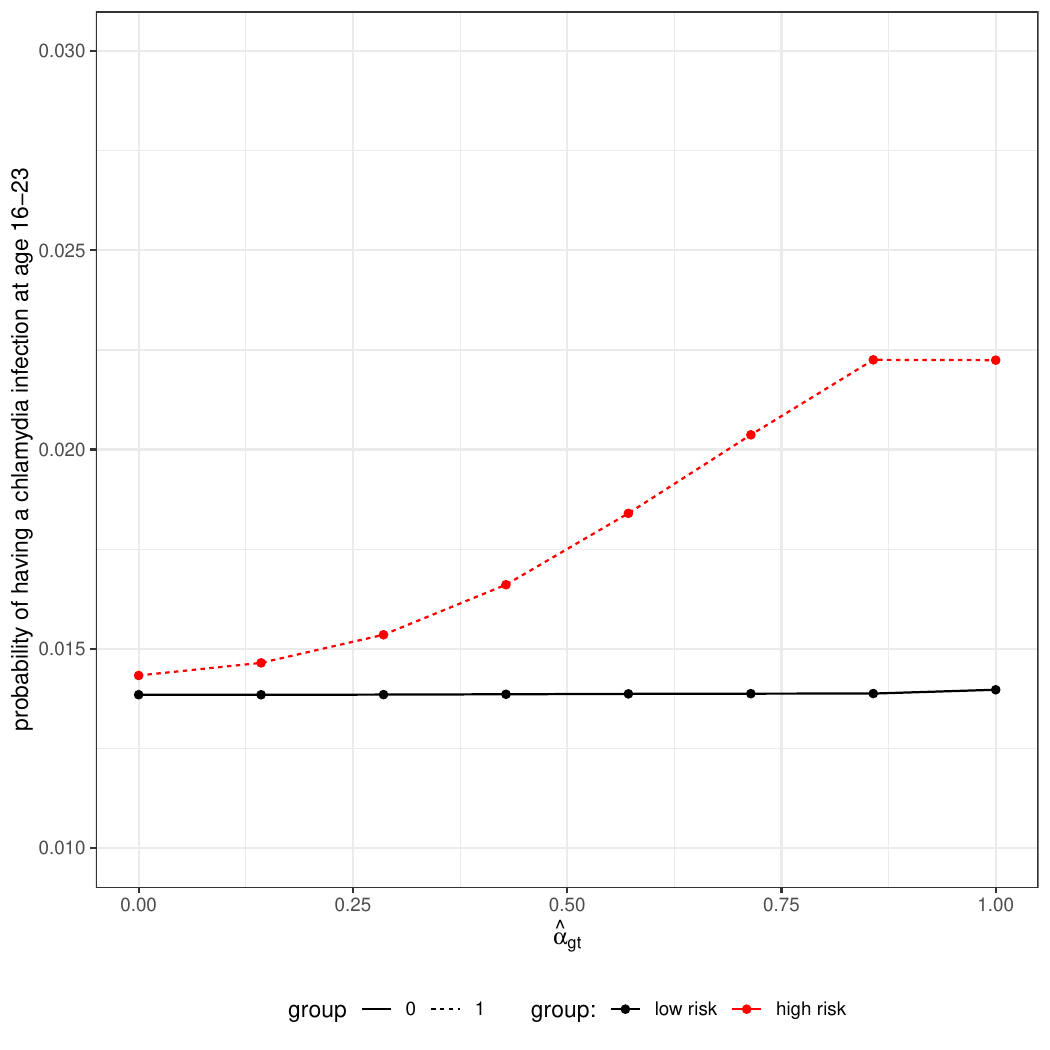}} 
		\subfigure[Excessive drinking]{\includegraphics[width=0.32\textwidth]{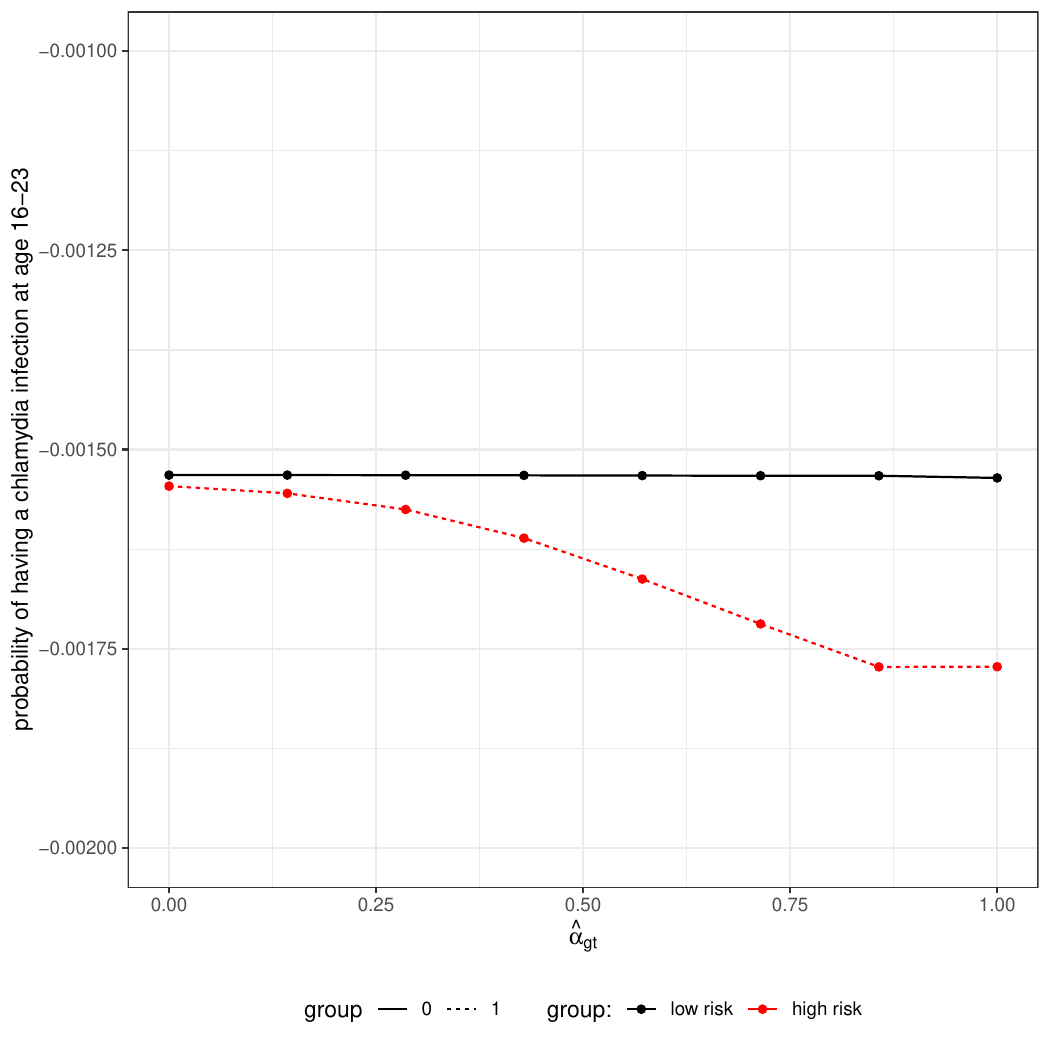}} }
	\caption{Predicted risk of being diagnosed with a certain health behavior by group-specific unobserved mental health risks.}
	\label{fig:predict_risk}
\end{figure}

We finally investigate whether the estimated profiles of unobserved mental health risk are correlated with other risky behaviors. We regress STD screenings, chlamydia infections, and excessive drinking on $\hat \alpha_{gt}$ and covariates and plot the group-specific predictions against $\hat \alpha_{gt}$. Figure \ref{fig:predict_risk} shows the predicted diagnosis risks for $G=2$.\footnote{The respective coefficient estimates can be found in Table \ref{app:predictRISK} in Appendix \ref{app_tab}.} In the high-risk group, the probability of STD screenings and chlamydia infections steeply increases with $\hat \alpha_{gt}$. By contrast, the probabilities are flat in the low-risk group. Alcohol intoxication is an exception. Here, group differences in predicted probabilities are very small and slightly negative for high-risk women.\footnote{Risky drinking typically happens at earlier ages than risky sex. \cite{marcus2015reducing} show that most hospitalizations from alcohol intoxication among women take place before age 16 and then sharply declines. We observe a similar decline in excessive drinking at age 16--20 (see Figure \ref{app:alcohol}, Appendix \ref{app_fig}).} Overall, Figure \ref{fig:predict_risk} shows that high-risk women have higher probabilities of risky sexual behavior, which confirms the suggested correlation between risky behaviors and unobserved heterogeneity. 

These findings strengthen our interpretation that time-varying unobserved heterogeneity captures choice processes for engaging in risky behavior. Researchers rarely observe these decisions but measure realized behaviors which are outcomes of these decisions. We have shown that controlling for such observed behaviors is insufficient to obtain an unbiased estimate in our application. Instead, the GFE estimator seems necessary to account for the unobserved decision-making processes.

\section{A framework of mental health and risky behavior}\label{sec:model}
The results obtained in Section \ref{subsec:risky_beh} suggest the following explanation: women differ in their decisions to engage in risky behaviors, which are reflected in differences in estimated group profiles. One reason for a large amount of group-specific heterogeneity could be that women have different preferences, leading to differences in dynamic decisions and thus to different mental health trajectories. \citet{o2001risky} discuss the role of time-inconsistent preferences for risky behaviors among youths, e.g., unprotected sex. Present-biased preferences make unprotected sex today more likely since teenagers weigh the benefits today much higher than potential future costs \citep{levine2001sexual}. This behavioral bias affects all dynamic behaviors, e.g., educational choices like school drop-outs. \citet{cobb2020depression} suggest that self-control problems explain differences in the correlation between depression and risky behaviors like a lack of exercise.

Based on this discussion, we formulate a theoretical model of endogenous mental health and risky choices. Risky behavior leads to short-term benefits but harms mental health development. At the same time, mental health problems change the preferences for risky behavior and thereby shape its time paths. To allow for heterogeneity across women, we introduce non-standard time preferences. Women in the high-risk group have a high degree of present bias, over-weighting current pleasure compared to future mental health risks. Women in the low-risk group have preferences that are close to time consistent. As such, our model closely follows the literature in behavioral economics.

Our model offers an interpretation for differences in inter-temporal decision-making across the two groups and the consequences for mental health development. Of course, it is not the only model that could explain the observed patterns. For instance, heterogeneity in decision-making and mental health could be driven by heterogeneity in impatience, i.e., by different time discounting without present bias. Through the present bias, we stress the importance of \textit{now} regardless of the future \citep{laibson1997golden,o2015present}. It seems plausible that a woman at a party who meets a handsome guy decides in the ``heat-of-the-moment'' to have unprotected sex even though she may be aware of future costs, e.g., in mental health costs. However, she might say no if you ask her whether she should behave this way at the next party. The behavioral literature discusses several other models that incorporate anomalies in discounted utility, such as ``visceral influences'', habit formation or projection bias \citep[for a discussion, see][]{frederick2002time}. Our exploratory theoretical analysis does not aim at differentiating between these models.

\subsection{A DGP for mental health, risky behavior and abortion}

We formulate a data generating process (DGP) of risky decision making, abortion, and mental health. For simplicity, we assume that latent mental health $M$ is generated by
\begin{flalign} \label{eq: dgp}
M_{it+1} = \psi M_{it}+ \zeta\rho_{it}  + \epsilon_{it}. 
\end{flalign}

Woman $i$'s mental health at age $t+1$, $M_{it+1}$, is determined by her mental health at age $t$, her risky choice, $\rho_{it}$, and an iid mental health production shock $\epsilon_{it}\sim N(0,\sigma_\epsilon)$. To keep the model tractable we ignore covariates.\footnote{In our empirical analysis we proxy this latent mental health status by observed diagnoses.} Abortion probabilities do not enter Equation \eqref{eq: dgp} directly but are correlated with risky choices. We model the probability of having an abortion $A$ at age $t$, $A_{it}$, as a function of unobserved risky choices $\rho_{it}$ (systematically varying with $A_{it}$), and an idiosyncratic error (e.g. $\eta_{it}\sim N(0,\sigma_{\eta})$),
\begin{flalign}
\label{eq: ab}
Pr(A_{it}=1)=\E\left[ \mathbb{I}(\rho_{it} + \eta_{it}> 0)\right] , ~~~\text{increasing in } \rho_{it}.
\end{flalign}

Together with Equation \eqref{eq: dgp}, Equation \eqref{eq: ab} implies that a regression of $M_{it}$ on $A_{it}$ would produce a spurious correlation even with individual-specific fixed-effects. Since we only observe the abortion but not women's decision to engage in unprotected sex, $\rho_{it}$, we could interpret the observed abortion as a signal for risky decision making.

\subsection{Preferences}

We assume that women are sophisticated decision-makers who know about their self-control problems when making choices \citep{o1999doing}. At each age $t$, a woman enjoys flow utility, $u(\rho_t, M_t)$, which is a function of mental health and chosen risky behavior.\footnote{We subsequently suppress the individual subscript $i$ for ease of notation.} We assume $u$ to be a constant relative risk aversion (CRRA) utility function with mental health dependent risk aversion minus quadratic mental health costs,\footnote{In the CRRA term, we multiply by the sign of the exponent rather than dividing by it which would be more common. This is useful when calibrating/estimating the model because in this formulation small changes in risk aversion do not have a strong impact on the levels of utility.}  
\begin{flalign}\label{eq: crra}
u(\rho_t,M_t)= \text{sgn}(1-a-c M_t)\cdot\rho^{(1-a-c M_t)}-b M_t^2 .
\end{flalign} 

The parameter $a$ is the baseline level of risk aversion, which is modified by the mental health dependent term $c \cdot M_t$. Women with positive $c$ become more risk-averse as their mental health problems increase, introducing additional preference heterogeneity.\footnote{By letting risk aversion depend on mental health, we follow an early version of \cite{cronin2020good}. In our application allowing for $c\neq 0$ leads to a considerably better model fit than $c=0$.} The second term captures direct costs of mental health problems, which are determined by $b$.

We formulate an infinite horizon decision problem and focus on the first eight periods, corresponding to ages 16--23 in our data. Realized future utility at age $t$ is given by
\begin{flalign*}
u(\rho_t,M_t) + \beta \sum_{\tau = t+1}^{\infty}\delta^{\tau-t}u(\rho_\tau,M_\tau),
\end{flalign*}
where the first term, $u(\rho_t,M_t)$ is the current flow utility. The second term aggregates future flow utility using $\beta-\delta$ discounting. $\delta$ is the usual exponential discount factor. The parameter $\beta$ induces the self-control problem. For $\beta=1$, the model is one with standard exponential discounting. For $0<\beta<1$, a woman exhibits some degree of present bias.

For a current level of mental health $M_t$, the problem a woman solves at age $t$ is
\begin{flalign}\label{eq: opt}
\max_{\rho_t} u(\rho_t,M_t) + \E_t\left[ \beta \sum_{\tau = t+1}^{\infty}\delta^{\tau-t}u(\rho^*_\tau(M^*_\tau),M^*_\tau)\right].
\end{flalign}

The $\rho^*_\tau$ are the optimal decisions of the future selves as functions of current mental health. $M^*_\tau$ is the mental health trajectory that starts at $M_t$ when choosing $\rho_t$ at age $t$ and choosing $\rho^*_\tau(M^*_\tau)$ at later ages. Solving this problem for every possible value of $M_t$ pins down the decision function $\rho^*_t$. $\E_t$ is the conditional expectation at age $t$.

We solve the model by backward recursion. This is a variation of classical dynamic programming that takes into account the time-inconsistency introduced through $\beta$.\footnote{Details about the model solution can be found in Appendix \ref{app_modelsolution}.} As in the empirical analysis, we allow for two groups of unobserved mental health risks. We assume two different degrees of present bias, defined by two parameters $\beta_1,\beta_2 \in (0,1)$ \citep{laibson1997golden}, while all other parameters are the same across groups.

To estimate the parameters of interest, we match model moments to observed moments for group-specific unobserved mental health trajectories. We perform a simulated annealing procedure \citep[e.g.][]{goffe1994global} to avoid being stuck in local minima of the mean-squared objective function and refine the solution using the Nelder-Mead algorithm.

\subsection{Results}
 
\begin{figure}[t]
	\caption{Mental health trajectories from the data and the model.}
	\mbox{%
		\subfigure[Data moments of mental health trajectory ]{\includegraphics*[width=0.49\textwidth]{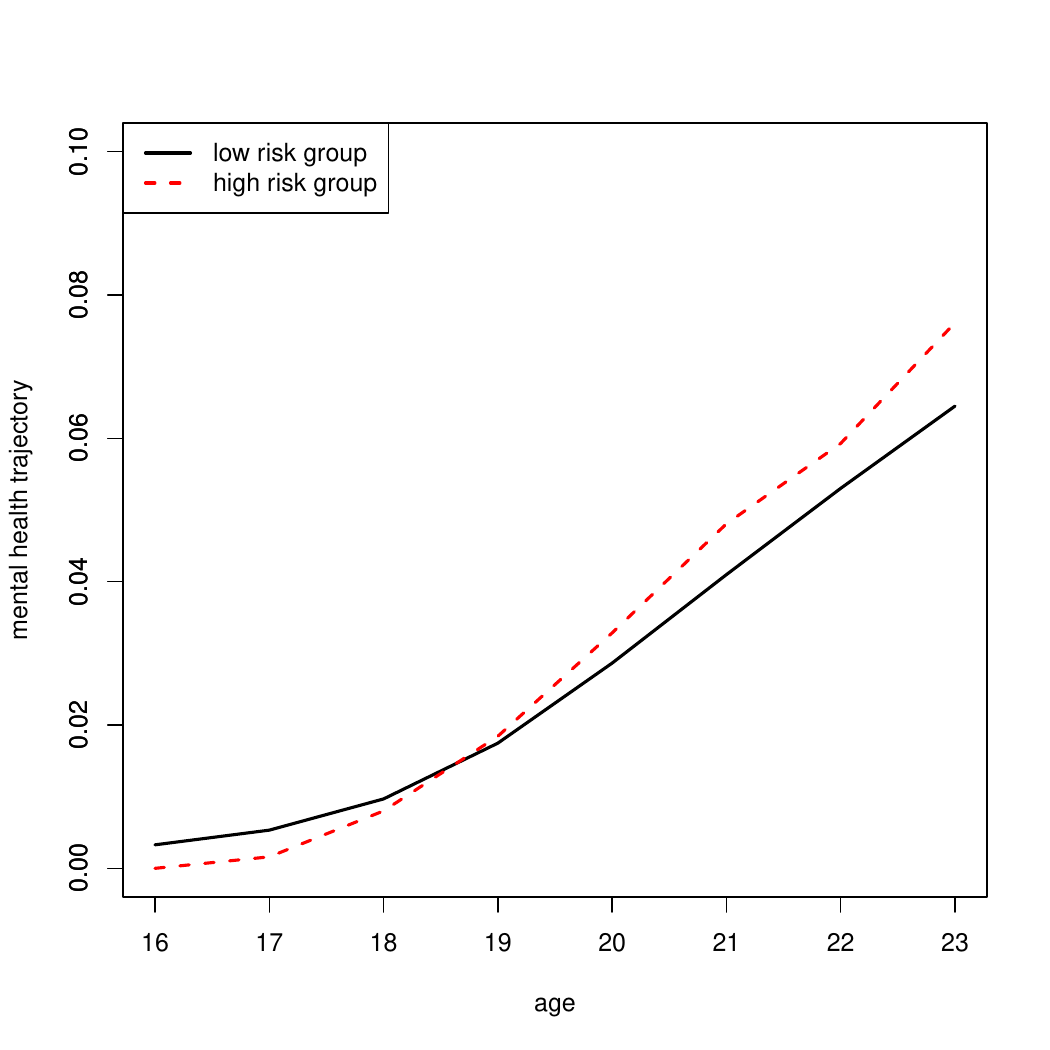}}
		\subfigure[Simulated moments of mental health trajectory]{\includegraphics*[width=0.49\textwidth]{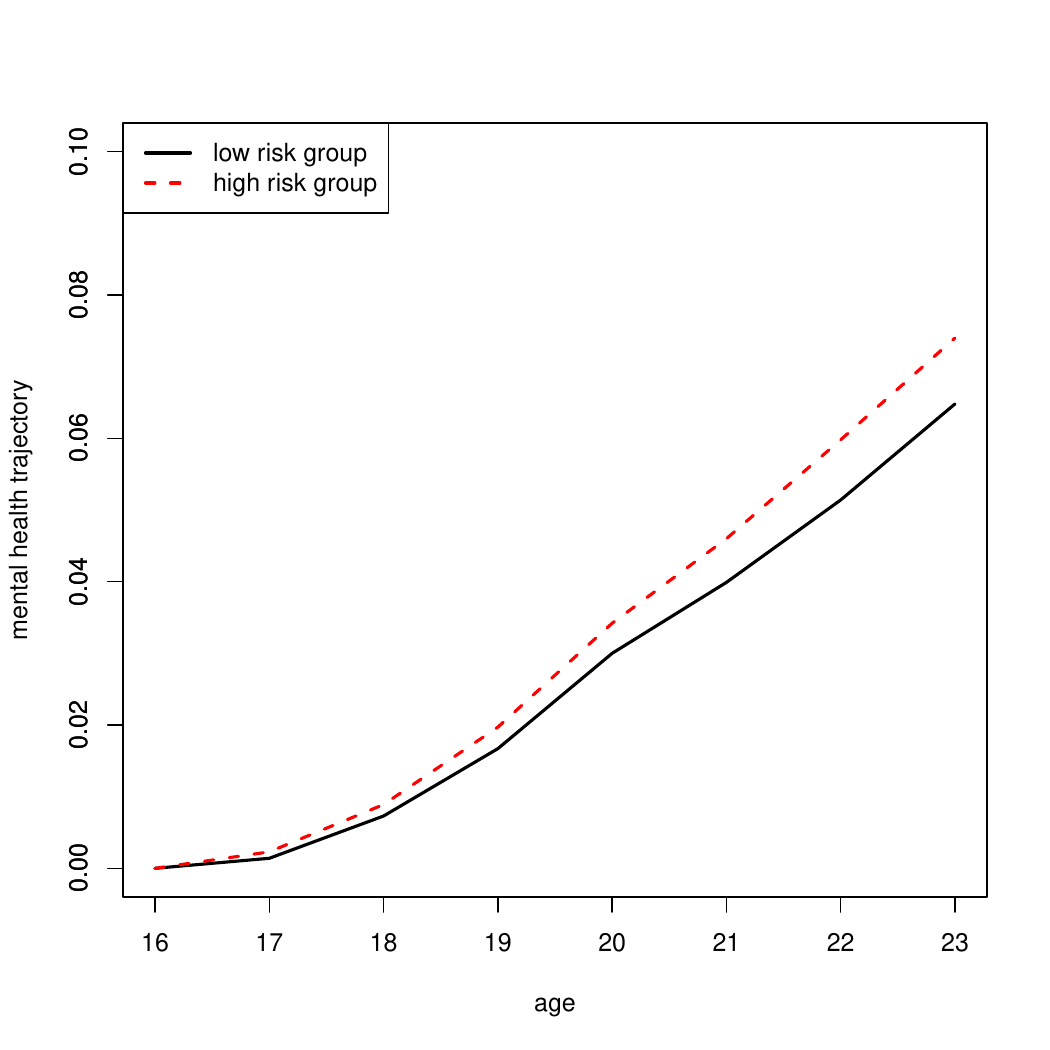}}
	}
	\label{fig:moments}
\end{figure}

Figure \ref{fig:moments}(a) plots the average mental health trajectories for the two groups in our sample. Women belonging to the high-risk group exhibit a steeper observed mental health trajectory than low-risk women. In the high-risk group, 7.6 of 100 women have been diagnosed with mental health problems by age 23. Low-risk women are on a somewhat lower mental health trajectory. About 6.4\% of them have received a mental health diagnosis by age 23. The difference in mental health problems across groups is 19\% at age 23.

Table \ref{smm_par} displays the estimated parameters obtained from moment matching. We find a clear difference in the estimated present bias between the two groups. For low-risk women, $\hat\beta_1$ is close to one, indicating that they have almost no present bias. The high-risk group exhibits a large degree of present bias, $\hat\beta_2 =0.598$. The estimated period (yearly) discount factor $\hat\delta$ is 0.925. For the low risk group, the estimated one-year discount factor is $\hat\beta_1 \cdot \hat \delta=0.944$. The corresponding factor for the high risk group is $\hat\beta_2 \cdot \hat\delta=0.553$. These values are well in the range found in the literature (e.g. \citet{laibson1997golden} or \cite{frederick2002time}). Overall, high-risk women discount the future much more strongly than low-risk women. Consequently, high-risk women are more prone to trading off short-term utility from risky behavior, e.g. immediate sexual pleasure, against long-run mental health deficits. As a result, these women face a more pronounced deterioration in mental health.

Figure \ref{fig:moments}(b) shows the group-specific mental health development obtained from the estimated parameters.\footnote{Since we do not observe risky choices, we cannot match moments to estimate risky choice trajectories.} While we cannot perfectly replicate the trajectories in the data i.e. the intersection at age 18--19, we do obtain a close match between the simulated trajectories and data moments. This suggests that heterogeneity in the present bias can generate most of the group-specific heterogeneity in observed mental health trajectories.

\begin{table}[t]
	\begin{tabularx}{\linewidth}{@{}X C{3cm}C{3cm}C{3cm}} \toprule
		&$\hat\beta_1$& $\hat\beta_2$ &$\hat\delta$\\
		\midrule
		Estimates time preferences & 1.021	&0.598		&0.925	\\\bottomrule
		\multicolumn{4}{@{}p{\linewidth}}{\scriptsize{Estimated parameters for time preferences obtained from final Nelder-Mead optimization after having applied simulated annealing (SA) for global optimization. For SA, we set the initial temperature to 1000 and the reduction of the temperature to 0.8. We set the number of inner loop iterations to 200. For more details about the SA procedure see \cite{husmann2017r}. The full set of parameter estimates for time preferences, flow utility and mental health dynamics can be found in Table \ref{app:ssm_allpar} in Appendix \ref{app_tab}.}}
	\end{tabularx}
	\caption{Estimated time preference parameters obtained from simulated method of moments}
	\label{smm_par}
\end{table}

Figure \ref{fig:moments}(b) does not only illustrate the mental health trajectories for high-risk and low-risk women. It also shows the counterfactual mental health trajectory for high-risk women if they did not exhibit self-control problems. If their self-control problems could be cured, mental health problems could be reduced by about 19\% by age 23. A back-of-the-envelope calculation using data on mental health costs suggests that the total mental health costs of all women would be reduced by 15.1\% if high-risk women had the same mental health trajectory as low-risk women.\footnote{From age 16--23, average mental health costs per woman are about 389 USD in the low-risk group and about 1,517 USD in the high-risk group. Given the share of women in the low-risk (93.9\%) and high-risk group (6.1\%), the average costs are about 458 USD.} \cite{alan2018fostering} investigate how a classroom intervention that aims at improving children's patience and self-control affects inter-temporal decision making. One result is that 9--10 year-old children who are present biased in the baseline benefit the most by delaying immediate gratification. Girls are particularly responsive to the intervention in the medium run. \citet{alan2018fostering} do not consider risky health behavior as an outcome. Yet, such an intervention fostering self-control could be a promising tool to reduce risky health behaviors among adolescents.

\section{Conclusion}\label{sec:conclusion}

In this study, we use individual-level administrative records from Sweden and the novel GFE estimator to quantify the causal impact of abortion on mental health in young women. The GFE estimator clusters individuals with similar unobserved characteristics in groups. Within these groups, unobserved heterogeneity is allowed to vary with age. Using this method, we estimate a precise null-effect of abortion on mental health. The result stands in contrast to the positive and significant associations between abortion and mental health obtained from several different identification strategies, not taking time-varying unobserved heterogeneity into account.

In our main specification with two groups, a small but significant share of women exhibits a high unobserved mental health risk while most women have a low unobserved risk. We show that the estimated profiles of unobserved heterogeneity likely capture decisions that result in risky health behaviors, such as unprotected sex or excessive drinking. These decisions are generally unobserved by researchers. Thus, the GFE estimator is necessary to obtain an unbiased estimate of the parameter of interest. Based on these considerations, we propose a model of risky choices and mental health. The estimated parameters from moment matching suggest a large degree of self-control problems among high-risk women. Our model can explain observed disparities in mental health trajectories across groups.

Our work has several implications. First, we show that an abortion from an unwanted pregnancy does not lead to more mental health problems, at least not in Sweden. In other countries, if the relationship between abortion and mental health would appear causal, this could be attributed to stigma rather than the abortion itself. Abortion opponents thus cannot use mental health problems as an argument for more restrictive abortion policies. Second, the estimated null effects imply no additional mental health care costs associated with abortion. With existing evidence on adverse economic outcomes, restrictive abortion policies thus are unlikely to be welfare-enhancing. Third, self-control problems and associated risky behaviors rather than abortions may trigger mental health problems. Thus, policymakers should find tools to identify and reduce self-control problems early rather than provide cost-intensive general mental health screenings.


\newpage
\appendix
\renewcommand{\thesubsection}{\Alph{subsection}}
\setcounter{table}{0}
\renewcommand{\thetable}{\Alph{subsection}.\arabic{table}}

\setcounter{figure}{0}
\renewcommand{\thefigure}{\Alph{subsection}.\arabic{figure}}
\setcounter{equation}{0}
\renewcommand{\theequation}{\Alph{subsection}.\arabic{equation}}

\section*{Appendix}

\subsection{Additional Tables}\label{app_tab}

\begin{table}[h]
	  \caption{Age, children born, abortions, AMD diagnoses per 1,000 women in Sk{\aa}ne and calendar year, age 16-23 for birth cohorts 1983-1985}
\begin{tabular}{@{}
		T{4.0, table-column-width=0.19\linewidth}@{}
		T{2.2, table-column-width=0.18\linewidth}@{}
		T{2.2, table-column-width=0.18\linewidth}@{}
		T{2.2, table-column-width=0.18\linewidth}@{}
		T{2.2, table-column-width=0.2\linewidth}
		@{}}
	\toprule 
	{Calendar year} & {Age}	& {Children born}  & {Abortions}   & {Mental health diagnoses} \\
	\midrule
		1999 &16.00		& 0.93 	& 7.46  	& 1.68\\  
		2000 &16.49		& 2.21 	& 13.34  	& 3.59\\  
		2001 &16.98		& 3.43 	& 15.54  	& 3.80\\  
		2002 &17.98		& 7.10  & 20.99 	& 4.33\\  
		2003 &18.99		& 11.04 & 22.43 	& 13.66\\  
		2004 &20.00		& 14.93 & 21.75 	& 22.81\\  
		2005 &20.99		& 17.67 & 21.02  	& 22.92\\  
		2006 &21.96		& 26.46 & 22.89  	& 27.31\\  
		2007 &22.50		&   	& 22.50 	& 29.18\\ 
		2008 &23.00		&		& 24.85		& 29.55\\
		\bottomrule
\multicolumn{5}{@{}p{\linewidth}@{}}{\scriptsize{The number of children born in the region of Sk{\aa}ne is not available in our data for 2007 and 2008.}}
\end{tabular}
\label{app:mean_year}
\end{table}

\begin{table}[h]
	\centering
	\caption{Estimated coefficients from the GFE estimator with $G=2,3,4$, OLS without and with individual fixed-effects (OLS FE)}
	\scalefont{0.92}
	\begin{tabularx}{\textwidth}{@{} X*{5}{T{1.3} @{}}}
		\toprule
		& \multicolumn{3}{c}{GFE} &  {OLS} & {OLS FE} \\
		\cmidrule(l{1em}r){2-4}
		& {$G=2$} & {$G=3$} & {$G=4$} &&\\
		\midrule
		Abort 			& -0.0006 	& -0.0010 	& -0.0003 	& 0.0306***	& 0.0092*** \\
		& (0.0015) 	& (0.0012) 	& (0.0009) 	& (0.0053) 	& (0.0036) \\
		\addlinespace
		Single  					& -0.0013 	& 0.0008 	& -0.0014 	& 0.0383***	& 0.0123 \\
		& (0.0053) 	& (0.0032) 	& (0.0025) 	& (0.0113) 	& (0.0117) \\

		Married 				& 0.0001 	& 0.0034 	& 0.0015 	& 0.0507***	& 0.0304** \\
		& (0.0063) 	& (0.0043) 	& (0.0031) 	& (0.0151) 	& (0.0138) \\

		College  				& 0.0018 	& 0.0019 	& 0.0006 	& 0.0010 	& 0.0051 \\
		& (0.0020) 	& (0.0014) 	& (0.0012) 	& (0.0022) 	& (0.0045) \\

		Employed 				& -0.0009 	& -0.0008 	& -0.0005 	& 0.0035 	& 0.0015 \\
		& (0.0013) 	& (0.0011) 	& (0.0009) 	& (0.0035) 	& (0.0025) \\

		Log earnings  			& -0.0001 	& -0.0001 	& -0.0001 	& -0.0017***& -0.0008*** \\
		& (0.0001) 	& (0.0001) 	& (0.0001) 	& (0.0004) 	& (0.0003) \\
		\addlinespace
		Mother: employed  		& 0.0025** 	& 0.0010 	& 0.0005 	& -0.0021 	& -0.0006 \\
		& (0.0012) 	& (0.0010) 	& (0.0007) 	& (0.0040) 	& (0.0045) \\

		Mother: married  		& -0.0013 	& -0.0015* 	& -0.0010 	& -0.0104***& 0.0037 \\
		& (0.0011) 	& (0.0008) 	& (0.0007) 	& (0.0023) 	& (0.0029) \\

		Mother: log earnings 	& 0.0000    & 0.0000    & -0.0001 	& -0.0011***& -0.0005 \\
		& (0.0001) 	& (0.0001) 	& (0.0001) 	& (0.0004) 	& (0.0004) \\

		Mother: college degree 	&       	&       	&       	& 0.0001 	&  \\
		&       	&       	&       	& (0.0018) 	&  \\
		\addlinespace
		Father: employed 		& 0.0023* 	& 0.0014 	& 0.0012 	& -0.0023 	& 0.0012 \\
		& (0.0013) 	& (0.0011) 	& (0.0008) 	& (0.0041) 	& (0.0029) \\

		Father: log earnings 	& -0.0001 	& 0.0000    & 0.0000    & -0.0011***& -0.0009*** \\
		& (0.0001) 	& (0.0001) 	& (0.0001) 	& (0.0004) 	& (0.0003) \\

		Father: college degree 	&       	&       	&       	& -0.0016 	&  \\
		&       	&       	&       	& (0.0018) 	&  \\
		\addlinespace
		Household: log disp. income 	& 0.0000    & 0.0002 	& 0.0000    & -0.0001 	& -0.0023*** \\
		& (0.0004) 	& (0.0002) 	& (0.0002) 	& (0.0010) 	& (0.0009) \\
		\addlinespace
		Censor: missing years 	& -0.0010 	& 0.0013 	& -0.0007 	& -0.0224* 	& -0.0225 \\
		& (0.0055) 	& (0.0034) 	& (0.0028) 	& (0.0127) 	& (0.0125) \\

		Censor: missing values  & 0.0019 	& 0.0031 	& -0.0014 	& -0.0042 	& -0.0201 \\
		& (0.0039) 	& (0.0022) 	& (0.0017) 	& (0.0081) 	& (0.0234) \\
		\addlinespace
		Constant 				&       	&       	&       	& 0.0894*** 	&  \\
		&       	&       	&       	& (0.0337) 	&  \\
		\bottomrule
		\multicolumn{6}{@{} p{\linewidth} @{}}{\scriptsize{Standard errors clustered on individual level; ***\,$p < 0.01$, **\,$p < 0.05$, *\,$p < 0.1$; Number of observations: 165,624; number women: 20,703; Additional controls: year fixed-effects, municipality FE, year of birth FE for woman/mother/father.}}
	\end{tabularx}%
	\label{app:allcoef}%
\end{table}%

\begin{table}
	\caption{Comparison of socioeconomic characteristics of women, mothers and fathers by low risk group and high risk group}
	\begin{tabularx}{1.00\textwidth}{@{} X T{1.3}T{1.3}T{1.3}T{1.5}@{}}
	\toprule
		& \multicolumn{3}{c}{Mean}  &               {$t$-statistic}\\
		\cmidrule(lr){2-4} 
		& {low risk group}& 	{high risk group}&{group differences}	& {($p$-value)}\\[0.2em]
		\midrule
			\multicolumn{5}{@{}l}{\textit{A. Sample of women all ages}} \\ 
		Mother: college degree   &       0.336&       0.330&       0.006&       1.064\, (0.287)\\
		Father: college degree   &       0.360&       0.371&      -0.011*&      -1.834\, (0.067)\\
		Mother: employed  &       0.843&       0.833&       0.010**&       2.041\,  (0.041)\\
		Father: employed  &       0.848&       0.834&       0.015***&       3.173\, (0.002)\\
		Mother: log earnings    &      10.764&      10.533&       0.231***&       4.420\, (0.000)\\
		Father: log earnings    &      11.060&      10.884&       0.176***&       3.281\, (0.001)\\
		Mother: married   &       0.671&       0.681&      -0.010&      -1.622 \,    (0.105)\\
		Father: married   &       0.687&       0.693&      -0.006&      -0.971  \,   (0.332)\\ \addlinespace
		
		Woman: single      &       0.990&       0.987&       0.003**&       2.185\,  (0.029)\\
		Woman: employed    &       0.207&       0.211&      -0.004&      -0.808 \, (0.419)\\
		Woman: college degree     &       0.307&       0.348&      -0.041***&      -6.928 \,(0.000)\\
		Woman: age at abortion&      19.617&      19.499&       0.118&       1.610 \, (0.108)\\
\addlinespace[2em]
		
	\multicolumn{5}{@{}l}{\textit{B. Sample of women at age 16}} \\ 
	Mother: college degree   &       0.333&       0.319&       0.014&       0.811  \,     (0.418)\\
	Father: college degree   &       0.352&       0.359&      -0.007&      -0.380  \,     (0.704)\\
	Mother: employed  &       0.844&       0.809&       0.035**&       2.452    \,   (0.014)\\
	Father: employed  &       0.863&       0.838&       0.025*&       1.863    \,  (0.063)\\
	Mother: log earnings  &      10.800&      10.461&       0.339**&       2.314  \,     (0.021)\\
	Father: log earnings    &      11.225&      11.049&       0.176&       1.212 \,  (0.226)\\
	Mother: married   &       0.698&       0.718&      -0.021&      -1.271     \,  (0.204)\\
	Father: married   &       0.708&       0.723&      -0.016&      -0.981   \,    (0.327)
		\\\bottomrule
	\multicolumn{5}{@{}p{\linewidth}@{}}{\scriptsize{***\,$p < 0.01$, **\,$p < 0.05$, *\,$p < 0.1$; results obtained from $t$-test comparing sample means of the high risk group and the low risk group; group variances are assumed to be unequal; $p$-values refer to the alternative hypothesis that group differences are not equal.}}	
	\end{tabularx}\label{app:group_char_socio}
\end{table}

\begin{table}
	\caption{The effect of abortion on mental health by age, OLS with individual-specific fixed-effects} \label{app:agedependentAbortion}
	\begin{tabularx}{\textwidth}{@{} X T{1.4, table-column-width=0.5\linewidth} @{}}\toprule
		& {OLS FE}  \\ \midrule
		Abortion & 0.005  \\
		& (0.013)  \\
		Abortion $\times$ Age 17   & -0.010   \\
		& (0.018)  \\
		Abortion $\times$ Age 18   & -0.001   \\
		& (0.016)   \\
		Abortion $\times$ Age 19   & -0.005   \\
		& (0.015)   \\
		Abortion $\times$ Age 20  & -0.003  \\
		& (0.015)   \\
		Abortion $\times$ Age 21&   -0.002   \\
		& (0.016)   \\
		Abortion $\times$ Age 22   & -0.007   \\
		& (0.016)   \\
		Abortion $\times$ Age 23   & 0.021   \\
		& (0.019)   \\
		
		Age 17 &  0.008***   \\
		& (0.002)   \\
		Age 18   & 0.018***   \\
		& (0.003)   \\
		Age 19   & 0.034***   \\
		& (0.004)   \\
		Age 20  & 0.057***   \\
		& (0.006)   \\
		Age 21  & 0.079***   \\
		& (0.007)   \\
		Age 22  & 0.098***   \\
		& (0.007)   \\
		Age 23  & 0.115***   \\
		& (0.008)   \\ \addlinespace		
		Observations & {17,584} \\ \bottomrule
		\multicolumn{2}{@{}p{\linewidth}@{}}{\scriptsize{Standard errors clustered on the individual level; ***\,$p < 0.01$, **\,$p < 0.05$, *\,$p < 0.1$; OLS regression of cumulative mental health diagnoses on age-dependent abortion, controlling for individual-specific fixed-effects; Control variables: woman: relationship status, log earnings, college degree, employed; mother: log earnings, employed, college degree, relationship status; father: log earnings, employed, college degree; log household disposable income; municipality FE, year of birth FE for woman/mother/father; indicator missing observations. }}					
	\end{tabularx}
\end{table}

\begin{table}[h]
	\caption{Estimated impact of risky health behaviors in the past on mental health development}
	\begin{tabularx}{\textwidth}{@{}X T{1.2}T{1.2}T{1.2}T{1.2}T{1.3}@{}} \toprule
		&\multicolumn{5}{c}{Woman has mental health problems at 16--23}\\
		\cmidrule(l){2-6}
		&\multicolumn{4}{c}{OLS FE} & {GFE}\\
		\cmidrule(lr){2-5} \cmidrule(l){6-6}
		{Before age $t$, woman had}& {(1)} & {(2)} & {(3)} & {(4)} & {(5)} \\
		\midrule
		Abortion 					& 0.009*** 	& 0.009**  	& 0.009** 	&  0.009** & -0.0006\\
		& (0.004) 	& (0.004) 	& (0.004) 	&  (0.004) & (0.0015)\\		
		Acute drunkenness 			&  0.120***	& 			&  			& 0.118*** & -0.0034 \\
		& (0.022)	& 			&     		& (0.022)  & (0.0034)	\\
		Chlamydia infection 		&  			& 0.018***	&  			& 0.013**  & -0.0003\\
		&  			& (0.007)	&    		& (0.007)  & (0.0021)\\
		STD screening  				&  			&  			& 0.017***	& 0.016*** &  0.0001\\
		&  			&  			& (0.004)	& (0.004)  & (0.0014) \\ \addlinespace		
		Number women				&\multicolumn{5}{c}{20,703}  \\					
		Observations &\multicolumn{5}{c}{165,624}\\ \bottomrule
		\multicolumn{6}{@{} p{\linewidth} @{}}{\scriptsize{Standard errors clustered on the individual level; ***\,$p < 0.01$, **\,$p < 0.05$, *\,$p < 0.1$; Columns (1)-(4): OLS regression of cumulative mental health diagnoses on abortion and past risky health behavior, controlling for individual-specific fixed-effects. Column (5): GFE estimation with $G=2$ and individual-specific fixed-effects. Control variables: woman: relationship status, log earnings, college degree, employed; mother: log earnings, employed, college degree, relationship status; father: log earnings, employed, college degree; log household disposable income; year fixed-effects, municipality FE, year of birth FE for woman/mother/father; indicator missing observations. }}					
	\end{tabularx}
	\label{app:mhrisk}
\end{table}

\begin{table}[h] 
	\caption{Correlations between profiles of age-dependent unobserved heterogeneity and risky health behaviors}
	\begin{tabularx}{\textwidth}{@{} X T{1.3}T{1.3}T{1.3}T{1.3}@{}} \toprule
		& {unwanted pregnancy}	& {STD screening}	& {chlamydia infection} & {acute drunkenness} \\\midrule
		\multicolumn{5}{@{}l}{\textit{A. OLS without individual-specific fixed-effects}} \\\addlinespace
		$\hat \alpha_g$    & 0.029***  & 0.010***  & 0.008***  & -0.0002      \\
						& (0.004)      & (0.004)      & (0.003)      & (0.0012)      \\
		Constant 		 & 0.010        & 0.068***  & 0.014*    & -0.0015       \\
						& (0.012)      & (0.016)      & (0.007)      & (0.0025)      \\ \addlinespace[2em]
		\multicolumn{5}{@{}l}{\textit{B.OLS with individual-specific fixed-effects}} \\ \addlinespace
		$\hat \alpha_g$         & 0.026***  & 0.009**   & 0.007***  & -0.0001       \\
						& (0.004)      & (0.004)      & (0.002)      & (0.001)      \\\addlinespace		
		Number women				&\multicolumn{4}{c}{20,703}  \\					
		Observations &\multicolumn{4}{c}{165,624}\\\bottomrule
		\multicolumn{5}{@{}p{\linewidth}@{}}{\scriptsize{Standard errors clustered on the individual level; ***\,$p < 0.01$, **\,$p < 0.05$, *\,$p < 0.1$. OLS regression of current risky health behaviors on estimated profiles of unobserved mental health risk; Estimated profiles of unobserved heterogeneity $\hat \alpha_g$ for $G=2$; Control variables: woman: relationship status, log earnings, college degree, employed; mother: log earnings, employed, college degree, relationship status; father: log earnings, employed, college degree; log household disposable income; municipality FE, year of birth FE for woman/mother/father; indicator missing observations. }}					
	\end{tabularx}
	\label{app:predictRISK}
\end{table}

\begin{table}[h]
	\caption{Estimated parameters obtained from simulated method of moments (SMM)}
	\begin{tabularx}{\textwidth}{@{} C{1cm}C{1cm}C{1cm}C{0.45cm}C{1cm}C{1cm}C{1cm}C{0.45cm}C{1cm}C{1cm}C{1cm}C{1cm} @{}}
		\toprule
		\multicolumn{3}{c}{Time preferences} && \multicolumn{3}{c}{Flow utility} && \multicolumn{4}{c}{Mental health dynamics}\\
		\cmidrule{1-3} \cmidrule{5-7} \cmidrule{9-12}
		$\hat\beta_1$& $\hat\beta_2$ &$\hat\delta$	&& $\hat a$&$\hat b$&$\hat c$&&$\hat\psi$ & $\hat\sigma_{\epsilon}$ & $\hat\zeta$ & $\hat\kappa$\\
		\midrule
		1.021	&0.598		&0.925		&&0.864&0.186&0.058&& 0.962&0.613&0.447 & 2.543\\
		\bottomrule
		\multicolumn{12}{@{}p{\linewidth}}{\scriptsize{Estimated parameters for preferences, flow utility and mental health dynamics obtained from final Nelder-Mead optimization after having applied simulated annealing (SA) for global optimization; SA parameters: initial temperature $=1000$, reduction of temperature $=0.8$, number inner loop iterations $=200$; More details about SA procedure can be found in \cite{husmann2017r}.}}
	\end{tabularx}
	\label{app:ssm_allpar}
\end{table}

\newpage \clearpage

\subsection{Additional figures}\label{app_fig}
\setcounter{subfigure}{0}
\begin{figure}[h]
	\mbox{
		\subfigure[Abortions by gestation week, age -19]{\includegraphics[width=0.49\textwidth]{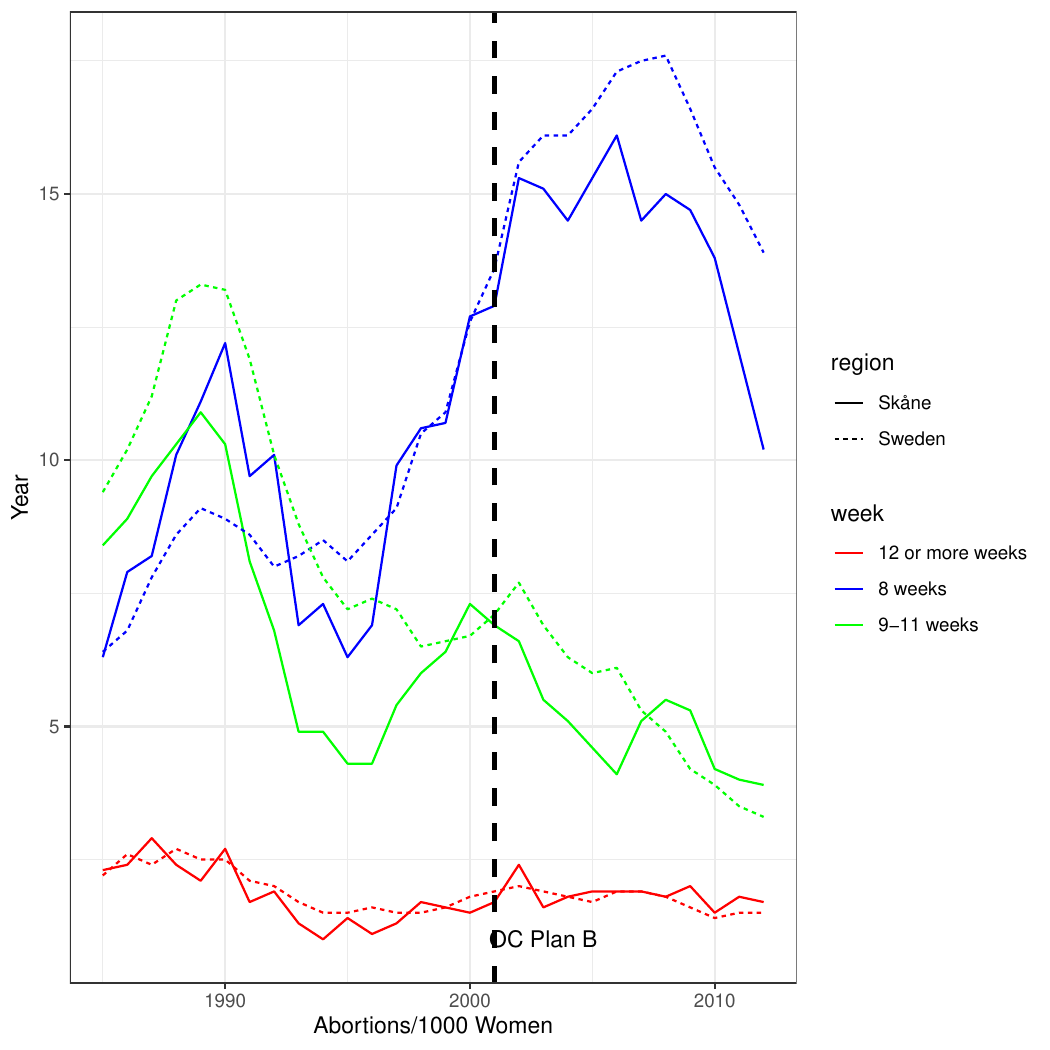}} 
		\subfigure[Abortions by gestation week, age 20-24]{\includegraphics[width=0.49\textwidth]{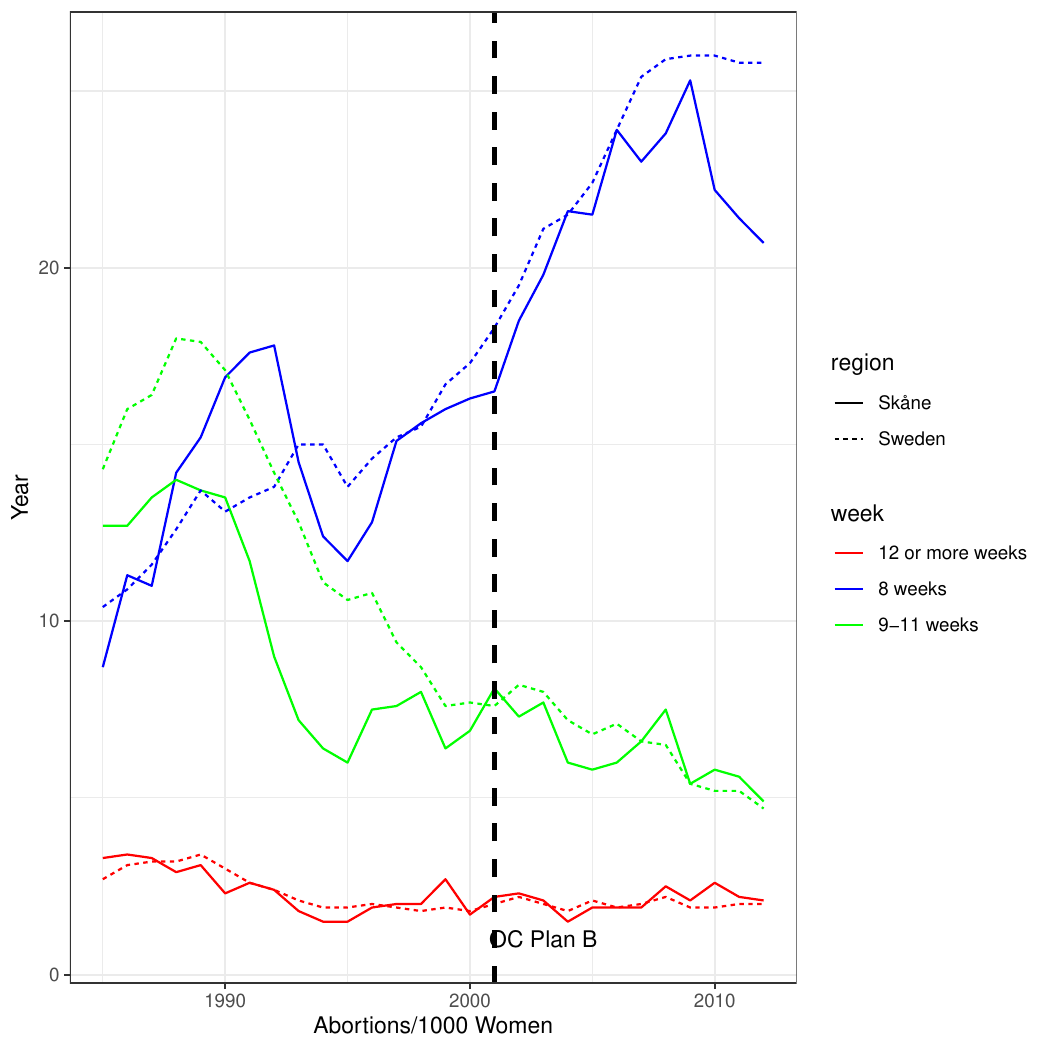}}
	}		
	\caption{Aggregate abortion trends by gestation week around the introduction of the OCP.}
	\label{app:agg_ab}
\end{figure}

\begin{figure}[h]
	\mbox{
	\includegraphics[width=0.49\textwidth]{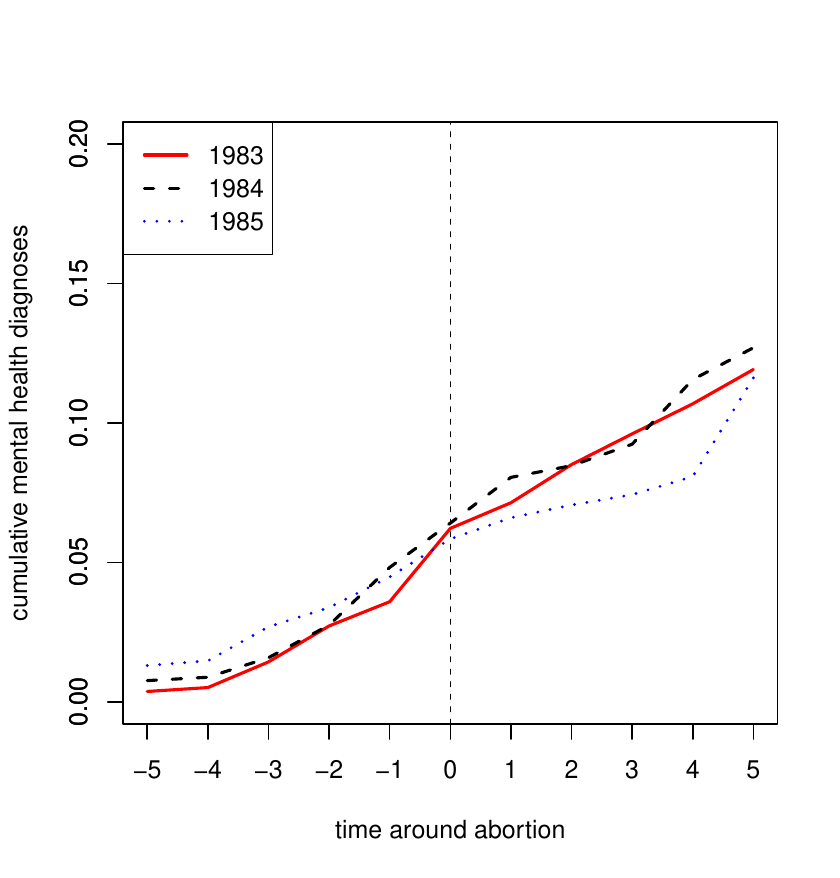}
	}		
	\caption{Mental health development around time of abortion by birth cohort.}
	\label{app:eventtime_bc}
\end{figure}

\begin{figure}[h]
	\mbox{
	\includegraphics[width=0.49\textwidth]{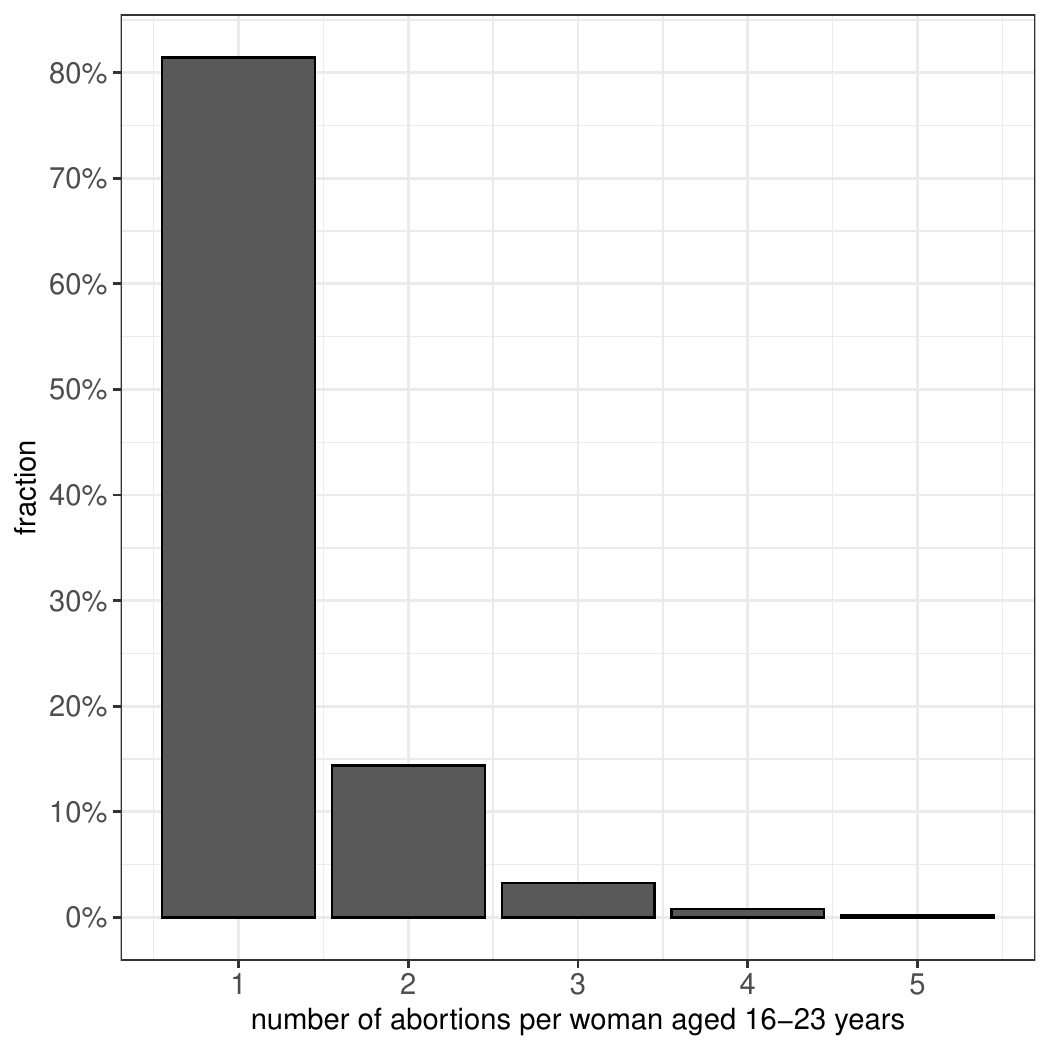}
	}		
	\caption{Number of abortions after an unwanted pregnancy per woman, aged 16-23 years.}
	\label{app:nr_abortions}
\end{figure}

\begin{figure}[h]
	\mbox{
	\subfigure[$\hat{\alpha}_{g_t}$, $G=3$]{\includegraphics[width=0.49\textwidth]{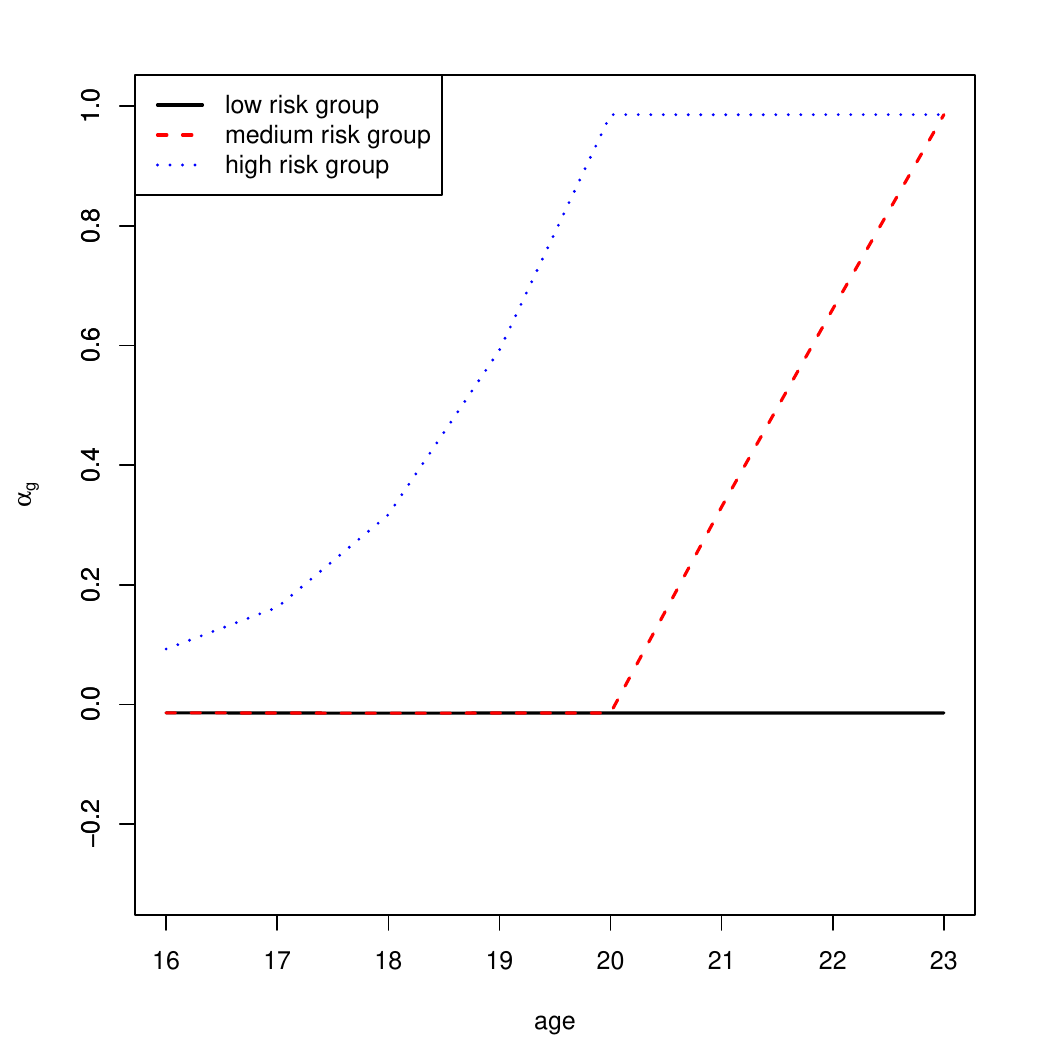}} 
\subfigure[$\hat{\alpha}_{g_t}$, $G=4$]{\includegraphics*[width=0.49\textwidth]{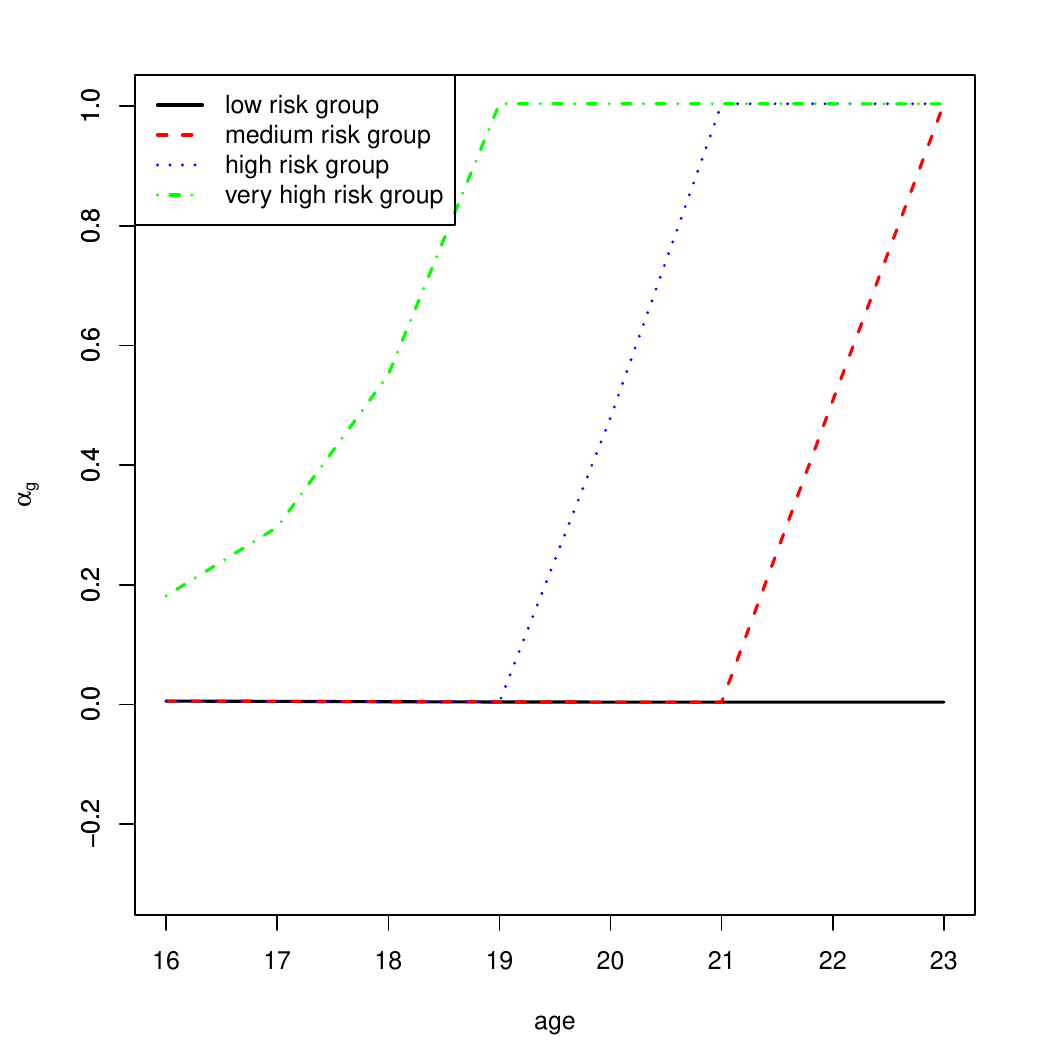}}}

\caption{Estimated profiles of group-specific unobserved heterogeneity for $G=3$ and $G=4$.}
\label{app:profG34}
\end{figure}

\begin{figure}[h]
\mbox{
	\subfigure[$G=2$]{\includegraphics[width=0.33\textwidth]{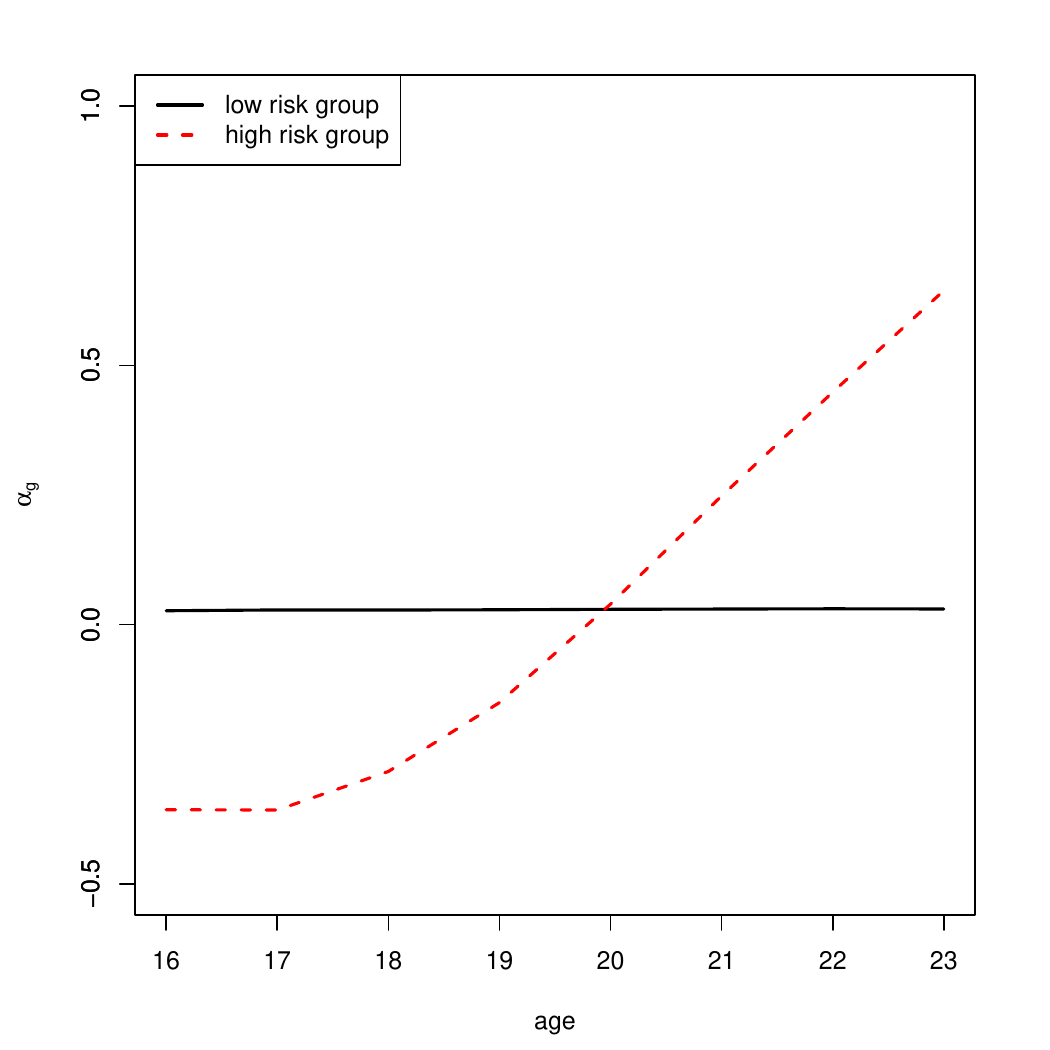}}
	\subfigure[$G=3$]{\includegraphics[width=0.33\textwidth]{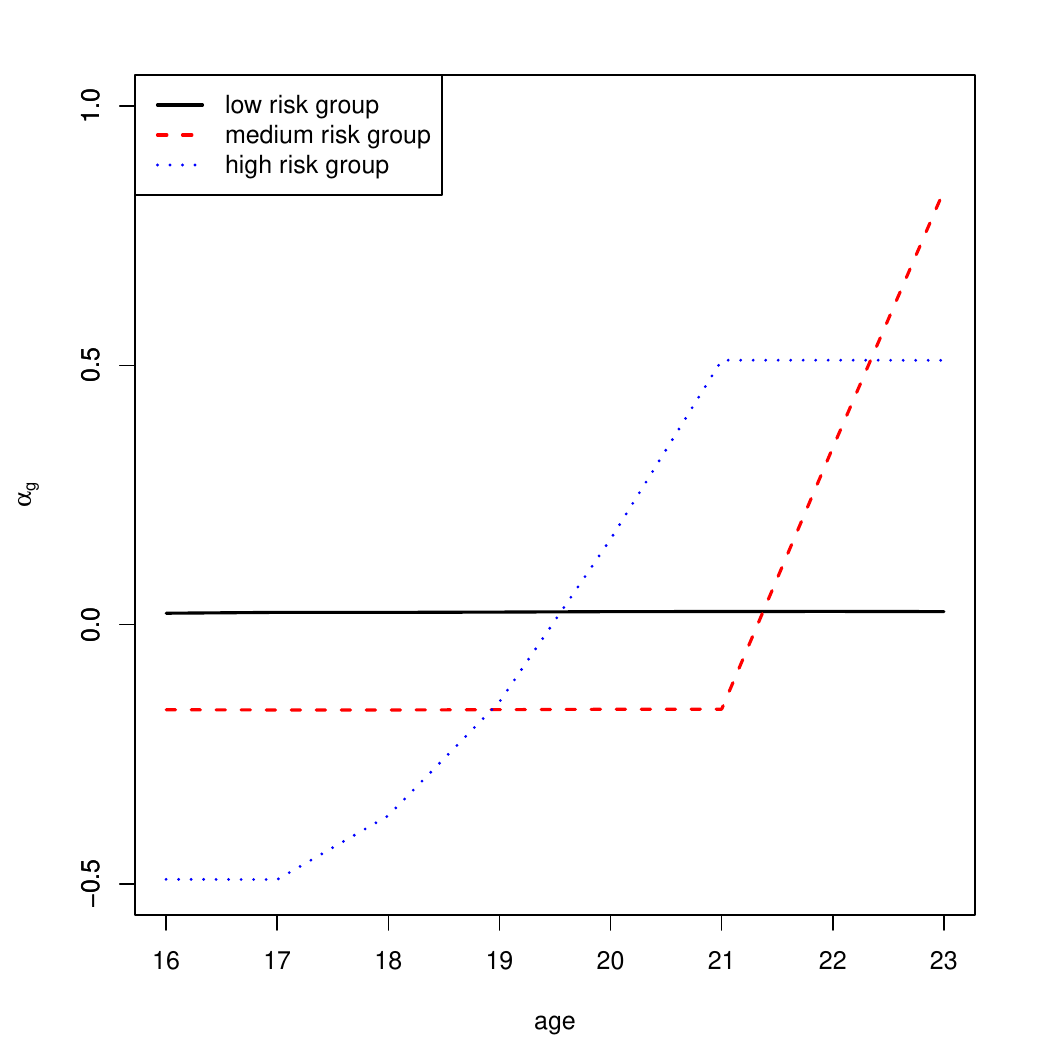}} 
	\subfigure[$G=4$]{\includegraphics[width=0.33\textwidth]{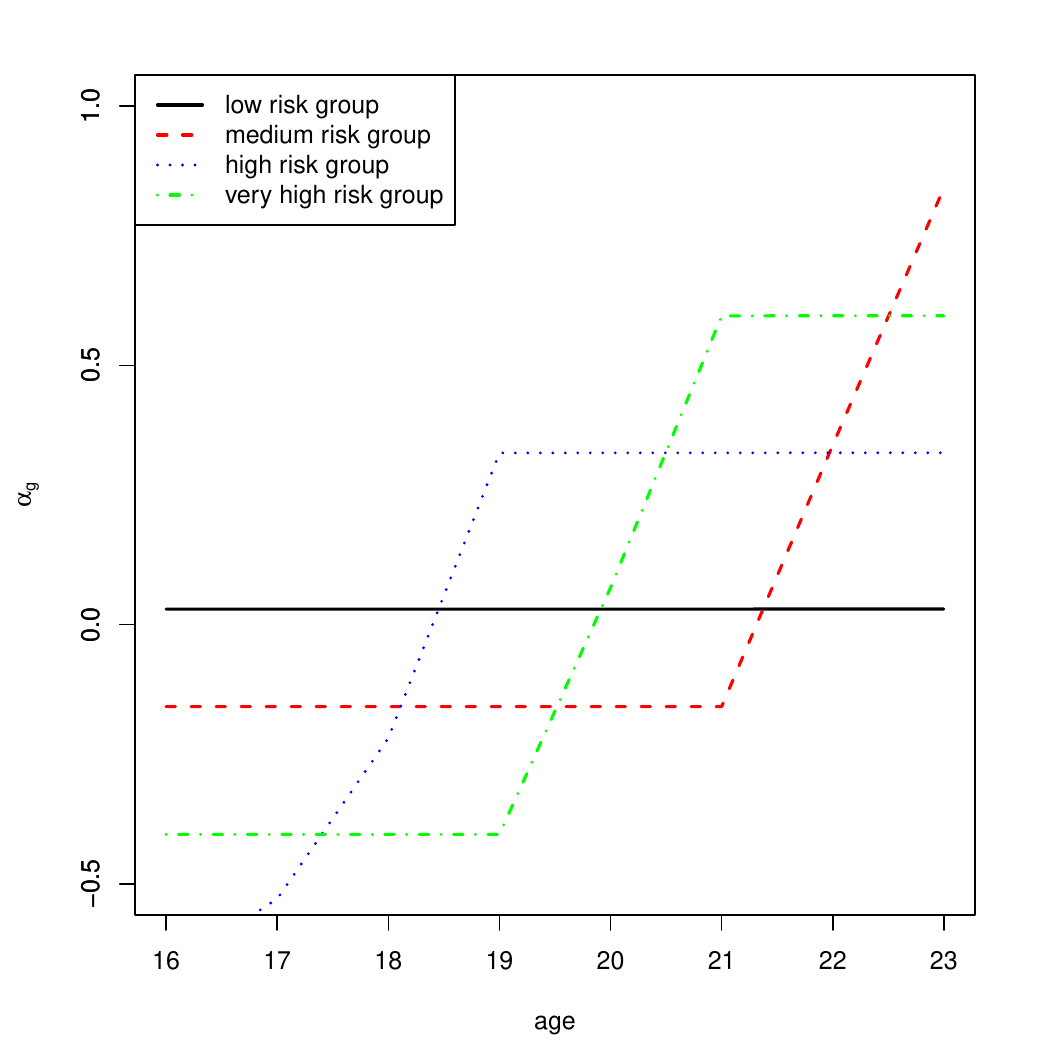}} 
}
\caption{Estimated time profiles of group-specific unobserved heterogeneity with individual-specific fixed-effects for $G=2,3,4$}
			\label{app:profFE}
\end{figure}

\begin{figure}[h]
	\includegraphics*[width=0.55\textwidth]{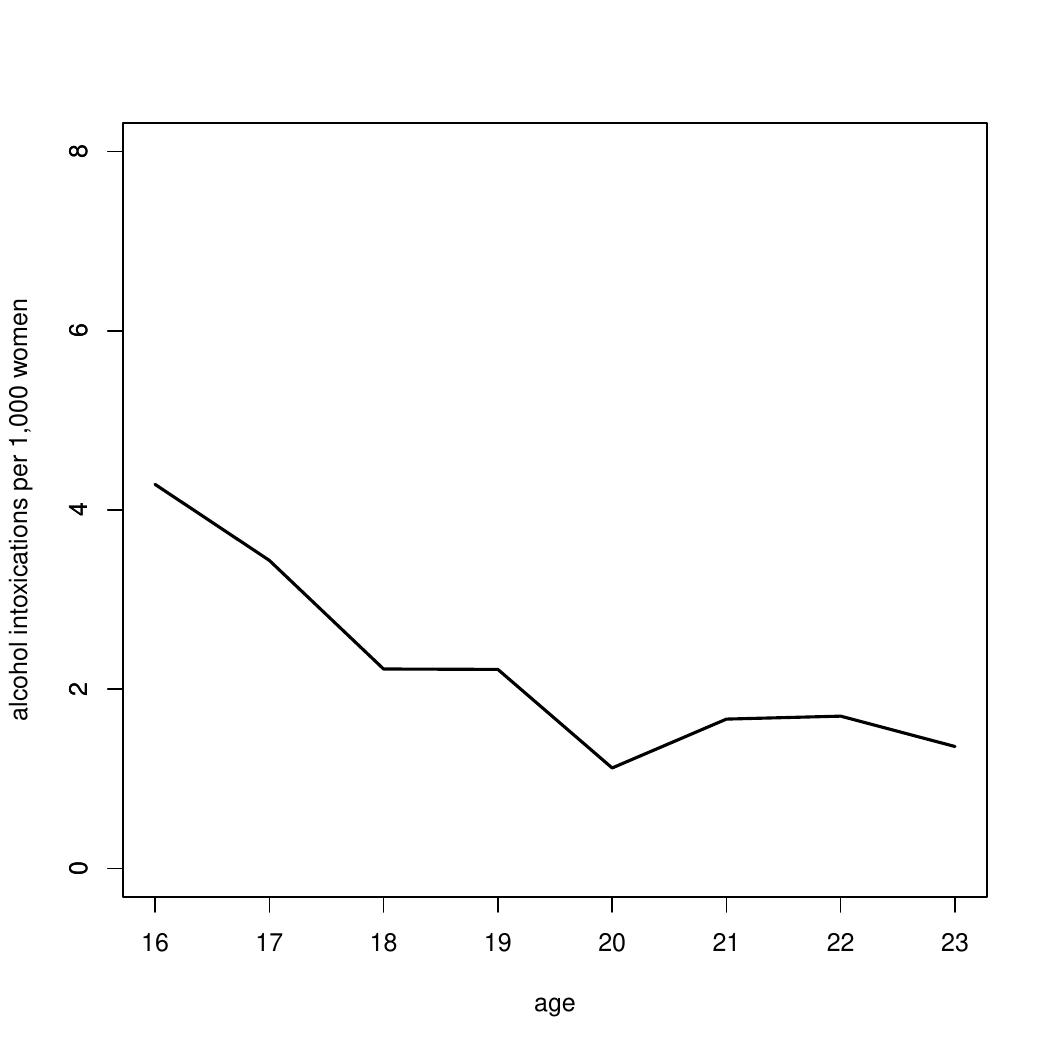}
	\caption{Number of alcohol intoxication per 1,000 women, age 16-23.}
	\label{app:alcohol}
\end{figure}

\newpage \clearpage

\setcounter{table}{0}
\renewcommand{\thetable}{\Alph{subsection}.\arabic{table}}

\setcounter{figure}{0}
\renewcommand{\thefigure}{\Alph{subsection}.\arabic{figure}}

\setcounter{equation}{0}
\renewcommand{\theequation}{\Alph{subsection}.\arabic{equation}}

\subsection{Simulation Example}\label{app_sim}

We build a general simulation set-up to check for known problems, such as dependence on starting values. This simulation set-up also aids us with the interpretation of our results, as well as with validating some specification choices, specifically determining the optimal number of groups. All replication files for this simulation are available on \url{https://github.com/LJanys/Mental_Health_Abortions_Risky_Behaviors}.

Our data is generated by the general data generating process as outlined in Equation \eqref{eq:bm3}, except that we disregard the cumulative properties with the following specific values and varying over four different sample sizes: 

\begin{itemize}
	\item true number of groups: $G=3$
	\item true parameter value of interest: $\xi=0$ 
	\item number of cross-sectional observations: $N=1000, 1500, 2000, 10000$
	\item number of time periods: $T=10$
\end{itemize}

\begin{figure}[t]
	\centering
	\mbox{
		\subfigure[True unobserved mental health profiles, $\alpha_{g_t}$]{\includegraphics*[width=0.49\textwidth]{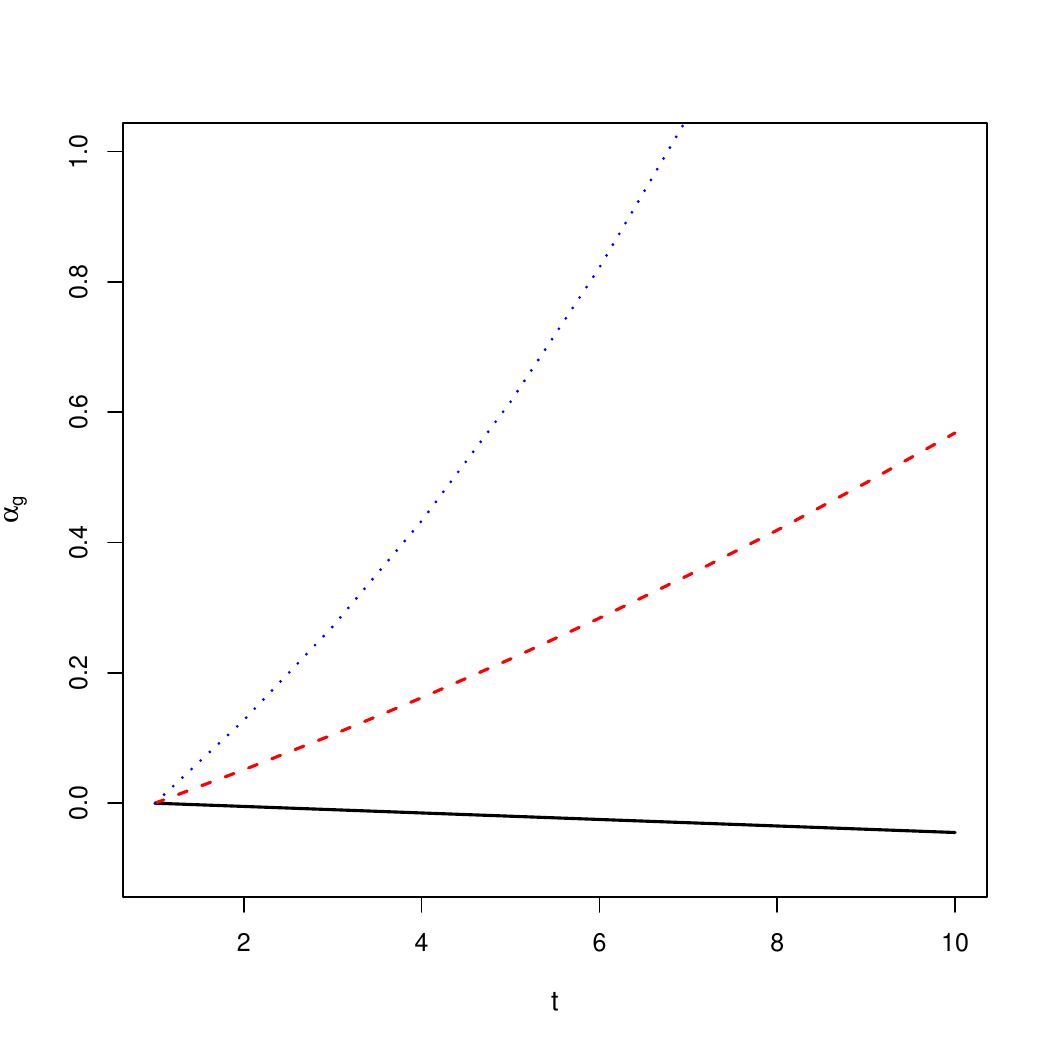}} 
		\subfigure[Estimated unobserved mental health profiles, $\hat{\alpha}_{g_t}$]{\includegraphics*[width=0.49\textwidth]{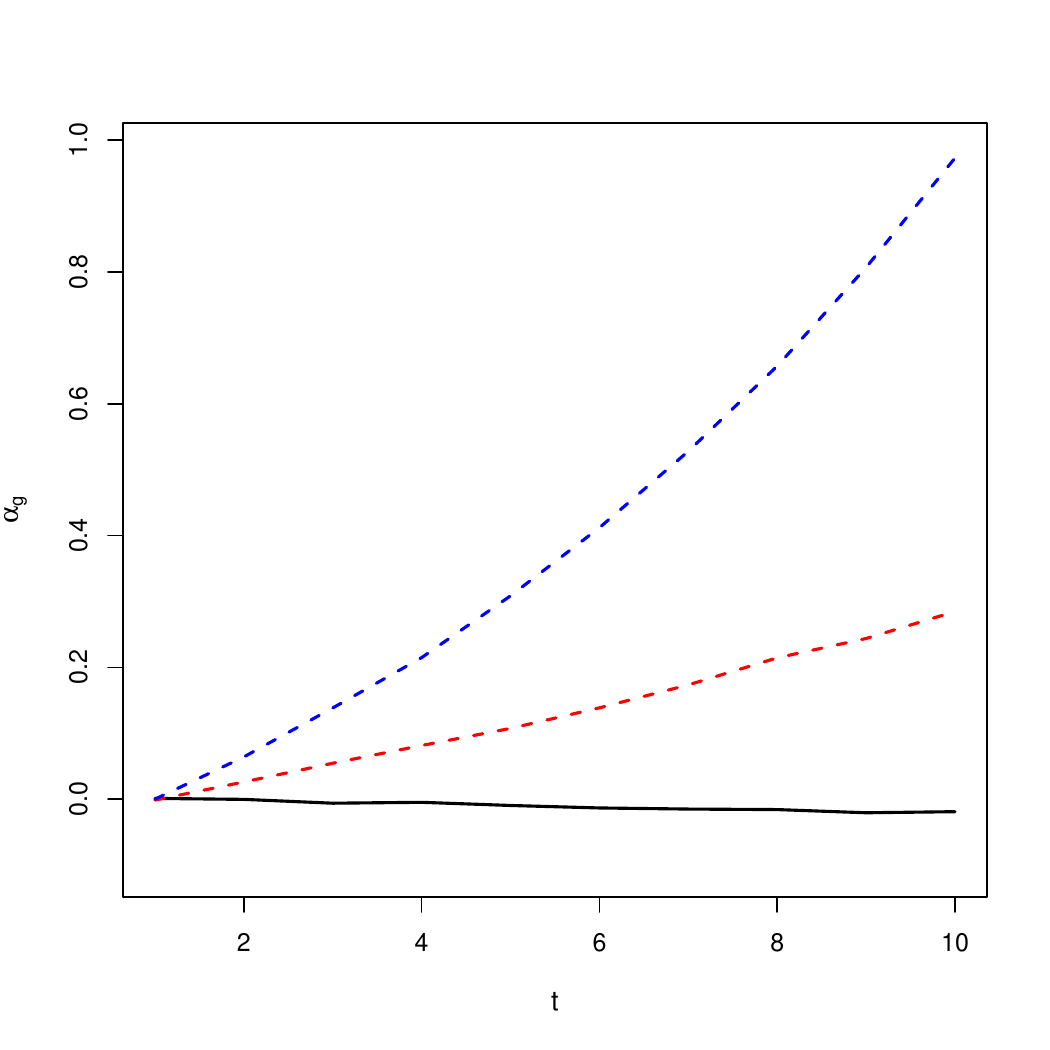}}}
		\caption{True and estimated mental health profiles over time}\label{fig_profile}
\end{figure}

$\alpha_{g_i}$, the unobserved grouped fixed-effects trajectories, are correlated with the contemporaneous probability of having an abortion, as well as the mental health diagnosis, inducing an omitted variable bias in the OLS and FE estimates. The true unobserved heterogeneity profile curves are depicted in the left-hand panel of Figure \ref{fig_profile}. Their analytical expressions are given by 
\begin{flalign}
	\alpha_{g_1}&= t(0.002)\\
\alpha_{g_2}&=-1+e^{(t/10)^{1/2}}\\
\alpha_{g_3}&=-1+e^{(t/10)^{1.2}}
\end{flalign}

The group membership is determined by the value of the unobserved, individual-specific fixed-effects $\alpha_i$: For each individual, we draw from a binomial distribution whether or not the abortion takes place in each period, with a vector of probabilities for the three groups of $p_{g_1}=0$, $p_{g_2}=0.1$ and $p_{g_3}=0.3$. This results in a contemporaneous correlation of the unobserved $\alpha_g$ and the abortion probability of 0.11, which is the source of the omitted variable bias. To match the characteristics in our real data, we define the groups to not be of equal size: The largest group is group one (``low-risk group''), which comprises 70\% of individuals; group two (``medium-risk group'') comprises 20\% of individuals; and group three (``high-risk group'') is the smallest group, with 10\% of individuals. 
\setcounter{subfigure}{0}

\bigskip

 With this DGP, we compare the results of the simulations along three margins: 
 \begin{itemize}
 	\item[(1)] we ascertain that the estimated curves of the unobserved heterogeneity are comparable to the true ones and to investigate adding ``superfluous'' groups.
 	\item[(2)] the estimated parameters for the OLS estimator ($\widehat{\xi}^\textup{OLS}_{sim}$), the individual-specific fixed-effects estimator ($\widehat{\xi}^\textup{FE}_{sim}$) and the grouped fixed-effects estimator ($\widehat{\xi}^\textup{GFE}_{sim,G}$) behave similar to the pattern we observe in our empirical analysis.
 	\item[(3)] the chosen information criterion is reliably minimized at the correct number of groups. 
 	 \end{itemize}

The right-hand side of Figure \ref{fig_profile} shows the estimated unobserved heterogeneity profiles obtained from the GFE estimator for $N=10,000$ observations. The profiles look very similar, indicating that the GFE can reliably estimate the group-specific profiles of unobserved heterogeneity.
 
\begin{figure}
\mbox{
		\subfigure[$N=1000$]{\includegraphics[width=0.49\textwidth]{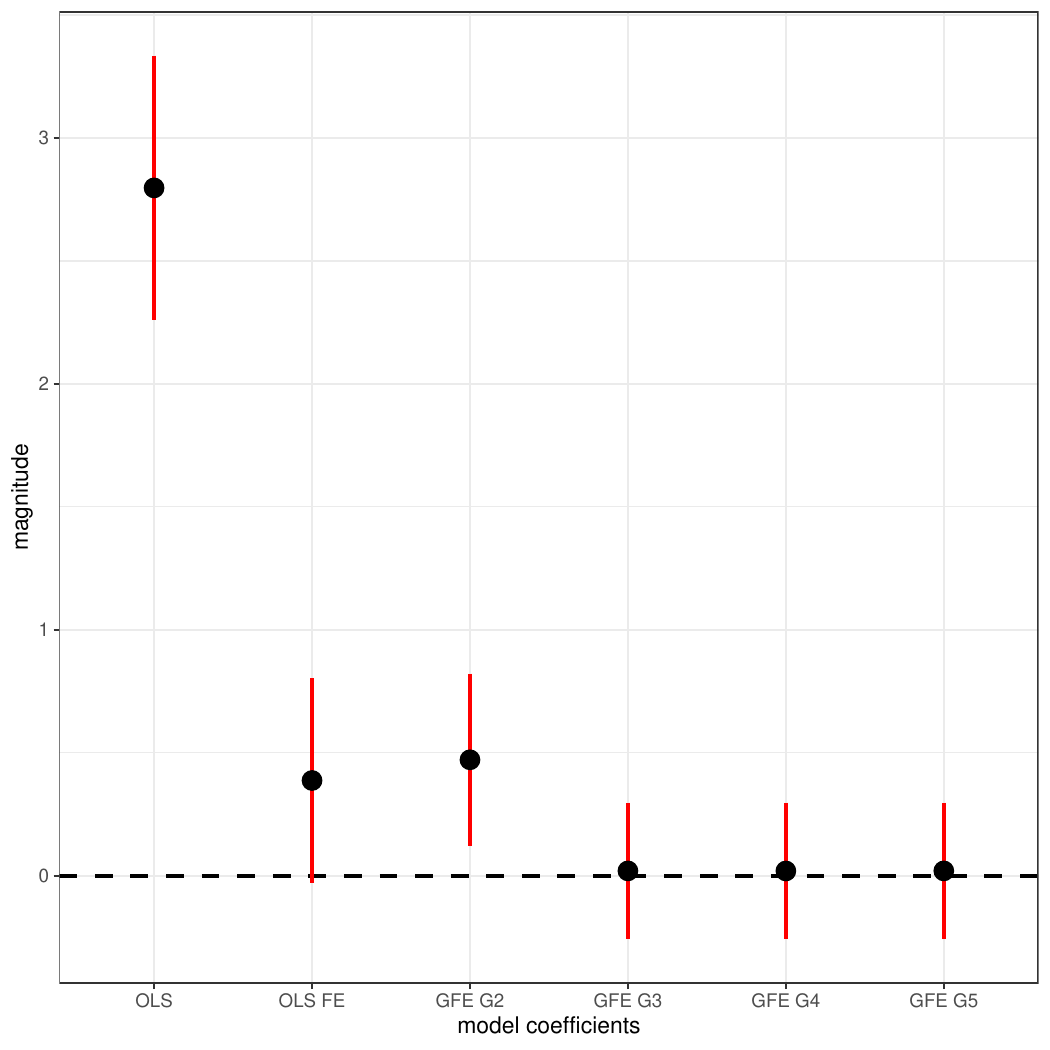}}
		\subfigure[$N=1500$]{\includegraphics[width=0.49\textwidth]{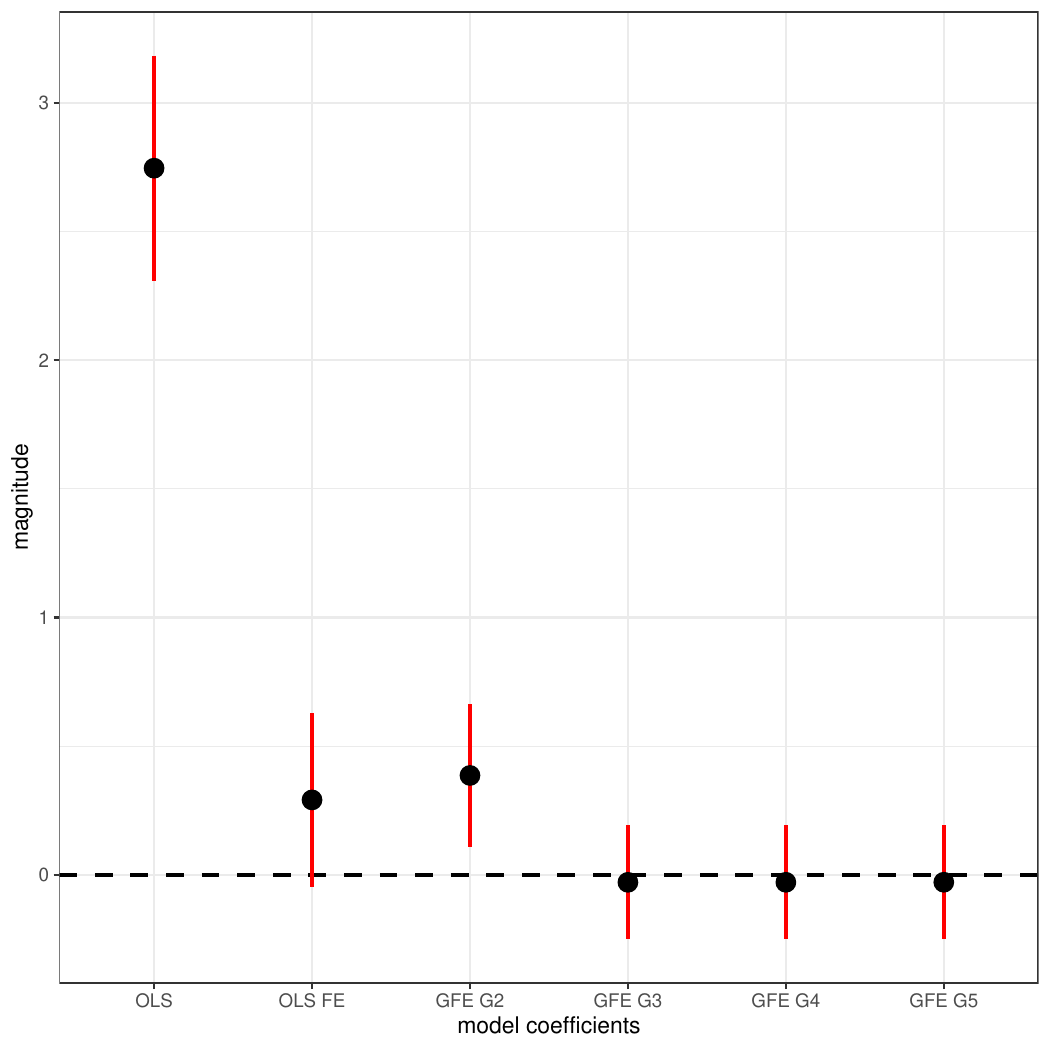}} 
}
\mbox{		
		\subfigure[$N=2000$]{\includegraphics[width=0.49\textwidth]{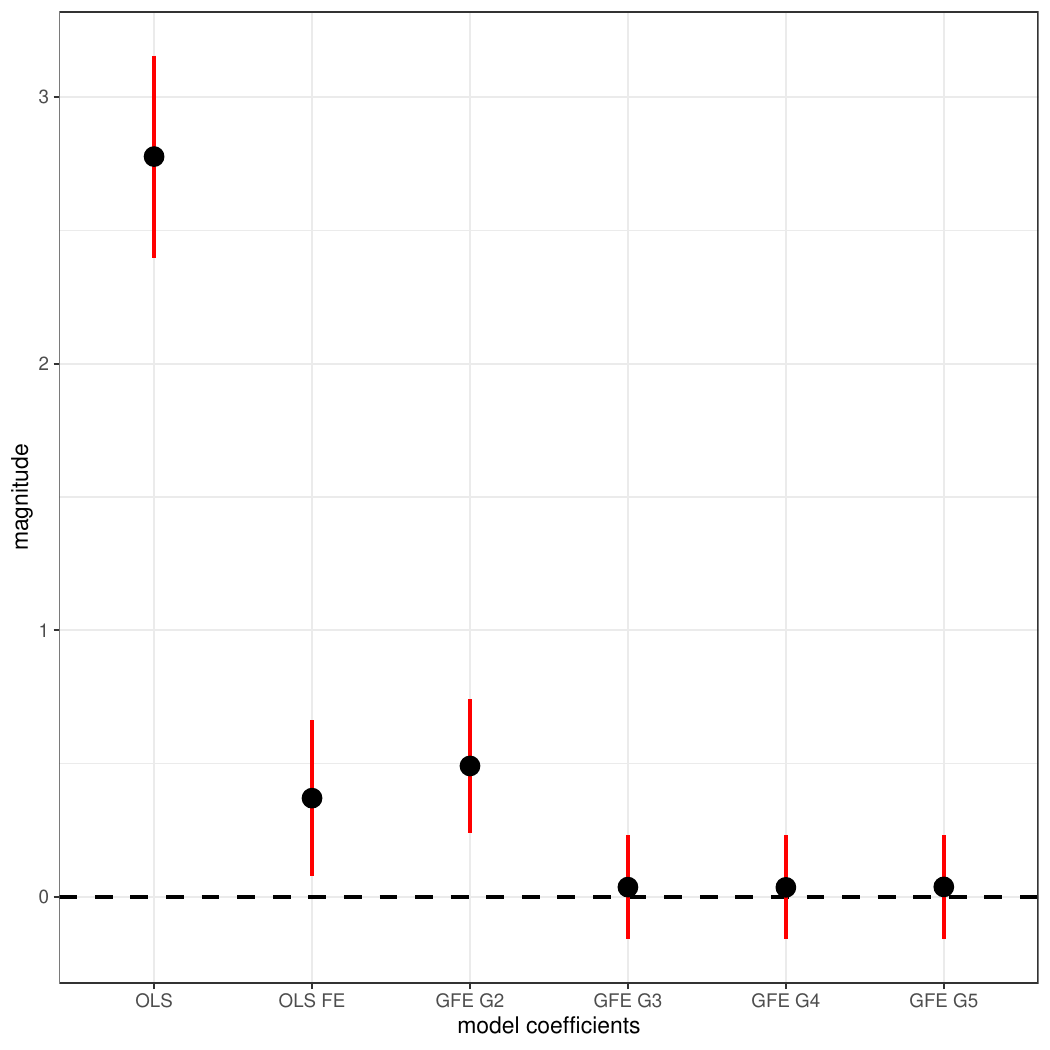}}
		\subfigure[$N=10000$]{\includegraphics[width=0.49\textwidth]{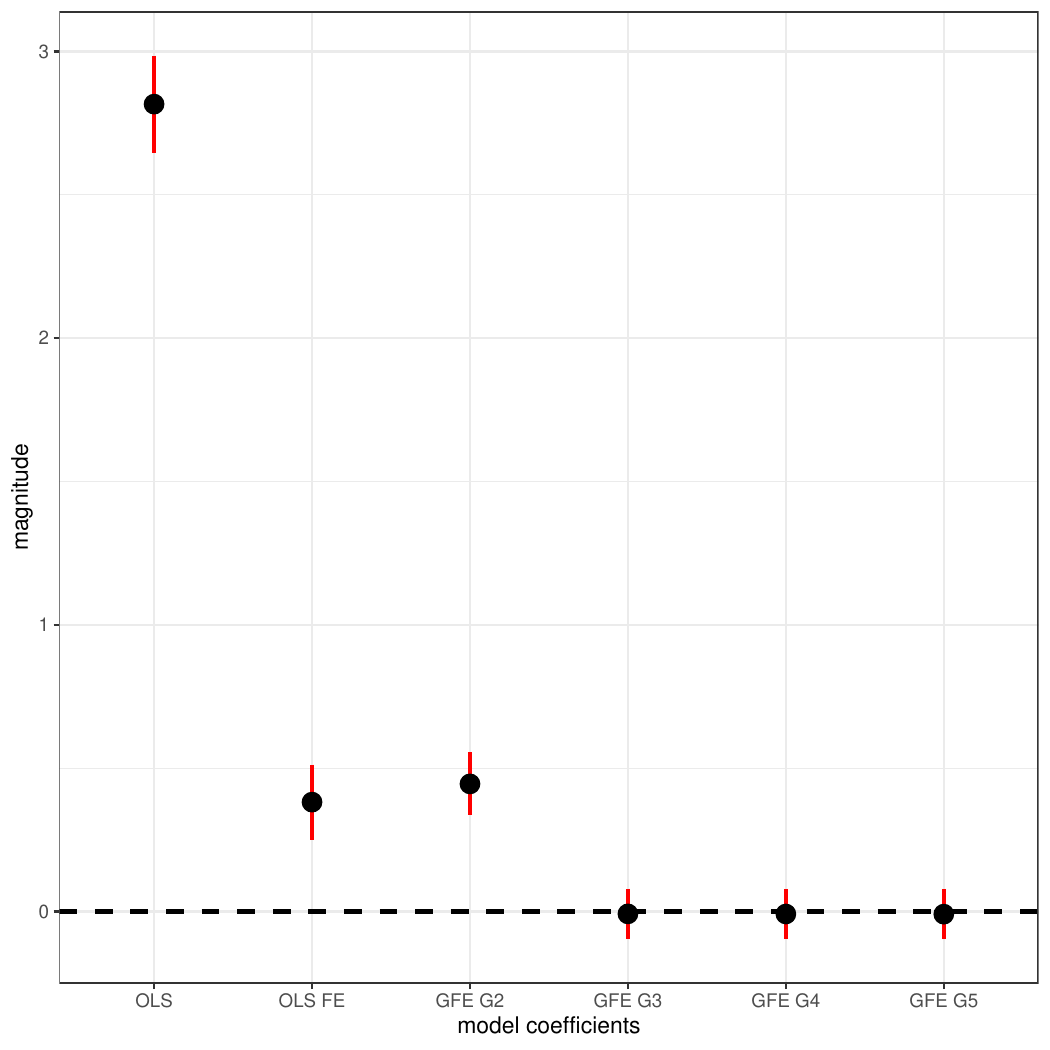}}
}
\caption{Estimated coefficient $\xi$ for different model specifications from left to right: (1) OLS, (2) individual-specific fixed-effects (OLS FE), (3) GFE estimator with two groups, (4) GFE estimator with three groups, (5) GFE estimator with four groups, (6) GFE estimator with five groups. Confidence intervals are depicted in red and are calculated using analytical standard errors.}\label{fig:sim_coef}	
\end{figure}

The resulting estimates for the parameter of interest in the different specifications for the effect of mental health are displayed in Figure \ref{fig:sim_coef}. The OLS estimator overestimates the effect by a significant amount due to the omitted variable bias, but even in OLS with individual-specific fixed-effects, the effect estimate remains sizable and significant for sample sizes similar to ours, although we reduced $N$ by half to reduce computation time. When we control for dynamic grouped fixed-effects, the estimate $\widehat{\xi}$ shrinks toward zero and the confidence interval includes zero.  

\begin{figure}\vspace{-1cm}
\mbox{
	\subfigure[$G=1$]{\includegraphics[width=0.45\textwidth]{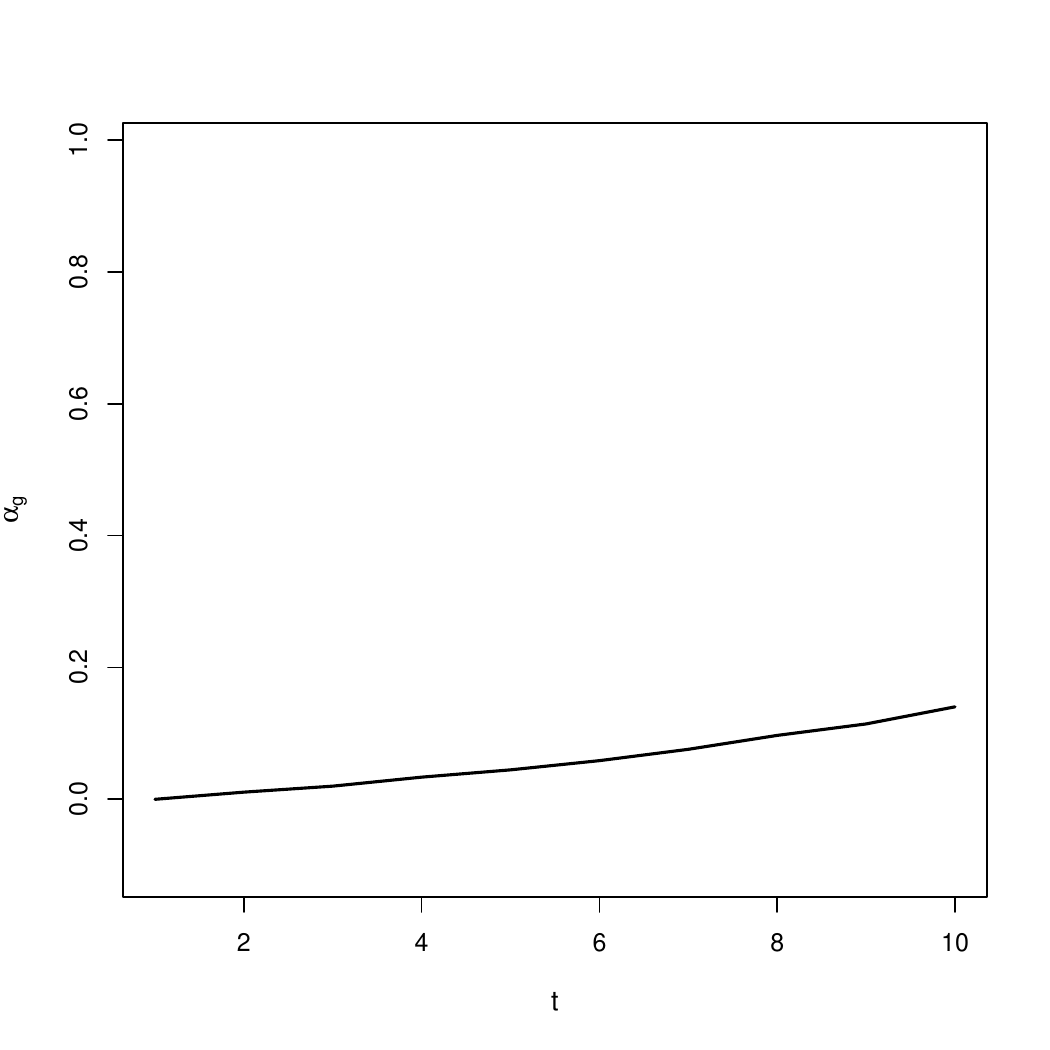}}
\subfigure[$G=2$]{\includegraphics[width=0.45\textwidth]{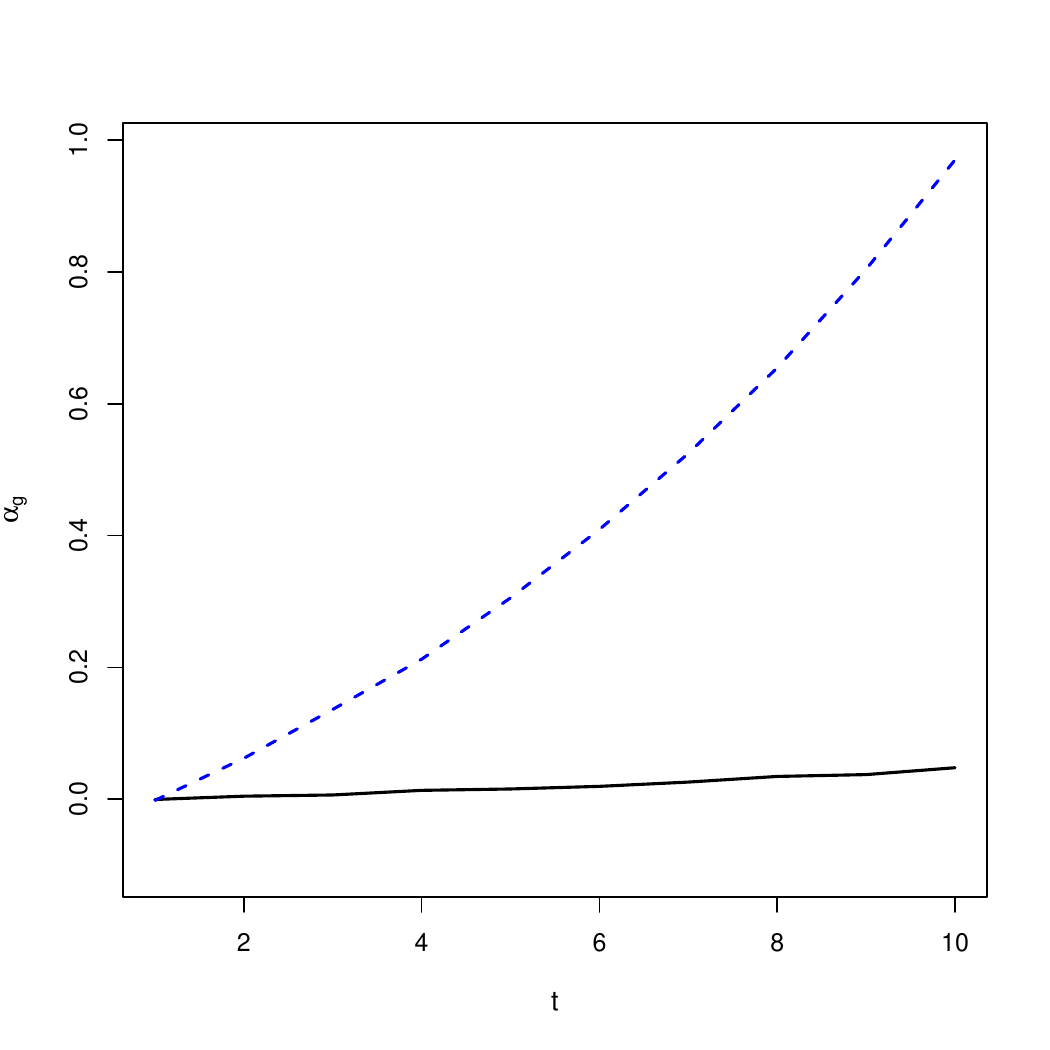}}
}
\\[-1.5em]
\mbox{
\subfigure[$G=3$]{\includegraphics[width=0.45\textwidth]{10000g3.pdf}}
\subfigure[$G=4$]{\includegraphics[width=0.45\textwidth]{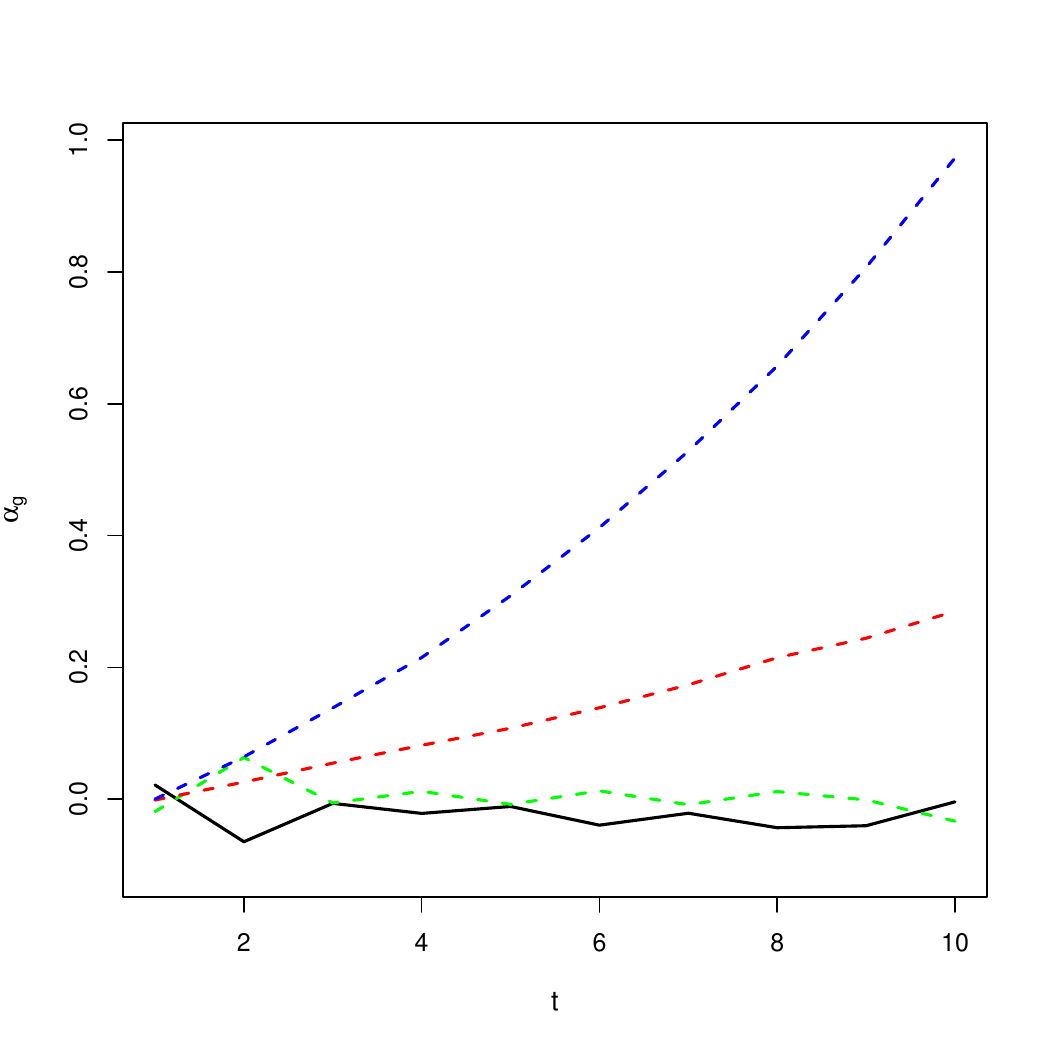}}
} 
\\[-1.5em]
\subfigure[$G=5$]{\includegraphics[width=0.45\textwidth]{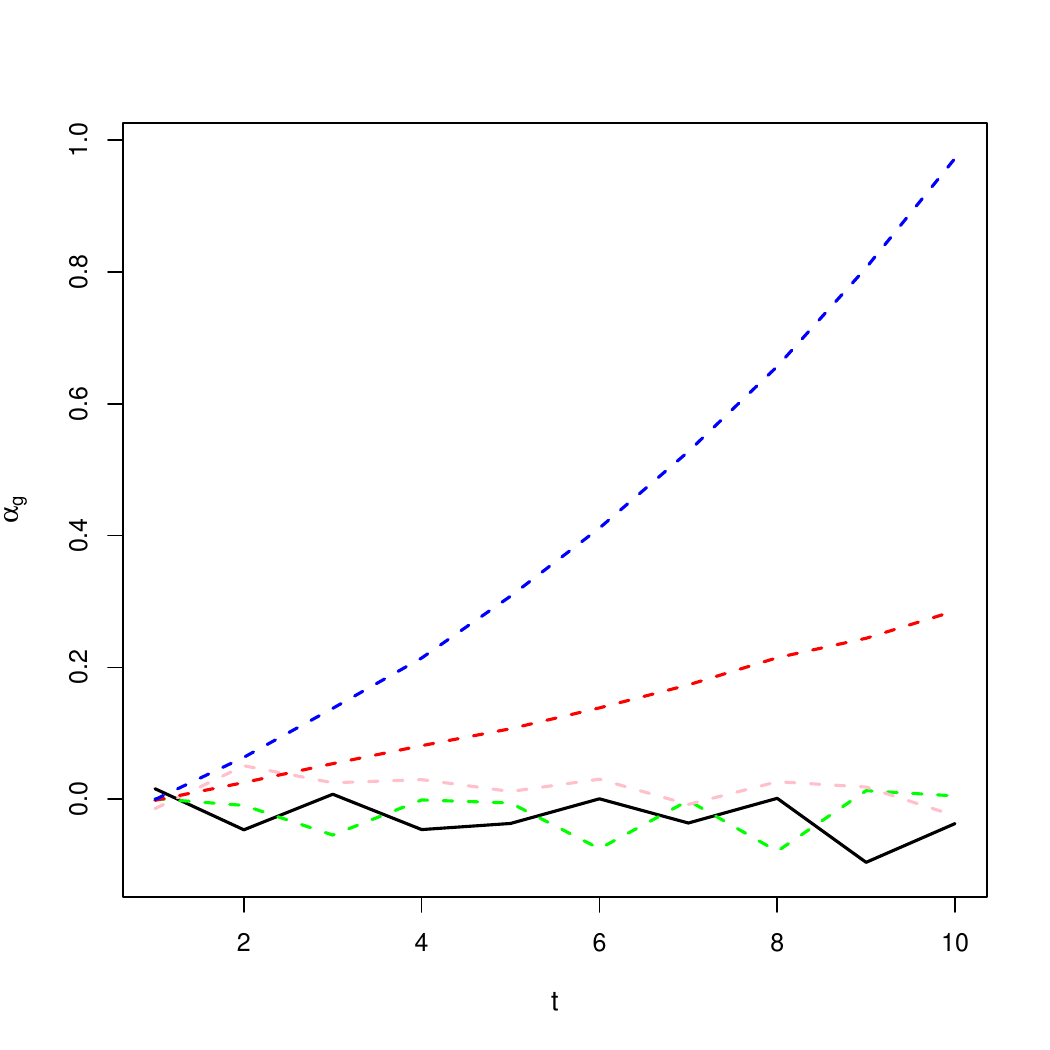}}
\caption{Estimated grouped fixed-effects profiles $\alpha_g$, for $G=1,...,5$, for one simulation run.}\label{fig:sim_profile}
\end{figure}

The GFE estimator is not ``identified'' in the sense that it requires the number of groups to be known, i.e. chosen by the researcher. Note that the optimal number of groups in our simulation example is three. As shown in Figure \ref{fig:sim_coef}, the GFE correctly estimates a zero effect when the correct number of groups is chosen. However, for $G=2$ the estimated coefficient is heavily upward biased to a similar amount as the OLS estimator with individual-specific fixed-effects. By contrast, selecting too many groups does not bias the estimated coefficients. This indicates that the GFE estimator consistently estimates the true effect, once the number of groups corresponds at least to the optimal one, at least for our data generating process. 

Figure \ref{fig:sim_profile} displays the estimate profiles of unobserved heterogeneity for $N=10,000$ observations and a varying number of groups. Figures \ref{fig:sim_profile}(a) and \ref{fig:sim_profile}(b) show that the time profiles do not exhibit a sufficient amount of unobserved heterogeneity which results in biased coefficient estimates (see Figure \ref{fig:sim_coef}). By contrast, adding more groups than necessary does not imply that the estimator does not assign any individual observations to these superfluous groups. Rather, the GFE splits existing groups which leads to an ``overfitting'' of the time profiles (see Figures \ref{fig:sim_profile}(d) and \ref{fig:sim_profile}(e)). This behavior is similar to what we observe in our empirical, real data application. Adding more groups splits up the existing groups and the generated trajectories of unobserved mental health profiles for the additional groups are similar to the group that was split up.

\begin{figure}
\mbox{
	\subfigure[$N=1000$]{\includegraphics[width=0.49\textwidth]{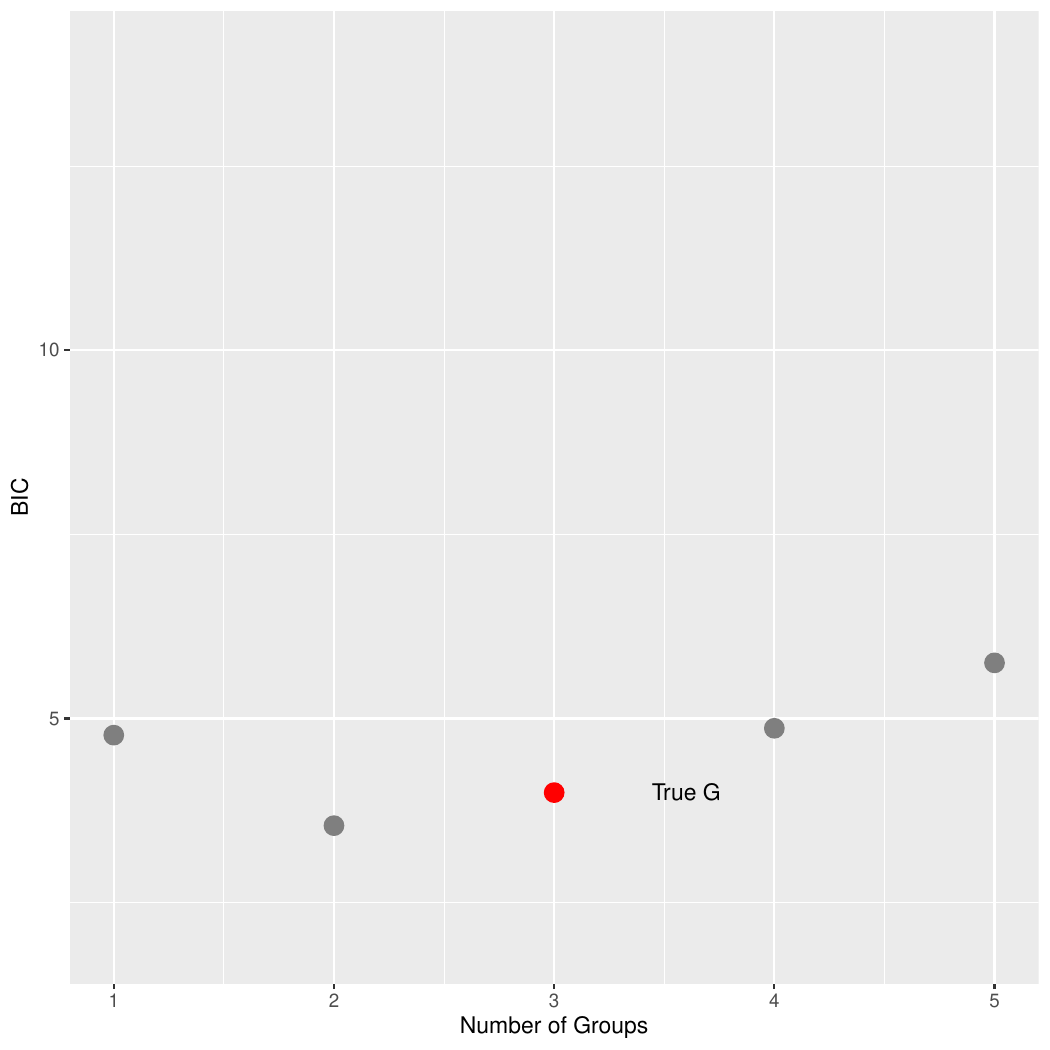}}
	\subfigure[$N=1500$]{\includegraphics[width=0.49\textwidth]{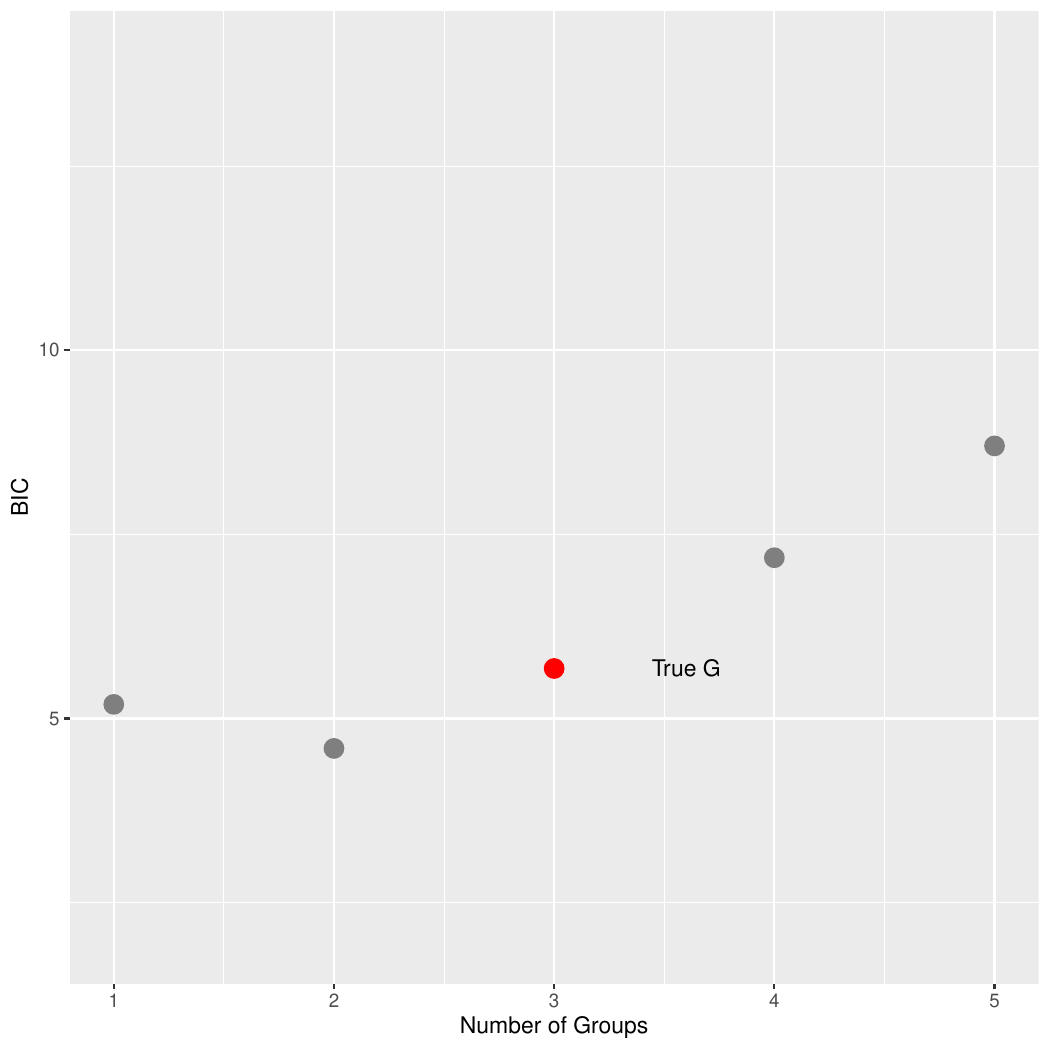}}
}
\mbox{
	\subfigure[$N=2000$]{\includegraphics[width=0.49\textwidth]{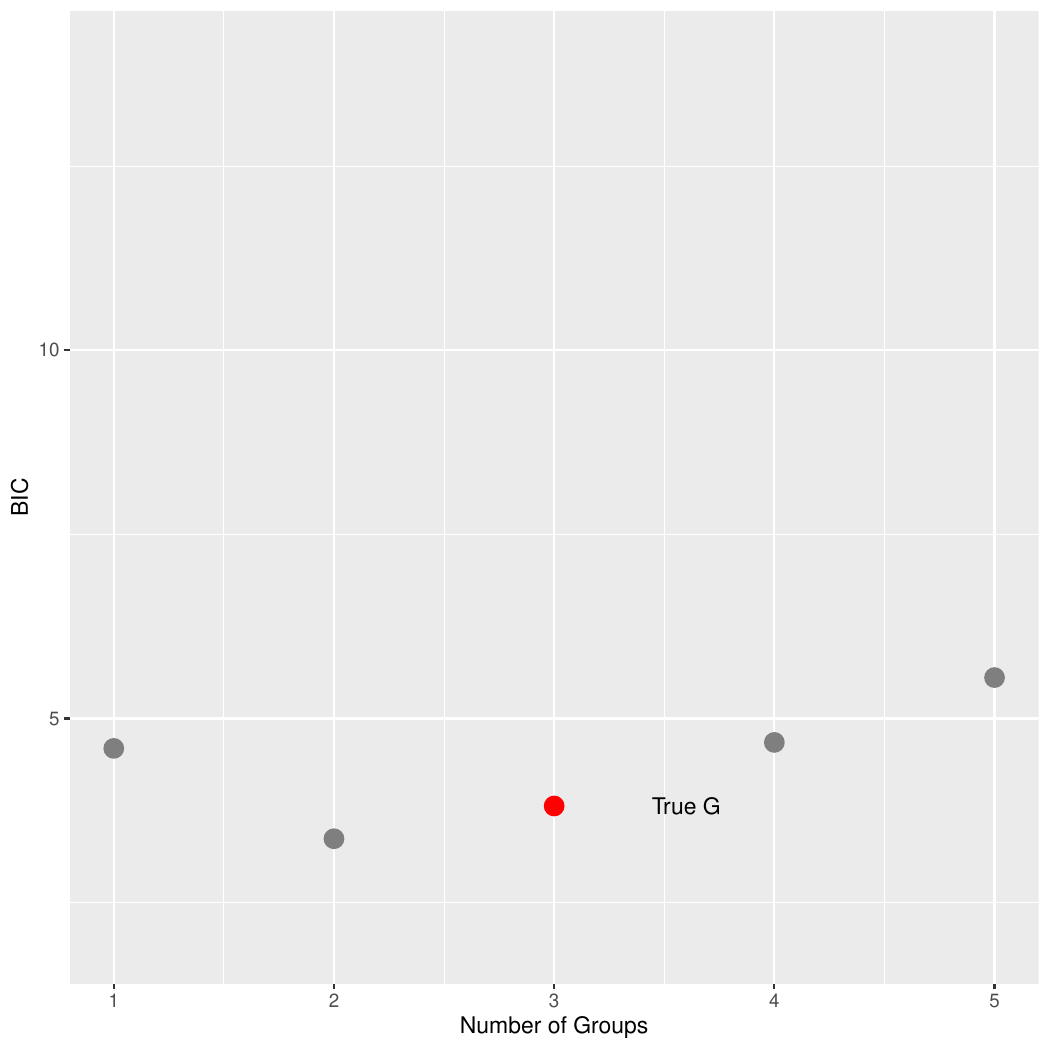}}
	\subfigure[$N=10000$]{\includegraphics[width=0.49\textwidth]{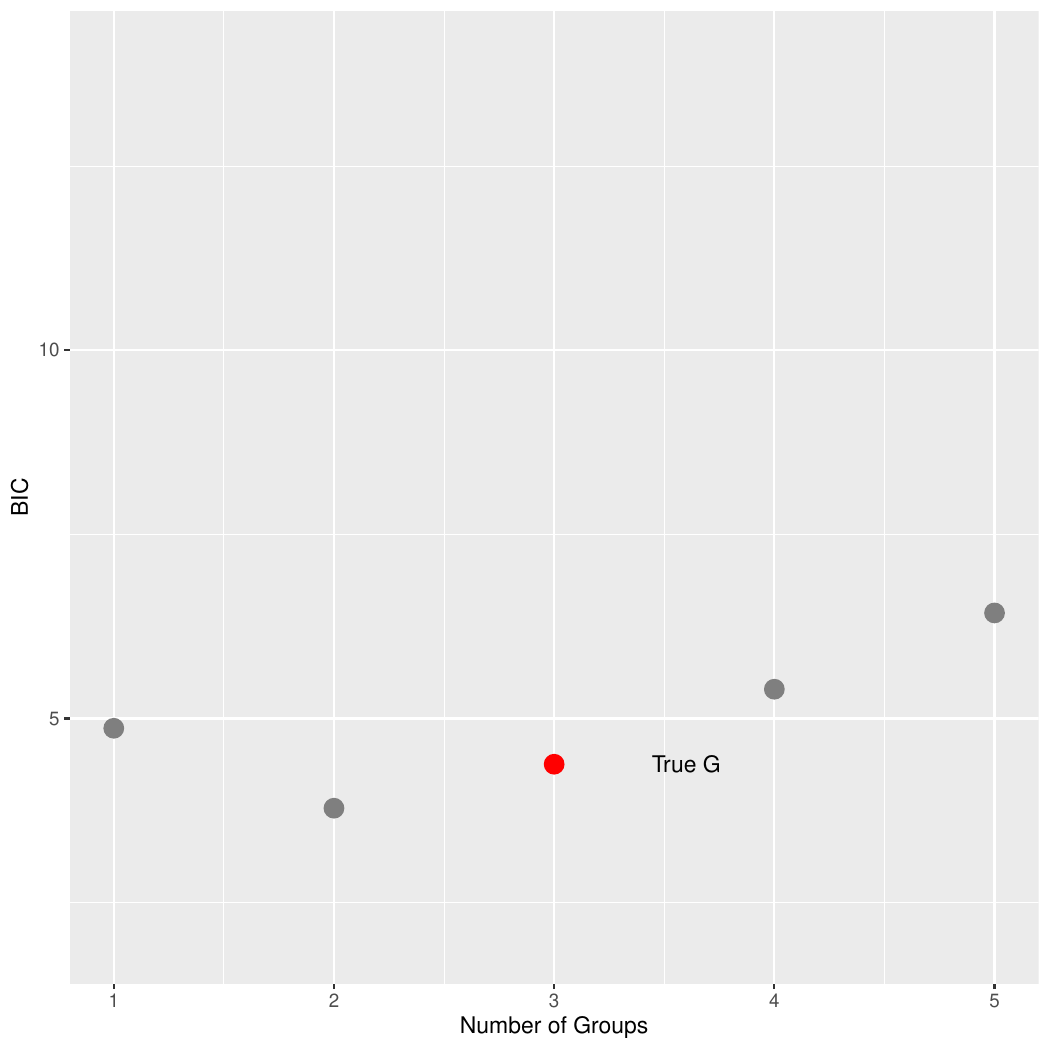}}
}
\caption{Results for the BIC with the steeper penalty in terms of $G$. }\label{fig:sim_BIC}	
\end{figure}

\begin{figure}
\mbox{
	\subfigure[$N=1000$]{\includegraphics[width=0.49\textwidth]{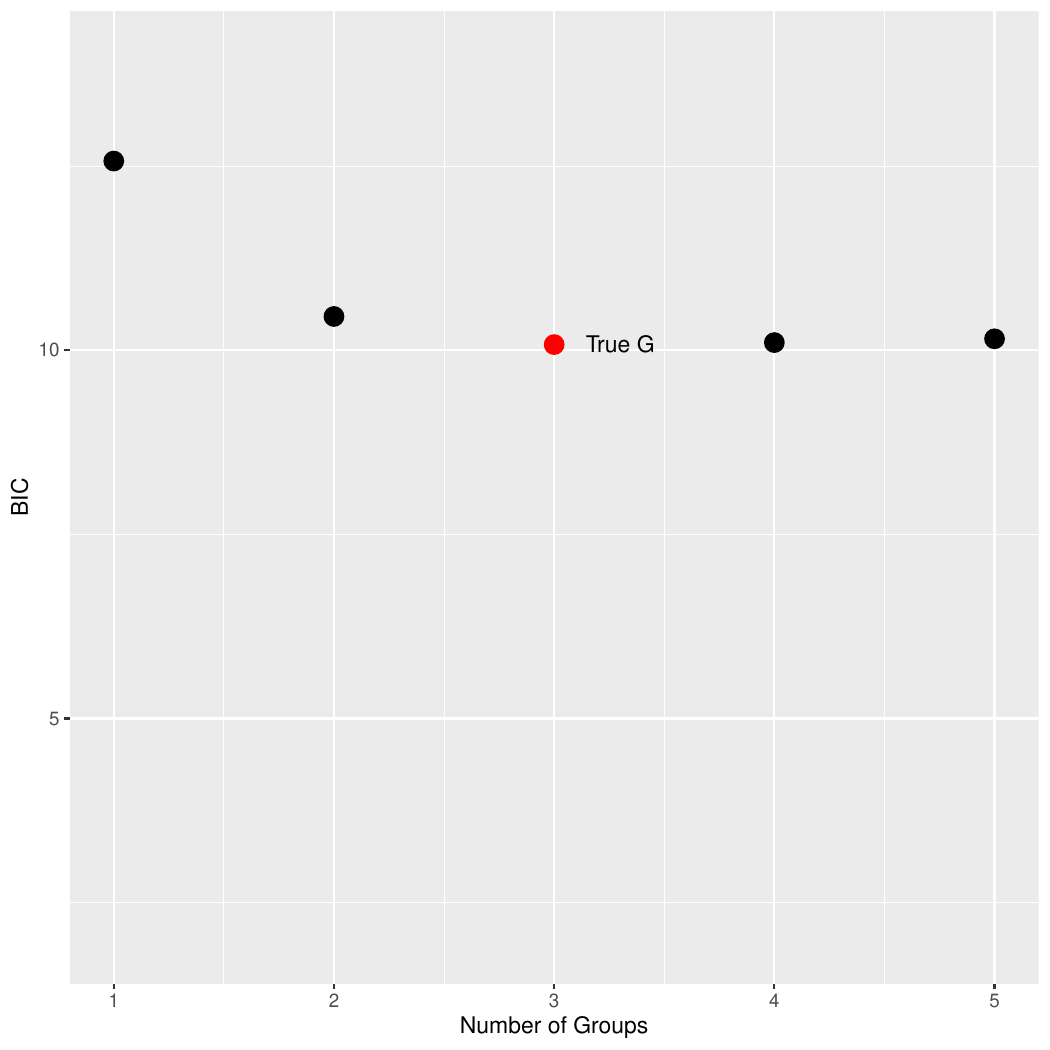}}
	\subfigure[$N=1500$]{\includegraphics[width=0.49\textwidth]{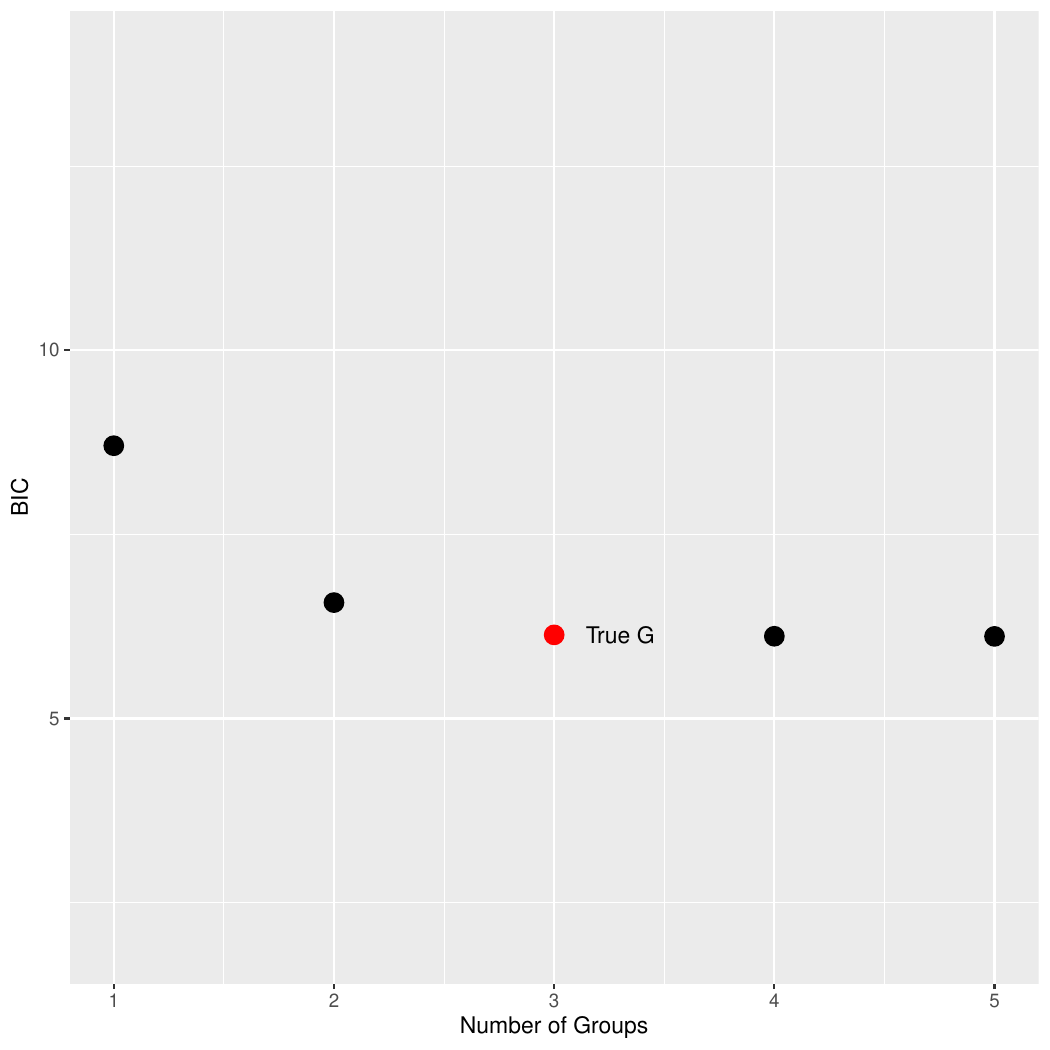}}
}
\mbox{
	\subfigure[$N=2000$]{\includegraphics[width=0.49\textwidth]{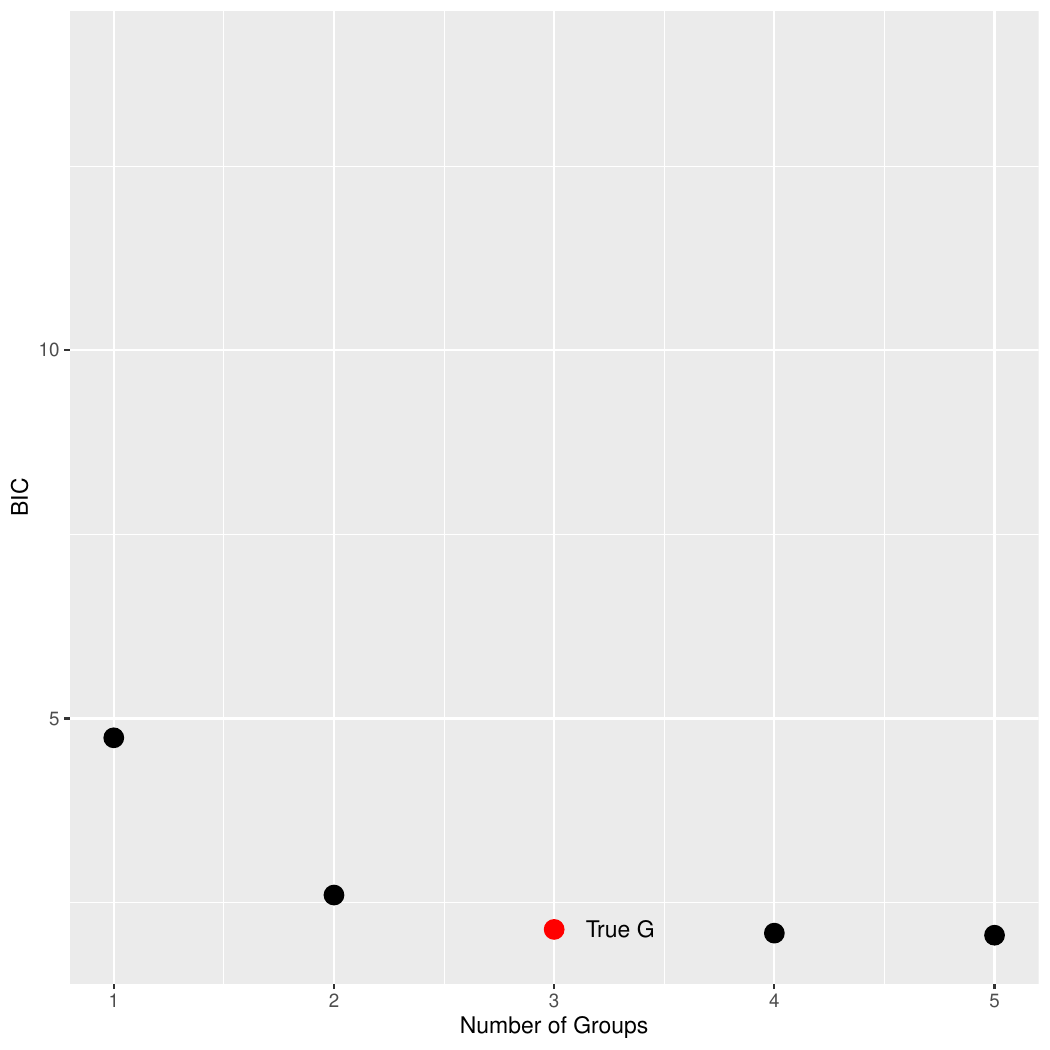}}
	\subfigure[$N=10000$]{\includegraphics[width=0.49\textwidth]{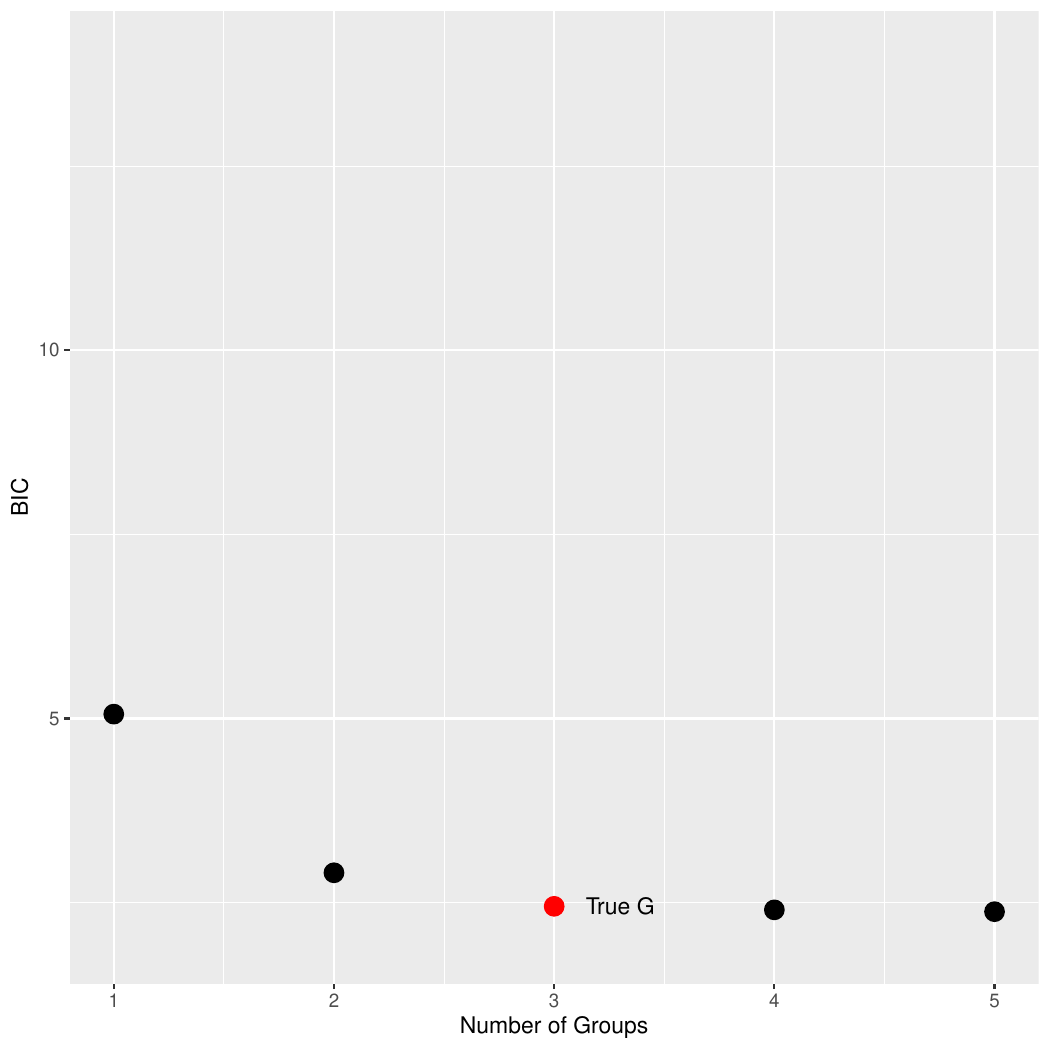}}
}
\caption{Results for the BIC with the less steep penalty used in \cite{bonhomme2015grouped}.}\label{fig:sim_BIC1}	
\end{figure}

Finally, we investigate the finite sample behavior of the BIC criterion with two different penalty terms (Figures \ref{fig:sim_BIC} and \ref{fig:sim_BIC1}) in a setting with large $N$ and fixed $T$. As discussed in Section \ref{sec:res_alpha} the BIC preferred by \cite{bonhomme2015grouped} (BIC standard) does not discriminate sufficiently for all $G\geq$ in our our application. Our simulation exercise clearly shows that the number of groups selected by the BIC standard depends on the number of observations $N$ relative to the number of time periods $T$. As shown in Figure \ref{fig:sim_BIC1}, the BIC selects the correct number of groups, $G=3$, for $1,000$ observations. However, when we increase $N$, the number of groups selected by this BIC increases, indicating that the penalization used in this BIC is not steep enough. As in our application, the BIC standard remains practically unchanged when increasing the number of groups once $G>1$.

By contrast, the BIC with a steeper penalty term (in $G$) always chooses two groups regardless of the number of observations. As indicated by the steep increase in the value of this BIC, the penalization with respect to the number of groups is too strong (Figure \ref{fig:sim_BIC}). We observe a similar behavior in our application.

%
%
%

\newpage \clearpage
\setcounter{table}{0}
\renewcommand{\thetable}{\Alph{subsection}.\arabic{table}}

\setcounter{figure}{0}
\renewcommand{\thefigure}{\Alph{subsection}.\arabic{figure}}

\setcounter{equation}{0}
\renewcommand{\theequation}{\Alph{subsection}.\arabic{equation}}

\subsection{Model solution}\label{app_modelsolution}
Numerically, the decision function $\rho^*_t$ can be computed by backward induction over $t$, starting with a guess in the far future which does not affect behavior in initial periods. The backward recursion is a variation of classical dynamic programming that takes into account the time-inconsistency introduced through $\beta$. To facilitate formulating the dynamic program, we denote by $F_t(M_t)$ the (classically, without $\beta$) discounted sum of future flow utilities given the decisions $\rho^*$ at age $\tau >t$
\begin{flalign}
F_t(M_t) =  u(\rho^*_t(M_t),M_t) + \E_t\left[ \sum_{\tau = t+1}^{\infty}\delta^{\tau-t}u(\rho^*_\tau(M^*_\tau),M^*_\tau)\right] .
\end{flalign}

Every woman's optimization problem specified by Equation \eqref{eq: opt} of computing $\rho^*_t(M_t)$ can be written compactly in terms of $F$ as
\begin{flalign}\label{eq: opt2}
\rho^*_t(M_t)= \text{arg}\max_{\rho_t} \left\lbrace  u(\rho_t,M_t) + \beta \delta \E_t[F_{t+1}(M^*_{t+1})] \right\rbrace 
\end{flalign}

Moreover, the functions $F_t$ satisfy the recursion
\begin{flalign}\label{eq: rec}
F_t(M_t)=u(\rho^*_t(M_t),M_t) + \delta \E_t[F_{t+1}(M^*_{t+1})] 
\end{flalign}

Our numerical approach via backward induction thus looks as follows. We initialize by guessing the terminal condition  $F_T(M_T)=\bar{u}(M_T)=0$ for $T=100$.\footnote{This guess is incorrect but can be expected not to affect behavior in early time periods $t=1,\ldots 8$. We verify this by checking that results do not change if we initialize instead at $T=200$.}
In order to sequentially compute the decision function $\rho^*_t$ and the functions $F_t$, we alternate two steps backwards in time. Assume that $F_{t+1}$ is already known. Then, we can compute $\rho^*_t(M_t)$ by solving the problem in Equation \eqref{eq: opt2} for every value of $M_t$.  Once $\rho_t^*(M_t)$ is known, we can compute $F_t$ from Equation \eqref{eq: rec} and go back one more step in time.
When solving this problem computationally, we first discretize the state variables $M$ and $\rho$ over a suitable grid.\footnote{We assume that the choice of $\rho$ is discrete and the possible value are $0,0.05,0.1,...,0.95,1$. For $M$ we simulate 1,000 trajectories for the two most extreme values of $\rho$, $\rho_t \equiv 0$ and $\rho_t \equiv 1$. We use the maximum and minimum of resulting mental health trajectories to determine the boundaries of the grid. We choose 200 equidistant levels in between.} Thus, the maximizations do not have to be performed for every possible value of $M$ and $\rho$, but only for every value on the grid. The resulting discrete functions $F_t$ are interpolated using monotone Hermite splines. To compute the conditional expectations at each age, we take a Monte Carlo average over 1,000 possible scenarios for $M_{t+1}$ for the next step given each combination $(M_t, \rho_t)$. In this way, the functions $F_t$ and $\rho^*_t$ can be computed backward in time one by one. With the resulting decision functions $\rho^*_t$, we then simulate 10,000 optimal mental health trajectories $M^*_t$ and the associated risky behavior $\rho^*_t(M^*_t)$.

\clearpage
\bibliography{Abortion_Mental_Health_Janys_Siflinger} 

\begin{thebibliography}{92}
\newcommand{\enquote}[1]{``#1''}
\expandafter\ifx\csname natexlab\endcsname\relax\def\natexlab#1{#1}\fi

\bibitem[\protect\citeauthoryear{Aizer}{Aizer}{2017}]{aizer2017review}
\textsc{Aizer, A.} (2017): \enquote{A Review Essay on Isabel Sawhill's
  Generation Unbound: Drifting into Sex and Parenting without Marriage and
  Laurence Steinberg's Age of Opportunity: Lessons from the New Science of
  Adolescence,} \emph{Journal of Economic Literature}, 55, 592--608.

\bibitem[\protect\citeauthoryear{Alan and Ertac}{Alan and
  Ertac}{2018}]{alan2018fostering}
\textsc{Alan, S. and S.~Ertac} (2018): \enquote{Fostering Patience in the
  Classroom: Results from Randomized Educational Intervention,} \emph{Journal
  of Political Economy}, 126, 1865--1911.

\bibitem[\protect\citeauthoryear{Ananat, Gruber, and Levine}{Ananat
  et~al.}{2007}]{ananat2007abortion}
\textsc{Ananat, E.~O., J.~Gruber, and P.~Levine} (2007): \enquote{Abortion
  Legalization and Life-Cycle Fertility,} \emph{Journal of Human Resources},
  42, 375--397.

\bibitem[\protect\citeauthoryear{Ananat, Gruber, Levine, and Staiger}{Ananat
  et~al.}{2009}]{ananat2009abortion}
\textsc{Ananat, E.~O., J.~Gruber, P.~B. Levine, and D.~Staiger} (2009):
  \enquote{Abortion and Selection,} \emph{The Review of Economics and
  Statistics}, 91, 124--136.

\bibitem[\protect\citeauthoryear{Anderson and Hsiao}{Anderson and
  Hsiao}{1982}]{anderson1982formulation}
\textsc{Anderson, T.~W. and C.~Hsiao} (1982): \enquote{Formulation and
  estimation of dynamic models using panel data,} \emph{Journal of
  Econometrics}, 18, 47--82.

\bibitem[\protect\citeauthoryear{Bai and Ng}{Bai and
  Ng}{2002}]{bai2002determining}
\textsc{Bai, J. and S.~Ng} (2002): \enquote{Determining the Number of Factors
  in Approximate Factor Models,} \emph{Econometrica}, 70, 191--221.

\bibitem[\protect\citeauthoryear{Biasi, Dahl, and Moser}{Biasi
  et~al.}{2021}]{biasi2018career}
\textsc{Biasi, B., M.~S. Dahl, and P.~Moser} (2021): \enquote{Career effects of
  mental health,} NBER Working Paper 29031, NBER.

\bibitem[\protect\citeauthoryear{Biggs, Brown, and Foster}{Biggs
  et~al.}{2020}]{biggs2020perceived}
\textsc{Biggs, M.~A., K.~Brown, and D.~G. Foster} (2020): \enquote{Perceived
  abortion stigma and psychological well-being over five years after receiving
  or being denied an abortion,} \emph{PloS one}, 15, e0226417.

\bibitem[\protect\citeauthoryear{Biggs, Upadhyay, McCulloch, and Foster}{Biggs
  et~al.}{2017}]{biggs2017women}
\textsc{Biggs, M.~A., U.~D. Upadhyay, C.~E. McCulloch, and D.~G. Foster}
  (2017): \enquote{Women’s Mental Health and Well-Being 5 Years after
  Receiving or Being Denied an Abortion: A Prospective, Longitudinal Cohort
  Study,} \emph{JAMA Psychiatry}, 74, 169--178.

\bibitem[\protect\citeauthoryear{Bonhomme and Manresa}{Bonhomme and
  Manresa}{2015}]{bonhomme2015grouped}
\textsc{Bonhomme, S. and E.~Manresa} (2015): \enquote{Grouped Patterns of
  Heterogeneity in Panel Data,} \emph{Econometrica}, 83, 1147--1184.

\bibitem[\protect\citeauthoryear{Borghans, Duckworth, Heckman, and
  Ter~Weel}{Borghans et~al.}{2008}]{borghans2008economics}
\textsc{Borghans, L., A.~L. Duckworth, J.~J. Heckman, and B.~Ter~Weel} (2008):
  \enquote{The Economics and Psychology of Personality Traits,} \emph{Journal
  of Human Resources}, 43, 972--1059.

\bibitem[\protect\citeauthoryear{Borusyak and Jaravel}{Borusyak and
  Jaravel}{2017}]{borusyak2017revisiting}
\textsc{Borusyak, K. and X.~Jaravel} (2017): \enquote{Revisiting Event Study
  Designs,} \emph{Available at SSRN 2826228}.

\bibitem[\protect\citeauthoryear{Callaway and Sant'Anna}{Callaway and
  Sant'Anna}{2021}]{callaway2021difference}
\textsc{Callaway, B. and P.~H. Sant'Anna} (2021):
  \enquote{Difference-in-differences with multiple time periods,} \emph{Journal
  of Econometrics}, 225, 200--230.

\bibitem[\protect\citeauthoryear{Cawley and Ruhm}{Cawley and
  Ruhm}{2011}]{cawley2011economics}
\textsc{Cawley, J. and C.~J. Ruhm} (2011): \enquote{The Economics of Risky
  Health Behaviors,} in \emph{Handbook of Health Economics}, Elsevier, vol.~2,
  95--199.

\bibitem[\protect\citeauthoryear{{Centers for Disease Control and
  Prevention}}{{Centers for Disease Control and Prevention}}{2019}]{STD2018US}
\textsc{{Centers for Disease Control and Prevention}} (2019): \enquote{Sexually
  Transmitted Disease Surveillance 2018,} \emph{Atlanta: U.S. Department of
  Health and Human Services}.

\bibitem[\protect\citeauthoryear{Chesson, Leichliter, Zimet, Rosenthal,
  Bernstein, and Fife}{Chesson et~al.}{2006}]{chesson2006discount}
\textsc{Chesson, H.~W., J.~S. Leichliter, G.~D. Zimet, S.~L. Rosenthal, D.~I.
  Bernstein, and K.~H. Fife} (2006): \enquote{Discount Rates and Risky Sexual
  Behaviors among Teenagers and Young Adults,} \emph{Journal of Risk and
  Uncertainty}, 32, 217--230.

\bibitem[\protect\citeauthoryear{Choi and Jeong}{Choi and
  Jeong}{2019}]{choi2019model}
\textsc{Choi, I. and H.~Jeong} (2019): \enquote{Model Selection for Factor
  Analysis: Some new Criteria and Performance Comparisons,} \emph{Econometric
  Reviews}, 38, 577--596.

\bibitem[\protect\citeauthoryear{Clarke and M{\"u}hlrad}{Clarke and
  M{\"u}hlrad}{2021}]{clarke2018abortion}
\textsc{Clarke, D. and H.~M{\"u}hlrad} (2021): \enquote{Abortion Laws and
  Women’s Health,} \emph{Journal of Health Economics}, 76, 102413.

\bibitem[\protect\citeauthoryear{Cobb-Clark, Dahmann, and
  Kettlewell}{Cobb-Clark et~al.}{2020}]{cobb2020depression}
\textsc{Cobb-Clark, D.~A., S.~C. Dahmann, and N.~Kettlewell} (2020):
  \enquote{Depression, Risk Preferences and Risk-Taking Behavior,}
  \emph{Journal of Human Resources}, 0419--10183R1.

\bibitem[\protect\citeauthoryear{Cronin, Forsstrom, and Papageorge}{Cronin
  et~al.}{2020}]{cronin2020good}
\textsc{Cronin, C.~J., M.~P. Forsstrom, and N.~W. Papageorge} (2020):
  \enquote{What Good Are Treatment Effects without Treatment? Mental Health and
  the Reluctance to Use Talk Therapy,} NBER Working Paper 27711, NBER.

\bibitem[\protect\citeauthoryear{Cuddy and Currie}{Cuddy and
  Currie}{2020}]{cuddy2020rules}
\textsc{Cuddy, E. and J.~Currie} (2020): \enquote{Rules vs. Discretion:
  Treatment of Mental Illness in US Adolescents,} NBER Working Paper 27890,
  NBER.

\bibitem[\protect\citeauthoryear{Currie}{Currie}{2020}]{currie2020child}
\textsc{Currie, J.} (2020): \enquote{Child Health as Human Capital,}
  \emph{Health Economics}, 29, 452--463.

\bibitem[\protect\citeauthoryear{Currie, Nixon, and Cole}{Currie
  et~al.}{1996}]{currie1996restrictions}
\textsc{Currie, J., L.~Nixon, and N.~Cole} (1996): \enquote{Restrictions on
  Medicaid Funding of Abortion: Effects on Birth Weight and Pregnancy
  Resolutions,} \emph{Journal of Human Resources}, 31, 159--188.

\bibitem[\protect\citeauthoryear{Currie, Stabile, Manivong, and Roos}{Currie
  et~al.}{2010}]{currie2010child}
\textsc{Currie, J., M.~Stabile, P.~Manivong, and L.~L. Roos} (2010):
  \enquote{Child Health and Young Adult Outcomes,} \emph{Journal of Human
  Resources}, 45, 517--548.

\bibitem[\protect\citeauthoryear{Curry, Silva, Rohde, Ginsburg, Kratochvil,
  Simons, Kirchner, May, Kennard, Mayes et~al.}{Curry
  et~al.}{2011}]{curry2011recovery}
\textsc{Curry, J., S.~Silva, P.~Rohde, G.~Ginsburg, C.~Kratochvil, A.~Simons,
  J.~Kirchner, D.~May, B.~Kennard, T.~Mayes, et~al.} (2011): \enquote{Recovery
  and recurrence following treatment for adolescent major depression,}
  \emph{Archives of General Psychiatry}, 68, 263--269.

\bibitem[\protect\citeauthoryear{Danielsson, Berglund, Forsberg, Larsson,
  Rogala, and Tyd{\'e}n}{Danielsson et~al.}{2012}]{danielsson2012sexual}
\textsc{Danielsson, M., T.~Berglund, M.~Forsberg, M.~Larsson, C.~Rogala, and
  T.~Tyd{\'e}n} (2012): \enquote{Sexual and Reproductive Health: Health in
  Sweden: The National Public Health Report 2012. Chapter 9,}
  \emph{Scandinavian Journal of Public Health}, 40, 176--196.

\bibitem[\protect\citeauthoryear{Eaton, Shao, Nestadt, Lee, Bienvenu, and
  Zandi}{Eaton et~al.}{2008}]{eatonprev}
\textsc{Eaton, W.~W., H.~Shao, G.~Nestadt, B.~H. Lee, O.~J. Bienvenu, and
  P.~Zandi} (2008): \enquote{Population-Based Study of First Onset and
  Chronicity in Major Depressive Disorder,} \emph{Archives of General
  Psychiatry}, 65, 513--520.

\bibitem[\protect\citeauthoryear{Elkington, Bauermeister, and
  Zimmerman}{Elkington et~al.}{2010}]{elkington2010psychological}
\textsc{Elkington, K.~S., J.~A. Bauermeister, and M.~A. Zimmerman} (2010):
  \enquote{Psychological Distress, Substance Use, and HIV/STI Risk Behaviors
  among Youth,} \emph{Journal of Youth and Adolescence}, 39, 514--527.

\bibitem[\protect\citeauthoryear{{European Centre for Disease Prevention and
  Control}}{{European Centre for Disease Prevention and
  Control}}{2020}]{STD2018EU}
\textsc{{European Centre for Disease Prevention and Control}} (2020):
  \enquote{Chlamydia Infection,} \emph{Annual Epidemiological Report for 2018.
  Stockholm: ECDC}.

\bibitem[\protect\citeauthoryear{Felkey and Lybecker}{Felkey and
  Lybecker}{2018}]{felkey2018restrictions}
\textsc{Felkey, A.~J. and K.~M. Lybecker} (2018): \enquote{Do Restrictions
  Beget Responsibility? The Case of US Abortion Legislation,} \emph{The
  American Economist}, 63, 59--70.

\bibitem[\protect\citeauthoryear{Fischer, Royer, and White}{Fischer
  et~al.}{2018}]{fischer2018impacts}
\textsc{Fischer, S., H.~Royer, and C.~White} (2018): \enquote{The Impacts of
  Reduced Access to Abortion and Family Planning Services on Abortions, Births,
  and Contraceptive Purchases,} \emph{Journal of Public Economics}, 167,
  43--68.

\bibitem[\protect\citeauthoryear{Fletcher}{Fletcher}{2010}]{fletcher2010adolescent}
\textsc{Fletcher, J.~M.} (2010): \enquote{Adolescent depression and educational
  attainment: results using sibling fixed effects,} \emph{Health Economics},
  19, 855--871.

\bibitem[\protect\citeauthoryear{Foster, Steinberg, Roberts, Neuhaus, and
  Biggs}{Foster et~al.}{2015}]{foster2015comparison}
\textsc{Foster, D.~G., J.~R. Steinberg, S.~C. Roberts, J.~Neuhaus, and M.~A.
  Biggs} (2015): \enquote{A comparison of depression and anxiety symptom
  trajectories between women who had an abortion and women denied one,}
  \emph{Psychological Medicine}, 45, 2073--2082.

\bibitem[\protect\citeauthoryear{Frederick, Loewenstein, and
  O'donoghue}{Frederick et~al.}{2002}]{frederick2002time}
\textsc{Frederick, S., G.~Loewenstein, and T.~O'donoghue} (2002):
  \enquote{{Time Discounting and Time Preference: A Critical Review},}
  \emph{Journal of Economic Literature}, 40, 351--401.

\bibitem[\protect\citeauthoryear{Glass}{Glass}{1938}]{glass1938effectiveness}
\textsc{Glass, D.~V.} (1938): \enquote{The Effectiveness of Abortion
  Legislation in Six Countries,} \emph{The Modern Law Review}, 2, 97--125.

\bibitem[\protect\citeauthoryear{Goffe, Ferrier, and Rogers}{Goffe
  et~al.}{1994}]{goffe1994global}
\textsc{Goffe, W.~L., G.~D. Ferrier, and J.~Rogers} (1994): \enquote{{Global
  Optimization of Statistical Functions with Simulated Annealing},}
  \emph{Journal of Econometrics}, 60, 65--99.

\bibitem[\protect\citeauthoryear{Gruber}{Gruber}{2001}]{gruber2000risky}
\textsc{Gruber, J.} (2001): \enquote{Risky Behavior among Youths:
  Introduction,} in \emph{Risky Behavior among Youths: An Economic Analysis},
  University of Chicago Press, 1--28.

\bibitem[\protect\citeauthoryear{Gruber and K{\"o}szegi}{Gruber and
  K{\"o}szegi}{2001}]{gruber2001addiction}
\textsc{Gruber, J. and B.~K{\"o}szegi} (2001): \enquote{Is Addiction
  “Rational”? Theory and Evidence,} \emph{The Quarterly Journal of
  Economics}, 116, 1261--1303.

\bibitem[\protect\citeauthoryear{Gruber, Levine, and Staiger}{Gruber
  et~al.}{1999}]{gruber1999abortion}
\textsc{Gruber, J., P.~Levine, and D.~Staiger} (1999): \enquote{Abortion
  Legalization and Child Living Circumstances: Who is the “Marginal
  Child”?} \emph{The Quarterly Journal of Economics}, 114, 263--291.

\bibitem[\protect\citeauthoryear{Guleria, Munk, Elfstr{\"o}m, Hansen,
  Sundstr{\"o}m, Liaw, Nyg{\aa}rd, and Kjaer}{Guleria
  et~al.}{2020}]{guleria2020emergency}
\textsc{Guleria, S., C.~Munk, K.~M. Elfstr{\"o}m, B.~T. Hansen,
  K.~Sundstr{\"o}m, K.-L. Liaw, M.~Nyg{\aa}rd, and S.~K. Kjaer} (2020):
  \enquote{Emergency Contraceptive Pill Use among Women in Denmark, Norway and
  Sweden: Population-based Survey,} \emph{Acta Obstetricia et Gynecologica
  Scandinavica}, 99, 1214--1221.

\bibitem[\protect\citeauthoryear{{Guttmacher Institute}}{{Guttmacher
  Institute}}{2020}]{guttmacher2020}
\textsc{{Guttmacher Institute}} (2020): \enquote{An Overview of Abortion Laws,}
  \url{https://www.guttmacher.org/state-policy/explore/overview-abortion-laws},
  last access: 26/2/2021.

\bibitem[\protect\citeauthoryear{Haegele}{Haegele}{2005}]{haegele2005ab}
\textsc{Haegele, M.} (2005): \enquote{Sexual and Reproductive Health and Rights
  in the {E}uropean {U}nion,} in \emph{Entre Nous - The European Magazine for
  Sexual and Reproductive Health}, WHO Regional Office for Europe, vol.~59,
  26--29.

\bibitem[\protect\citeauthoryear{Hallfors, Waller, Bauer, Ford, and
  Halpern}{Hallfors et~al.}{2005}]{hallfors2005comes}
\textsc{Hallfors, D.~D., M.~W. Waller, D.~Bauer, C.~A. Ford, and C.~T. Halpern}
  (2005): \enquote{Which Comes First in Adolescence--Sex and Drugs or
  Depression?} \emph{American Journal of Preventive Medicine}, 29, 163--170.

\bibitem[\protect\citeauthoryear{Heckman, Stixrud, and Urzua}{Heckman
  et~al.}{2006}]{heckman2006effects}
\textsc{Heckman, J.~J., J.~Stixrud, and S.~Urzua} (2006): \enquote{The Effects
  of Cognitive and Noncognitive Abilities on Labor Market Outcomes and Social
  Behavior,} \emph{Journal of Labor Economics}, 24, 411--482.

\bibitem[\protect\citeauthoryear{Hotz, McElroy, and Sanders}{Hotz
  et~al.}{2005}]{hotz2005teenage}
\textsc{Hotz, V.~J., S.~W. McElroy, and S.~G. Sanders} (2005): \enquote{Teenage
  childbearing and its life cycle consequences exploiting a natural
  experiment,} \emph{Journal of Human Resources}, 40, 683--715.

\bibitem[\protect\citeauthoryear{Hotz, Mullin, and Sanders}{Hotz
  et~al.}{1997}]{hotz1997bounding}
\textsc{Hotz, V.~J., C.~H. Mullin, and S.~G. Sanders} (1997): \enquote{Bounding
  causal effects using data from a contaminated natural experiment: Analysing
  the effects of teenage childbearing,} \emph{The Review of Economic Studies},
  64, 575--603.

\bibitem[\protect\citeauthoryear{Husmann, Lange, and Spiegel}{Husmann
  et~al.}{2017}]{husmann2017r}
\textsc{Husmann, K., A.~Lange, and E.~Spiegel} (2017): \enquote{The R package
  optimization: Flexible Global Optimization with Simulated-Annealing,} .

\bibitem[\protect\citeauthoryear{Jones and Henshaw}{Jones and
  Henshaw}{2002}]{jones2002mifepristone}
\textsc{Jones, R.~K. and S.~K. Henshaw} (2002): \enquote{Mifepristone for Early
  Medical Abortion: Experiences in France, Great Britain and Sweden,}
  \emph{Perspectives on Sexual and Reproductive Health}, 34, 154--161.

\bibitem[\protect\citeauthoryear{Kearney and Levine}{Kearney and
  Levine}{2015}]{kearney2015investigating}
\textsc{Kearney, M.~S. and P.~B. Levine} (2015): \enquote{Investigating recent
  trends in the US teen birth rate,} \emph{Journal of Health Economics}, 41,
  15--29.

\bibitem[\protect\citeauthoryear{Kortsmit, Jatlaoui, Mandel, Reeves, Oduyebo,
  Petersen, and Whiteman}{Kortsmit et~al.}{2020}]{cdcabortion2016}
\textsc{Kortsmit, K., T.~C. Jatlaoui, M.~G. Mandel, J.~A. Reeves, T.~Oduyebo,
  E.~Petersen, and M.~K. Whiteman} (2020): \enquote{Abortion Surveillance --
  United States, 2018,} \emph{Morbidity and Mortality Weekly Report.
  Surveillance Summaries 2020}, 69, 1--29.

\bibitem[\protect\citeauthoryear{Kristj{\'a}nsd{\'o}ttir, Olsson, Sundelin, and
  Naessen}{Kristj{\'a}nsd{\'o}ttir et~al.}{2011}]{kristjansdottir2011could}
\textsc{Kristj{\'a}nsd{\'o}ttir, J., G.~I. Olsson, C.~Sundelin, and T.~Naessen}
  (2011): \enquote{Could SF-36 be used as a screening instrument for depression
  in a Swedish youth population?} \emph{Scandinavian Journal of Caring
  Sciences}, 25, 262--268.

\bibitem[\protect\citeauthoryear{Lager, Berlin, Heimerson, and
  Danielsson}{Lager et~al.}{2012}]{lager2012young}
\textsc{Lager, A., M.~Berlin, I.~Heimerson, and M.~Danielsson} (2012):
  \enquote{Young people’s health: health in Sweden: the national public
  health report 2012. Chapter 3,} \emph{Scandinavian Journal of Public Health},
  40, 42--71.

\bibitem[\protect\citeauthoryear{Laibson}{Laibson}{1997}]{laibson1997golden}
\textsc{Laibson, D.} (1997): \enquote{Golden Eggs and Hyperbolic Discounting,}
  \emph{The Quarterly Journal of Economics}, 112, 443--478.

\bibitem[\protect\citeauthoryear{Lechner et~al.}{Lechner
  et~al.}{2010}]{lechner2011estimation}
\textsc{Lechner, M. et~al.} (2010): \enquote{The Estimation of Causal Effects
  by Difference-in-Difference Methods,} \emph{Foundations and Trends in
  Econometrics}, 4, 165--224.

\bibitem[\protect\citeauthoryear{Levine}{Levine}{2001}]{levine2001sexual}
\textsc{Levine, P.~B.} (2001): \enquote{The Sexual Activity and Birth-Control
  Use of American Teenagers,} in \emph{Risky Behavior among Youths: An Economic
  Analysis}, University of Chicago Press, 167--218.

\bibitem[\protect\citeauthoryear{Lindo, Myers, Schlosser, and Cunningham}{Lindo
  et~al.}{2020}]{lindo2020far}
\textsc{Lindo, J.~M., C.~K. Myers, A.~Schlosser, and S.~Cunningham} (2020):
  \enquote{How Far is too Far? New Evidence on Abortion Clinic Closures,
  Access, and Abortions,} \emph{Journal of Human Resources}, 55, 1137--1160.

\bibitem[\protect\citeauthoryear{Lindo and Pineda-Torres}{Lindo and
  Pineda-Torres}{2021}]{lindo2019waiting}
\textsc{Lindo, J.~M. and M.~Pineda-Torres} (2021): \enquote{New evidence on the
  effects of mandatory waiting periods for abortion,} \emph{Journal of Health
  Economics}, 80, 102533.

\bibitem[\protect\citeauthoryear{Marcus and Siedler}{Marcus and
  Siedler}{2015}]{marcus2015reducing}
\textsc{Marcus, J. and T.~Siedler} (2015): \enquote{Reducing Binge Drinking?
  The Effect of a Ban on Late-Night Off-Premise Alcohol Sales on
  Alcohol-Related Hospital Stays in Germany,} \emph{Journal of Public
  Economics}, 123, 55--77.

\bibitem[\protect\citeauthoryear{Markowitz, Kaestner, and Grossman}{Markowitz
  et~al.}{2005}]{markowitz2005investigation}
\textsc{Markowitz, S., R.~Kaestner, and M.~Grossman} (2005): \enquote{An
  Investigation of the Effects of Alcohol Consumption and Alcohol Policies on
  Youth Risky Sexual Behaviors,} \emph{American Economic Review}, 95, 263--266.

\bibitem[\protect\citeauthoryear{Meghir and Palme}{Meghir and
  Palme}{2005}]{meghir2005educational}
\textsc{Meghir, C. and M.~Palme} (2005): \enquote{Educational Reform, Ability,
  and Family Background,} \emph{American Economic Review}, 95, 414--424.

\bibitem[\protect\citeauthoryear{Miller, Wherry, and Foster}{Miller
  et~al.}{2020{\natexlab{a}}}]{miller2020economic}
\textsc{Miller, S., L.~R. Wherry, and D.~G. Foster} (2020{\natexlab{a}}):
  \enquote{The Economic Consequences of Being Denied an Abortion,} NBER Working
  Paper 26662, NBER.

\bibitem[\protect\citeauthoryear{Miller, Wherry, and Foster}{Miller
  et~al.}{2020{\natexlab{b}}}]{miller2020happens}
---\hspace{-.1pt}---\hspace{-.1pt}--- (2020{\natexlab{b}}): \enquote{What
  Happens after an Abortion Denial? A Review of Results from the Turnaway
  Study,} \emph{AEA Papers and Proceedings}, 110, 226--230.

\bibitem[\protect\citeauthoryear{M{\o}lland}{M{\o}lland}{2016}]{molland16}
\textsc{M{\o}lland, E.} (2016): \enquote{Benefits from Delay? The Effect of
  Abortion Availability on Young Women and their Children,} \emph{Labour
  Economics}, 43, 6--28.

\bibitem[\protect\citeauthoryear{Moon and Weidner}{Moon and
  Weidner}{2015}]{moon2015linear}
\textsc{Moon, H.~R. and M.~Weidner} (2015): \enquote{Linear Regression for
  Panel with Unknown Number of Factors as Interactive Fixed Effects,}
  \emph{Econometrica}, 83, 1543--1579.

\bibitem[\protect\citeauthoryear{Mulligan}{Mulligan}{2016}]{mulligan16}
\textsc{Mulligan, K.} (2016): \enquote{Access to Emergency Contraception and
  its Impact on Fertility and Sexual Behavior,} \emph{Health Economics}, 25,
  455--469.

\bibitem[\protect\citeauthoryear{Munk-Olsen, Laursen, Pedersen, Lidegaard, and
  Mortensen}{Munk-Olsen et~al.}{2011}]{munk2011induced}
\textsc{Munk-Olsen, T., T.~M. Laursen, C.~B. Pedersen, {\O}.~Lidegaard, and
  P.~B. Mortensen} (2011): \enquote{Induced First-Trimester Abortion and Risk
  of Mental Disorder,} \emph{New England Journal of Medicine}, 364, 332--339.

\bibitem[\protect\citeauthoryear{Myers}{Myers}{2017}]{myers2017power}
\textsc{Myers, C.~K.} (2017): \enquote{The Power of Abortion Policy:
  Reexamining the Effects of Young Women’s Access to Reproductive Control,}
  \emph{Journal of Political Economy}, 125, 2178--2224.

\bibitem[\protect\citeauthoryear{{NIH}}{{NIH}}{2019}]{depressionUS2017}
\textsc{{NIH}} (2019): \enquote{Major depression-National Institute on Mental
  Health,}
  \url{https://www.nimh.nih.gov/health/statistics/major-depression.shtml#part_155031},
  last access: 10/3/2021.

\bibitem[\protect\citeauthoryear{Nilsson and Paul}{Nilsson and
  Paul}{2018}]{nilsson2018health}
\textsc{Nilsson, A. and A.~Paul} (2018): \enquote{Patient Cost-Sharing,
  Socioeconomic Status, and Children's Health Care Utilization,} \emph{Journal
  of Health Economics}, 59, 109--124.

\bibitem[\protect\citeauthoryear{O'Donoghue and Rabin}{O'Donoghue and
  Rabin}{1999}]{o1999doing}
\textsc{O'Donoghue, T. and M.~Rabin} (1999): \enquote{Doing It Now or Later,}
  \emph{American Economic Review}, 89, 103--124.

\bibitem[\protect\citeauthoryear{O'Donoghue and Rabin}{O'Donoghue and
  Rabin}{2001}]{o2001risky}
---\hspace{-.1pt}---\hspace{-.1pt}--- (2001): \enquote{Risky Behavior among
  Youths: Some Issues from Behavioral Economics,} in \emph{Risky Behavior among
  Youths: An Economic Analysis}, University of Chicago Press, 29--68.

\bibitem[\protect\citeauthoryear{O'Donoghue and Rabin}{O'Donoghue and
  Rabin}{2015}]{o2015present}
---\hspace{-.1pt}---\hspace{-.1pt}--- (2015): \enquote{Present Bias: Lessons
  Learned and to be Learned,} \emph{American Economic Review}, 105, 273--79.

\bibitem[\protect\citeauthoryear{Olsson and von Knorring}{Olsson and von
  Knorring}{1997}]{olsson1997beck}
\textsc{Olsson, G. and A.-L. von Knorring} (1997): \enquote{Beck's Depression
  Inventory as a screening instrument for adolescent depression in Sweden:
  gender differences,} \emph{Acta Psychiatrica Scandinavica}, 95, 277--282.

\bibitem[\protect\citeauthoryear{Olsson and Von~Knorring}{Olsson and
  Von~Knorring}{1999}]{olsson1999adolescent}
\textsc{Olsson, G. and A.-L. Von~Knorring} (1999): \enquote{Adolescent
  depression: prevalence in Swedish high-school students,} \emph{Acta
  Psychiatrica Scandinavica}, 99, 324--331.

\bibitem[\protect\citeauthoryear{Pop-Eleches}{Pop-Eleches}{2006}]{pop2006impact}
\textsc{Pop-Eleches, C.} (2006): \enquote{The Impact of an Abortion Ban on
  Socioeconomic Outcomes of Children: Evidence from Romania,} \emph{Journal of
  Political Economy}, 114, 744--773.

\bibitem[\protect\citeauthoryear{Reardon}{Reardon}{2018}]{reardon2018abortion}
\textsc{Reardon, D.~C.} (2018): \enquote{The Abortion and Mental Health
  Controversy: a Comprehensive Literature Review of Common Ground Agreements,
  Disagreements, Actionable Recommendations, and Research Opportunities,}
  \emph{SAGE Open Medicine}, 6, 2050312118807624.

\bibitem[\protect\citeauthoryear{Rellstab, Bakx, and
  Garc{\i}a-G{\'o}mez}{Rellstab et~al.}{2021}]{rellstab2021effect}
\textsc{Rellstab, S., P.~Bakx, and P.~Garc{\i}a-G{\'o}mez} (2021): \enquote{The
  Effect of a Miscarriage on Mental Health, Labour Market, and Family
  Outcomes,} \emph{Unpublished working paper, Erasmus University Rotterdam}.

\bibitem[\protect\citeauthoryear{Sant’Anna and Zhao}{Sant’Anna and
  Zhao}{2020}]{sant2020doubly}
\textsc{Sant’Anna, P.~H. and J.~Zhao} (2020): \enquote{Doubly robust
  difference-in-differences estimators,} \emph{Journal of Econometrics}, 219,
  101--122.

\bibitem[\protect\citeauthoryear{Santelli, Grilo, Lindberg, Speizer, Schalet,
  Heitel, Kantor, Ott, Lyon, Rogers, Heck, and Mason-Jones}{Santelli
  et~al.}{2017}]{santelli2017abstinence}
\textsc{Santelli, J., S.~A. Grilo, L.~D. Lindberg, I.~Speizer, A.~Schalet,
  J.~Heitel, L.~Kantor, M.~A. Ott, M.~Lyon, J.~Rogers, C.~J. Heck, and A.~J.
  Mason-Jones} (2017): \enquote{Abstinence-Only-Until-Marriage Policies and
  Programs: An Updated Position Paper of the Society for Adolescent Health and
  Medicine,} \emph{The Journal of adolescent health: official publication of
  the Society for Adolescent Medicine}, 61, 400.

\bibitem[\protect\citeauthoryear{Sedgh, Finer, Bankole, Eilers, and
  Singh}{Sedgh et~al.}{2015}]{sedgh2015adolescent}
\textsc{Sedgh, G., L.~B. Finer, A.~Bankole, M.~A. Eilers, and S.~Singh} (2015):
  \enquote{Adolescent Pregnancy, Birth, and Abortion Rates across Countries:
  Levels and Recent Trends,} \emph{Journal of Adolescent Health}, 56, 223--230.

\bibitem[\protect\citeauthoryear{{Socialstyrelsen Sweden}}{{Socialstyrelsen
  Sweden}}{2010}]{abortsweden09}
\textsc{{Socialstyrelsen Sweden}} (2010): \enquote{Aborter 2009,}
  \url{https://www.socialstyrelsen.se/globalassets/sharepoint-dokument/artikelkatalog/statistik/2010-5-12.pdf},
  last access: 26/2/2021.

\bibitem[\protect\citeauthoryear{{Socialstyrelsen Sweden}}{{Socialstyrelsen
  Sweden}}{2020}]{abortsweden}
---\hspace{-.1pt}---\hspace{-.1pt}--- (2020): \enquote{Statistik om aborter
  2019,}
  \url{https://www.socialstyrelsen.se/globalassets/sharepoint-dokument/artikelkatalog/statistik/2020-6-6806.pdf},
  last access: 26/2/2021.

\bibitem[\protect\citeauthoryear{{Statistics Sweden}}{{Statistics
  Sweden}}{2016}]{SCB2009}
\textsc{{Statistics Sweden}} (2016): \enquote{Background facts 2016:1
  integrated database for labour market research,}
  \url{https://www.scb.se/contentassets/f0bc88c852364b6ea5c1654a0cc90234/dokumentation-av-lisa.pdf},
  last access: 26/2/2021.

\bibitem[\protect\citeauthoryear{Steinberg, Laursen, Adler, Gasse, Agerbo, and
  Munk-Olsen}{Steinberg et~al.}{2018}]{steinberg2018examining}
\textsc{Steinberg, J.~R., T.~M. Laursen, N.~E. Adler, C.~Gasse, E.~Agerbo, and
  T.~Munk-Olsen} (2018): \enquote{Examining the Association of Antidepressant
  Prescriptions with First Abortion and First Childbirth,} \emph{JAMA
  Psychiatry}, 75, 828--834.

\bibitem[\protect\citeauthoryear{Steinberg and Russo}{Steinberg and
  Russo}{2008}]{steinberg2008abortion}
\textsc{Steinberg, J.~R. and N.~F. Russo} (2008): \enquote{Abortion and
  anxiety: what's the relationship?} \emph{Social Science \& Medicine}, 67,
  238--252.

\bibitem[\protect\citeauthoryear{Steingrimsdottir}{Steingrimsdottir}{2016}]{steingrimsdottir2016reproductive}
\textsc{Steingrimsdottir, H.} (2016): \enquote{Reproductive Rights and the
  Career Plans of US College Freshmen,} \emph{Labour Economics}, 43, 29--41.

\bibitem[\protect\citeauthoryear{Sutter, Kocher, Gl{\"a}tzle-R{\"u}tzler, and
  Trautmann}{Sutter et~al.}{2013}]{sutter2013impatience}
\textsc{Sutter, M., M.~G. Kocher, D.~Gl{\"a}tzle-R{\"u}tzler, and S.~T.
  Trautmann} (2013): \enquote{{Impatience and Uncertainty: Experimental
  Decisions Predict Adolescents' Field Behavior},} \emph{American Economic
  Review}, 103, 510--31.

\bibitem[\protect\citeauthoryear{Sydsj{\"o}, Sydsj{\"o}, Bladh, and
  Josefsson}{Sydsj{\"o} et~al.}{2014}]{sydsjo2014reimbursement}
\textsc{Sydsj{\"o}, A., G.~Sydsj{\"o}, M.~Bladh, and A.~Josefsson} (2014):
  \enquote{Reimbursement of hormonal contraceptives and the frequency of
  induced abortion among teenagers in Sweden,} \emph{BMC Public Health}, 14,
  1--7.

\bibitem[\protect\citeauthoryear{Tertilt and van~den Berg}{Tertilt and van~den
  Berg}{2015}]{tertilt2015association}
\textsc{Tertilt, M. and G.~J. van~den Berg} (2015): \enquote{The Association
  between own Unemployment and Violence Victimization among Female Youths,}
  \emph{Jahrb{\"u}cher f{\"u}r National{\"o}konomie und Statistik}, 235,
  499--516.

\bibitem[\protect\citeauthoryear{Turon, Carey, Boyes, Hobden, Dilworth, and
  Sanson-Fisher}{Turon et~al.}{2019}]{turon2019agreement}
\textsc{Turon, H., M.~Carey, A.~Boyes, B.~Hobden, S.~Dilworth, and
  R.~Sanson-Fisher} (2019): \enquote{Agreement between a single-item measure of
  anxiety and depression and the Hospital Anxiety and Depression Scale: A
  cross-sectional study,} \emph{PloS one}, 14, e0210111.

\bibitem[\protect\citeauthoryear{van~den Berg and Siflinger}{van~den Berg and
  Siflinger}{2021}]{vandenBerg2018}
\textsc{van~den Berg, G.~J. and B.~M. Siflinger} (2021): \enquote{The Effects
  of a Daycare Reform on Health in Childhood--Evidence from Sweden,}
  \emph{Journal of Health Economics}, 102577.

\bibitem[\protect\citeauthoryear{Young}{Young}{2021}]{young2021leverage}
\textsc{Young, A.} (2021): \enquote{Leverage, Heteroskedasticity and
  Instrumental Variables in Practical Application,} \emph{Unpublished working
  paper, London School of Economics}.

\end{thebibliography}
\bibliographystyle{ecta}
\clearpage

\end{document}